\documentclass[prd,preprint,showpacs,eqsecnum]{revtex4}

\usepackage{amsfonts}
\usepackage{amssymb}
\usepackage{graphicx}
\usepackage{pstricks}
\usepackage{axodraw}

\newcommand{\be}{\begin{equation}}
\newcommand{\bea}{\begin{eqnarray}}
\newcommand{\ee}{\end{equation}}
\newcommand{\eea}{\end{eqnarray}}
\newcommand{\bpi}{\begin{picture}}
\newcommand{\bce}{\begin{center}}
\newcommand{\epi}{\end{picture}}
\newcommand{\ece}{\end{center}}

\newcommand{\gp}{\Gamma^{\rm P}}

\newcommand{\Qpsm}{Q\hspace{-0.24cm}/\,'}
\newcommand{\Qsm}{Q\hspace{-0.24cm}/}
\newcommand{\ksm}{k\hspace{-0.24cm}/}
\newcommand{\qsm}{q\hspace{-0.22cm}/}

\newcommand{\sw}{s_{\rm{\scriptscriptstyle{W}}}}
\newcommand{\cw}{c_{\rm{\scriptscriptstyle{W}}}}
\newcommand{\gw}{g_{\rm{\scriptscriptstyle{W}}}}
\newcommand{\tw}{\theta_{\rm{\scriptscriptstyle{W}}}}
\newcommand{\Mw}{M_{\scriptscriptstyle{W}}}
\newcommand{\Mz}{M_{\scriptscriptstyle{Z}}}
\newcommand{\Mi}{M_{\scriptscriptstyle{{\cal V}^i}}}
\newcommand{\PL}{P_{\rm{\scriptscriptstyle{L}}}}
\newcommand{\PR}{P_{\rm{\scriptscriptstyle{R}}}}

\newcommand{\dw}{d_{\scriptscriptstyle{W}}}
\newcommand{\dnu}{d_{\scriptscriptstyle{{\cal V}^i}}}

\def\g{{\rm I}\hspace{-0.07cm}\Gamma}

\def\r#1{(\ref{#1})}

\begin{document}

\title{The Two-Loop Pinch Technique \\ 
in the Electroweak Sector}
\date{April 29, 2002}

\author{Daniele Binosi}
\author{Joannis Papavassiliou}
\affiliation{Departamento de F\'\i sica Te\'orica and IFIC, Centro Mixto, 
Universidad de Valencia-CSIC,
E-46100, Burjassot, Valencia, Spain}

\email{Daniele.Binosi@uv.es; Joannis.Papavassiliou@uv.es}

\begin{abstract}

The generalization of the  two-loop Pinch Technique to the Electroweak
Sector of the  Standard Model is presented.  We  restrict ourselves to
the  case  of conserved  external  currents,  and  provide a  detailed
analysis  of  both  the  charged  and neutral  sectors.   The  crucial
ingredient for  this construction is  the identification of  the parts
discarded   during   the    pinching   procedure   with   well-defined
contributions  to   the  Slavnov-Taylor  identity   satisfied  by  the
off-shell one-loop gauge-boson vertices;  the latter are nested inside
the conventional two-loop self-energies.   It is shown by resorting to
a  set  of  powerful  identities  that the  two-loop  effective  Pinch
Technique self-energies coincide  with the corresponding ones computed
in  the Background  Feynman gauge.  The aforementioned  identities are
derived in the context  of the Batalin-Vilkovisky formalism,
a fact which enables the  individual treatment of the self-energies of
the  photon   and  the  $Z$-boson.    Some  possible  phenomenological
applications are briefly discussed.

\end{abstract}


\preprint{FTUV-02-0429}
\preprint{IFIC-02-11}

\maketitle

\section{\label{sec:zero} Introduction}

In  strongly coupled  theories such  as  QCD the  need for  addressing
non-perturbative  phenomena  in the  continuum  through  the study  of
Schwinger-Dyson equations has motivated  the invention 
\cite{Cornwall:1982zr} and further development  \cite{Cornwall:1989gv} 
of  the diagrammatic  method known  as  the Pinch
Technique (PT), in an  attempt to device a self-consistent, physically
meaningful truncation  scheme.  The fundamental  underlying problem is
that  off-shell  Green's  function  are  in  general  unphysical,  and
reliable information  may be  extracted from them  only when  they are
combined  to form observables  order-by-order in  perturbation theory,
which  is  certainly not  the  case  when  dealing with  intrinsically
non-perturbative equations.  The PT reorganizes systematically a given
physical amplitude into sub-amplitudes,  which have the same kinematic
properties   as   conventional   $n$-point  functions,   (propagators,
vertices, boxes) but, in addition, are endowed with desirable physical
properties.  Most importantly, at  one-loop order they are independent
of  the   gauge-fixing  parameter  and   satisfy  naive,  (ghost-free)
tree-level  Ward  Identities (WIs),  
instead  of  the  usual  Slavnov-Taylor Identities (STIs) 
\cite{Slavnov:1972fg}.   
These,  and  other  important properties,  are  realized
diagrammatically by  exploiting the elementary Ward  identities of the
theory in order to enforce crucial cancellations.  Given their special
properties the  PT $n$-point functions  could thus serve, at  least in
principle, as the new building blocks of an improved set of manifestly
gauge-invariant  Schwinger-Dyson  equations,  a task  which,  however,
still remains incomplete.

On the  other hand,  the generalization of  the PT to  the Electroweak
sector            of           the            Standard           Model
\cite{Papavassiliou:1990zd,Degrassi:1992ue,Papavassiliou:1994pr}, 
has given rise to various
applications.  In  particular, the physics of  unstable particles, and
the  computation  of  resonant  transition  amplitudes  has  attracted
significant   attention  in   recent   years,  because   it  is   both
phenomenologically    relevant     and    theoretically    challenging
\cite{Pilaftsis:1989zt}.  At one-loop order, the resummation formalism
based  on  the  PT  \cite{Papavassiliou:1995fq} has  accomplished  the
simultaneous reconciliation  of crucial physical  requirements such as
gauge-fixing       parameter      independence,      gauge-invariance,
renormalization-group  invariance,  and  the optical  and  equivalence
theorems  \cite{Cornwall:1974km}.   Thus,  the  Breit-Wigner  type  of
propagators so constructed give  rise to Born-improved amplitudes free
of any unphysical artifacts.
Other applications include 
the correct definition of off-shell form-factors 
for extracting the anomalous $A W W$ and 
$ZWW$ couplings \cite{Papavassiliou:ex}, 
the derivation of a gauge-invariant and process-independent 
neutrino charged radius 
\cite{Papavassiliou:1990zd,Bernabeu:2000hf},  
the  gauge-invariant formulation 
\cite{Degrassi:1993kn} of the STU parameters \cite{Peskin:1991sw},
the 
manifestly gauge- and renormalization-group-invariant 
formulation of the precision electroweak corrections
\cite{Hagiwara:1994pw}, the unambiguous definition 
of the universal part of the 
two-loop 
$\rho$ parameter \cite{Papavassiliou:1995hj},
the gauge-invariant formulation of 
resonant CP violation 
\cite{Pilaftsis:1996ac}, and the 
resolution of issues related to gauge- and scheme-dependence
of mixing matrix renormalization 
\cite{Yamada:2001px,Pilaftsis:2002nc}. 

The generalization  of the  PT beyond one-loop  has been  presented in
\cite{Papavassiliou:1999az}  for  the  case  of  massless  Yang-Mills.
However,  its extension to  the Electroweak  Sector has  been pending,
mainly  due to  the following  two  reasons: First,  at the  technical
level, the direct application of the diagrammatic construction used in
the QCD  case to the Electroweak  Sector, due to  the proliferation of
Feynman diagrams  in the  latter, would lead  to a  major book-keeping
challenge. Second,  at the conceptual  level, the modification  of the
STIs  used  in   intermediate  steps,   and  in
particular  the non-transversality  of the  gauge-boson self-energies,
complicates   further  the   construction,  and   requires  additional
theoretical input, not needed in the QCD case.

Recently however significant progress has been made in formulating the
PT non-diagrammatically, leading to  an enormous simplification of the
operational aspects of the PT construction, by allowing the collective
treatment of entire  sets of Feynman graphs, instead  of the algebraic
manipulation of individual graphs \cite{Binosi:2002ez}.  There are two
basic facts which have enabled the aforementioned improvement: First,
it has been realized that the PT constructions amounts to the judicious
reallocation   of  well-defined   contributions   generated  when   the
longitudinal momenta  circulating inside the one-  and two-loop graphs
trigger the  STI of the  three-gluon vertex (tree-level  and one-loop,
respectively),   which   is  {\it   nested}   in  the   aforementioned
graphs. Thus, the parts of  the one-loop and two-loop Feynman diagrams
that   are   shuffled  around   during   the   pinching  process   are
systematically identified  in terms of  well-defined field-theoretical
objects,  namely  the ghost  Green's  functions  which  appear in  the
aforementioned  STI. 
Second, the  task of  comparing the  resulting PT
effective Green's functions to those  computed in the Feynman gauge of
the  Background Field  Method (BFM) \cite{Dewitt:ub}, in order to
verify whether the known correspondence 
\cite{Denner:1994nn,Papavassiliou:1999az}
persists in the two-loop Electroweak case,   
is significantly  facilitated by
resorting  to   a  set   of  non-trivial  identities,   the  so-called
Background-Quantum  Identities (BQIs), which  relate the  BFM $n$-point
functions  to  the   corresponding  conventional  $n$-point  functions
computed  in the  covariant renormalizable  gauges, to  all  orders in
perturbation  theory. These  BQIs are  derived  in the  context of  the
Batalin-Vilkovisky  (BV)  formalism \cite{Batalin:jr},  
and  contain  auxiliary  Green's
functions of ``anti-fields'' and background sources.

In this paper we generalize the two-loop PT in the case of the 
Electroweak  Sector by resorting to the aforementioned 
theoretical ingredients. In particular, we focus on the 
``intrinsic PT construction'', which represent a  
more economical alternative 
to the usual, more laborious, explicit ``S-matrix'' PT.  
We carry out the PT construction both for the charged and 
the neutral sector, thus defining the PT two-loop self-energies
for the $W$- and $Z$-bosons respectively. To simplify 
the construction, without compromising the novel features 
we want to address, we restrict ourselves to the case where 
the external (charged and neutral) currents are conserved, {\it i.e.},
the external on-shell fermions are considered to be massless. 
One of the main 
ingredients of the two-loop construction
are the STIs satisfied by the off-shell three-gauge-boson vertices 
appearing nested inside the $WW$, $ZZ$, $AA$, and $AZ$
self-energies, {\it i.e.}, the  vertices $WWZ$,  and $WWA$. 
These STI are triggered by longitudinal momenta originating 
from other {\it elementary}   
three-gauge-boson vertices, appearing inside the 
same Feynman graphs.
The STIs employed are directly derived in the framework of the 
BV formalism, which allows for an elegant unified 
treatment, and, in addition, facilitates the task of comparing the 
resulting PT gauge-boson self-energies to those of the BFM.
Specifically, the BV formalism applied in the Electroweak Sector,
and for the particular objectives we would like to achieve,
proves more suitable  
than the Zinn-Justin approach~\cite{ZJ}, usually employed in the literature.
The basic advantage of the BV approach in the present context 
is that, by treating on equal 
footing the photon and the $Z$, allows one to disentangle 
the BQI for the photon self-energy from the corresponding BQI for the 
$Z$-boson self-energy. Thus, one may compare  the PT and BFM expressions
for each of the two self-energies separately.  
Instead, the Zinn-Zustin approach 
yields a BQI involving both self-energies in a single expression, 
which, even though is sufficient for addressing issues of 
renormalization, is not particularly helpful to our purposes.

The paper is organized as follows: In Section~\ref{sec:one}
we present 
the BV formalism for the case of the Electroweak Sector
of the Standard Model. In Section~\ref{sec:BI} we derive the 
basic ingredients, which will allow us both the definition of 
the two-loop intrinsic PT self-energies, and their easy comparison
to the corresponding quantities defined in the Feynman gauge of 
the BFM. In particular, we derive the STI for the three-gauge-boson 
vertices, and the {\it disentangled} BQIs for the gauge-boson 
self-energies. In section~\ref{sec:1l} we use the known one-loop 
results derived in the context of the ``$S$-matrix'' PT
in order to 
familiarize ourselves with the correspondence
between the two formalisms. In Section~\ref{sec:IP} 
we prepare the stage for the two-loop construction 
by studying the one-loop intrinsic PT, by explaining  
the role of the STI, which in this case coincides with the
usual tree-level WI satisfied by the bare three-gauge-boson 
vertex. The next three sections are rather technical, 
and contain  the main result of our paper, namely the 
Electroweak two-loop intrinsic PT construction. In particular,  
in Section~\ref{sec:IP2} present the general philosophy 
and methodology, which is subsequently applied 
in detail in the charged  (Section~\ref{sec:2lcs}) 
and neutral (Section~\ref{sec:2lns}) sectors. 
In Section~\ref{sec:concl} we present our conclusions, whereas
Feynman rules for constructing perturbatively the auxiliary Green's 
functions appearing in the BQIs are listed in the final 
Appendix. 

\section{\label{sec:one} The BV Formalism in the Electroweak Sector}

In this section we will briefly review the most salient features of
the BV formalism \cite{Batalin:jr} 
as it applies to the Electroweak Sector. 
In order to define the relevant quantities and set up the
notation used throughout the paper, we begin by writing the classical
(gauge invariant) Standard Model Lagrangian as
\be
{\cal L}^{\rm cl}_{\rm SM}={\cal L}_{\rm YM}+{\cal L}_{\rm H}.
\ee
The gauge invariant 
${SU}(2)_{\scriptscriptstyle{W}}
\otimes{U}(1)_{\scriptscriptstyle{Y}}$ 
Yang-Mills
part ${\cal L}_{\rm YM}$ consists of an isotriplet $W^a_\mu$ (with
$a={1,2,3}$) associated with the weak isospin generators
$T^a_{\scriptscriptstyle{W}}$,
an isosinglet $W^4_\mu$ with weak hypercharge $Y_{\scriptscriptstyle{W}}$
associated to the group factor $U(1)_{\scriptscriptstyle{Y}}$; it reads
\bea
{\cal L}_{\rm YM}&=&-\frac14F^a_{\mu\nu}F^{a\,\mu\nu} \nonumber \\
&=& -\frac14\left(\partial_\mu W^a_\mu-\partial_\nu W^a_\mu+\gw
f^{abc}W^b_\mu W^c_\nu\right)^2-\frac14\left(\partial_\mu W^4_\nu
-\partial_\nu W^4_\mu\right)^2+{\cal L}_{\psi}. 
\eea 
The Higgs-boson part ${\cal L}_{\rm H}$ involves a complex 
${SU}(2)_{\scriptscriptstyle{W}}$
scalar doublet field $\varphi$ and its complex conjugate
$\varphi^\dagger$ given by
\be
\varphi=\left(
\begin{array}{c} 
\phi^+\\
\frac1{\sqrt2}\left(H+i\chi\right)
\end{array}
\right),
\hspace{2.5cm}
\varphi^\dagger=\left(
\begin{array}{c} 
\frac1{\sqrt2}\left(H-i\chi\right)\\
-\phi^-
\end{array}
\right).
\ee
Here $H$ denotes the physical Higgs field while $\phi^\pm$ and $\chi$
represents respectively charged and neutral unphysical degrees of freedom.
Then ${\cal L}_{\rm H}$ takes the form
\be
{\cal L}_{\rm
H}=\left(D_\mu\varphi\right)^\dagger\left(D^\mu\varphi\right)-V(\varphi)
\ee
with the covariant derivative $D_\mu$ defined as
\be
D_\mu=\partial_\mu-i\gw T^a_{\rm W}W^a_\mu+ig_1\frac{Y_{\rm
W}}2W^4_\mu
\ee
and the Higgs potential as
\be
V(\varphi)=\frac\lambda4\left(\varphi^\dagger\varphi\right)^2
-\mu^2\left(\varphi^\dagger\varphi\right).
\ee

The Higgs field $H$ will give mass to all the Standard Model
fields, by acquiring
a vacuum expectation value 
$v$; in particular the masses of the gauge fields are generated
after absorbing the massless would-be Goldstone bosons $\phi^\pm$ and
$\chi$. The physical massive gauge-bosons $W^\pm,\,Z$ and the photon $A$ 
are then obtained by diagonalizing the mass matrix, and reads
\be
W^\pm_\mu=\frac1{\sqrt2}\left(W^1_\mu\mp iW^2_\mu\right),
\hspace{1.5cm}
\left(
\begin{array}{c}
Z_\mu \\
A_\mu
\end{array}
\right)=
\left(
\begin{array}{cc}
\cw & \sw \\
-\sw & \cw
\end{array}\right)
\left(
\begin{array}{c}
W^3_\mu \\
W^4_\mu
\end{array}
\right),
\ee
where 
\be
\cw=\cos\tw=\frac{\gw}{\sqrt{g_1^2+\gw^2}}=
\frac{M_{\scriptscriptstyle{W}}}{M_{\scriptscriptstyle{Z}}},
\hspace{1.5cm}
\sw=\sin\tw=\sqrt{1-\cw^2},
\ee
and $\tw$ is the weak mixing angle. 
Finally, to the Lagrangian  ${\cal L}^{\rm cl}_{\rm SM}$
must be added 
the matter Lagrangian ${\cal L}_{\psi}$; its explicit form 
may be found in 
\cite{Denner:1995xt}, but is not important for what follows.  

For quantizing the theory, a gauge fixing term must be added to the
classical Lagrangian ${\cal L}^{\rm cl}_{\rm SM}$. 
To avoid tree-level mixing between gauge and
scalar fields, a renormalizable $R_\xi$ gauge of the 't Hooft type is
most commonly chosen; this is specified by one gauge parameter for
each gauge-boson, and defined through the linear gauge fixing
functions
\bea
F^\pm&=&\partial^\mu W^\pm_\mu\mp i\xi_{\scriptscriptstyle W}\Mw\phi^\pm,
\nonumber \\
F^{\scriptscriptstyle Z}&=&
\partial^\mu Z_\mu- i\xi_{\scriptscriptstyle Z}\Mz\chi,
\nonumber \\
F^{\scriptscriptstyle A}&=&
\partial^\mu A_\mu, 
\eea
yielding to the $R_\xi$ gauge fixing Lagrangian
\be
{\cal L}_{\rm GF}=\left(\xi_{\scriptscriptstyle
W}B^+B^-+B^+F^-+B^+F^-\right)
+\left[\frac12\xi_{\scriptscriptstyle Z} \left(B^{\scriptscriptstyle
Z}\right)^2+B^{\scriptscriptstyle Z}F^{\scriptscriptstyle Z}\right] 
+\left[\frac12\xi_{\scriptscriptstyle A} 
\left(B^{\scriptscriptstyle A}\right)^2+
B^{\scriptscriptstyle A}F^{\scriptscriptstyle A}\right]. 
\ee

The fields $B^\pm,\ B^{\scriptscriptstyle Z}$ and
$B^{\scriptscriptstyle A}$ represent auxiliary, non propagating
fields: they are the so called Nakanishy-Lautrup Lagrange multipliers
for the gauge condition, and they can be eliminated through their
equations of motion
\be
B^\pm= -\frac1{\xi_{\scriptscriptstyle W}}F^\pm, \qquad
B^{\scriptscriptstyle Z}= -\frac1{\xi_{\scriptscriptstyle Z}}
F^{\scriptscriptstyle Z}, \qquad
B^{\scriptscriptstyle A}= -\frac1{\xi_{\scriptscriptstyle A}}
F^{\scriptscriptstyle A},
\ee
leading to the usual gauge fixing Lagrangian
\be
{\cal L}_{\rm GF}=-\frac1{\xi_{\scriptscriptstyle W}}F^+F^--
\frac1{2\xi_{\scriptscriptstyle Z}}\left(F^{\scriptscriptstyle Z}\right)^2
-\frac1{2\xi_{\scriptscriptstyle A}}\left(F^{\scriptscriptstyle
A}\right)^2.
\ee

The corresponding Faddeev-Popov ghost sector reads then
\be
{\cal L}_{\rm FPG}=-\sum_{n,m}\bar u^n\frac{\delta
F^n}{\delta\theta^m}u^m,
\ee
where $n,m=\pm,Z,A$, while $\delta F^n/\delta\theta^m$ denotes the
variation of the gauge fixing functions under an infinitesimal gauge
transformation (with gauge parameters $\theta^m$).

The complete Standard Model Lagrangian in the $R_\xi$ gauges reads then
\be
{\cal L}_{\rm SM}={\cal L}^{\rm cl}_{\rm SM}+
{\cal L}_{\rm GF}+{\cal L}_{\rm FPG}.
\label{SMlag}
\ee
The full set of Feynman rules derived from this Lagrangian
(together with the BFM gauge fixing procedure and the corresponding
Feynman rules) can be found in \cite{Denner:1995xt}, and will be used
throughout the paper. 

The starting point of the BV formalism
is the introduction of an {\it external}
field -- called anti-field --
$\Phi^{*,n}$ for {\it each} 
field  $\Phi^n$ appearing in the Lagrangian, regardless of its 
transformation properties under the Becchi-Rouet-Stora-Tyutin (BRST)
symmetry \cite{Becchi:1974md}.  
This is to be contrasted with the approach proposed 
by Zinn-Justin \cite{ZJ}, where one introduces anti-fields 
only for fields transforming non-linearly under the 
BRST transformation.
This would mean that one would introduce the anti-field
$W^{*,3}$ for the $W^3$ combination of the physical fields $Z$ and
$A$, but no anti-field $W^{*,4}$ for the $W^4$ combination, which
transforms linearly.
Moreover, no Weinberg rotation for the anti-fields should be
introduced. While the Zinn-Justin approach encodes the 
necessary information for addressing issues of 
renormalization in 
the Standard Model, the STIs and BQIs of the neutral 
sector become entangled, 
and it is not evident how to extract
the identities needed  
for constructing the {\it individual}  PT two-loop
self-energies (see Section \ref{sec:2lns}), 
and for subsequently comparing them  
with those of the BFM. The great advantage
of the BV formalism in the present context 
is that it treats on the same footing all the
gauge fields, avoiding the inconvenient 
entanglement of both the STIs as well as the BQIs. 
 
The anti-fields $\Phi^{*,n}$ will carry the same Bose/Fermi
statistic of the corresponding field $\Phi^n$ and a ghost number such that
\be
gh\,\{\Phi^{*,n}\}=-gh\,\{\Phi^{n}\}-1.
\ee
Thus, since the
ghost number is equal to $1$ for the ghost fields $u^n$, 
to $-1$ for the anti-ghost 
fields $\bar u^n$, 
and zero for the other fields, one has the assignment  
\be
gh\,\{V^{*,n}_\mu,u^{*,n},\bar u^{*,n},G^*,
\psi^{*,I},\bar\psi^{*,I}\}=\{-1,-2,0,-1,-1,-1\},
\ee
where we introduced the short-hand notation
\bea
V^n_\mu=(W^+_\mu,W^-_\mu,Z_\mu,A_\mu), &\qquad& 
V^{*,n}_\mu=(W^{*,+}_\mu,W^{*,-}_\mu,Z^*_\mu,A^*_\mu), \nonumber \\
u^n=(u^+,u^-,u^Z,u^A), &\qquad& u^{*,n}=(u^{*,+},u^{*,-},u^{*,Z},u^{*,A}),
\nonumber \\
G^n=(\phi^+,\phi^-,\chi,H), &\qquad& G^{*,n}=(\phi^{*,+},\phi^{*,-},
\chi^*,H^*). 
\eea
The original gauge invariant Lagrangian is then supplemented by a term 
coupling the fields $\Phi^{n}$ to the corresponding 
anti-fields $\Phi^{*,n}$, giving the modified Lagrangian
\bea
{\cal L}_{\rm BV}&=&{\cal L}_{\rm YM}+{\cal L}_{\rm H}+{\cal L}_{\rm
BRST} 
\nonumber \\
&=& {\cal L}_{\rm YM}+{\cal L}_{\rm H}
+\sum_{n}\Phi^{*,n}s\Phi^n, 
\eea
where $s$ is the BRST operator. The BRST transformations of
all the Standard Model fields can be found in \cite{Kraus:1998bi}.

The action $\g^{(0)}[\Phi,\Phi^*]$ which is built up from the new Lagrangian 
${\cal L}_{\rm BV}$ will 
satisfy the {\it master equation}
\be
\sum_n
\int\!d^4x\left[\frac{\delta\g^{(0)}}{\delta\Phi^{*,n}}\frac{\delta\g^{(0)}}  
{\delta\Phi^n}\right]=0,
\label{mecl}
\ee
which is just a consequence of the BRST invariance of the action and of the 
nilpotency of the BRST operator. 

Since the  anti-fields are  external fields we  must constrain  them to
suitable  values  before we  can  use  the  action $\g^{(0)}$  in  the
calculation of $S$-matrix elements. To this purpose one introduces an
arbitrary fermionic functional $\Psi[\Phi]$ (with
$gh\,\{\Psi[\Phi]\}=-1$) such that            
\be
\Phi^{*,n}=\frac{\delta\Psi[\Phi]}{\delta\Phi^n}.  
\ee
Then the action
becomes                                                            
\bea
\g^{(0)}[\Phi,\delta\Psi/\delta\Phi]&=&\g^{(0)}[\Phi]+\sum_n(s\Phi^n)
\frac{\delta\Psi[\Phi]}{\delta\Phi^n}\nonumber \\
&=&\g^{(0)}[\Phi]+s\Psi[\Phi], 
\eea 
{\it i.e.}, it is equivalent to the
gauge fixed action of the  Yang-Mills theory under scrutiny, since we
can choose the fermionic functional $\Psi$ to satisfy
\be
s\Psi[\Phi]=\int\!d^4x\left({\cal L}_{\rm GF}+{\cal L}_{\rm
FPG}\right).  
\ee
The fermionic functional $\Psi$ is often referred to
as the gauge fixing fermion. 

Moreover, since the anti-ghost anti-fields $\bar u^{*,n}$ and the
auxiliary fields $B^n$ have linear BRST transformations, they form
the so called {\it trivial pairs}: they enter, 
together with their anti-fields, bilinearly in the action
\be
\g^{(0)}[\Phi,\Phi^*]=\g^{(0)}_{\rm min}[V^n_\mu,u^n,G^n,V^{*,n}_\mu,u^{*,n},
G^{*,n}]-B^n\bar u^{*,n}.
\ee
The last term has no effect on the master equation, which will be in fact
satisfied by the {\it minimal} action $\g^{(0)}_{\rm min}$ alone. 
In what follows we will restrict our considerations to the minimal 
action (which depends on the {\it minimal variables}  
${V^n_\mu,u^n,G^n,V^{*,n}_\mu,u^{*,n},G^{*,n}}$), 
dropping the corresponding subscript. 

It is well  known that the BRST symmetry is  crucial for providing the
unitarity  of the $S$-matrix  and the  gauge independence  of physical
observables; thus it must be  implemented in the theory at all orders,
not only at the classical  level. This is provided by establishing the
quantum corrected version  of Eq.\r{mecl}, in the form  of the 
STI functional       
\bea      
{\cal
S}(\g)[\Phi,\Phi^*]&=&\sum_n\int\!d^4x\left[
\frac{\delta\g}{\delta\Phi^{*,n}}\frac{\delta\g} {\delta\Phi^n}\right]
\nonumber \\ 
&=&\sum_n\int\!d^4x \left[\frac{\delta\g}{\delta V^{*,n}_\mu}
\frac{\delta\g}{\delta V^{n}_\mu}
+\frac{\delta\g}{\delta u^{*,n}}
\frac{\delta\g}{\delta u^{n}}+
\frac{\delta\g}{\delta G^{*,n}}
\frac{\delta\g}{\delta G^{n}}\right. \nonumber \\
&+&\left.\sum_I\left(
\frac{\delta \g}{\delta\psi^{*,I}}\frac{\delta\g}{\delta\bar\psi^I}
+\frac{\delta \g}{\delta\psi^{I}}\frac{\delta\g}{\delta\bar\psi^{*,I}}
\right)\right] \nonumber \\
&=&\int\!d^4x \left[ 
\frac{\delta \g}{\delta W^{*,+}_\mu}\frac{\delta\g}{\delta  W^-_\mu}
+\frac{\delta \g}{\delta W^{*,-}_\mu}\frac{\delta\g}{\delta  W^+_\mu}
+\frac{\delta \g}{\delta u^{*,+}}\frac{\delta\g}{\delta  u^-}
+\frac{\delta \g}{\delta u^{*,-}}\frac{\delta\g}{\delta  u^+}
\right.\nonumber \\
&+&\frac{\delta  \g}{\delta  Z^*_\mu}\frac{\delta\g}{\delta   Z_\mu}
+\frac{\delta  \g}{\delta  u^{*,Z}}\frac{\delta\g}{\delta   u^Z}
+\frac{\delta  \g}{\delta  A^*_\mu}\frac{\delta\g}{\delta   A_\mu} 
+\frac{\delta  \g}{\delta  u^{*,A}}\frac{\delta\g}{\delta   u^A}
\nonumber \\
&+&\frac{\delta \g}{\delta\phi^{*,+}}\frac{\delta\g}{\delta\phi^-}
+\frac{\delta \g}{\delta\phi^{*,-}}\frac{\delta\g}{\delta\phi^+}
+\frac{\delta  \g}{\delta  \chi^*}\frac{\delta\g}{\delta\chi}
+\frac{\delta  \g}{\delta  H^*}\frac{\delta\g}{\delta   H}
\nonumber \\
&+&\left.\sum_I\left(
\frac{\delta \g}{\delta\psi^{*,I}}\frac{\delta\g}{\delta\bar\psi^I}
+\frac{\delta \g}{\delta\psi^{I}}\frac{\delta\g}{\delta\bar\psi^{*,I}}
\right)\right]
\nonumber \\
&=&0
\label{mequ}
\eea
where $\g[\Phi,\Phi^*]$ is now the effective action, and the sum is extended 
over all the Standard Model fermions. Eq.\r{mequ} gives rise to
the complete set of
 non linear STIs at all orders in the perturbative theory, via 
the repeated 
application of functional differentiation. Notice that $gh\,\{{\cal S}(\g)\}
=+1$, and that Green functions with non-zero ghost charge vanish, since it is 
a conserved quantity. This implies that for obtaining
 non-trivial identities it is 
necessary to differentiate the expression \r{mequ} with respect to one ghost 
field (ghost charge~+1), or with respect 
to two ghost fields and one anti-field 
(ghost charge $+2-1=+1$ again). For example, for deriving the STI satisfied by
the three-gauge-boson vertex, one has to differentiate Eq.\r{mequ}
with respect to two gauge-boson fields and one ghost field (see
Section~\ref{subSTI} below).

A technical remark is in order here. Recall that we have chosen to
work with the minimal generating functional $\g$, from which the
trivial pairs $({B^n,\bar u^{*,n}})$ has been removed 
\cite{Barnich:2000zw}. In the case of a linear gauge
fixing, such as the one at hand, this is equivalent to
working with the ``reduced'' functional $\g$, defined by subtracting
from the complete generating functional $\g^{\scriptscriptstyle{\rm
C}}$ the local term $\int\!d^4x{\cal L}_{\rm GF}$ corresponding to the
gauge fixing part of the Lagrangian. 
One should then keep in mind that the Green's functions
generated by the minimal effective action $\g$, or the complete one
$\g^{\rm C}$, are {\it not} equal \cite{Gambino:1999ai}. 
At tree-level, one has for example that
\bea
\g_{W^\pm_\mu W^\mp_\nu}^{(0)}(q)&=&\g^{{\rm C}\,(0)}_{W^\pm_\mu
W^\mp_\nu}(q)+\frac1{\xi_{\rm{\scriptscriptstyle{W}}}}q_\mu q_\nu \nonumber \\
&=&-i\left[\left(q^2-\Mw^2\right)g_{\mu\nu}-q_\mu q_\nu\right],
\nonumber  \\
\g_{\phi^\pm\phi^\mp}^{(0)}(q)&=&\g_{\phi^\pm\phi^\mp}^{{\rm C}\,(0)}+
\xi_{\rm{\scriptscriptstyle{W}}}\Mw^2. 
\label{uns}
\eea  
At higher orders the difference depends only on
the renormalization of the $W$ field and of the gauge parameter (and,
as such, is immaterial for our purposes). 
It should be noticed that, since we have eliminated the classical
gauge-fixing fermion from the generating functional $\g$, we allow for
tree-level mixing between the scalar and the gauge-boson sector: this
means that $\g_{\phi^\pm W^\mp_\mu}$ and $\g_{\chi Z}$ do not vanish
at tree-level. 
However the aforementioned mixing is not present at the propagator
level, {\it i.e.}, when these particles circulate in
loops~\cite{Barnich:2000zw,Gambino:1999ai,Grassi:1999tp}; therefore
loops must be computed using the usual Feynman rules of the $R_\xi$
gauges~\cite{Fujikawa:fe,Abers:qs}.

Another important ingredient of the construction we carry out in what follows 
is to write down the STI functional in the BFM. 
For doing this we introduce the set of classical vector and scalar fields
$\Omega^n_\mu$ and $\Omega^{G^n}$
\be
\Omega^n_\mu=(\Omega^+_\mu,\Omega^-_\mu,\Omega^Z_\mu,\Omega^A_\mu),
\qquad \Omega^{G^n}=(\Omega^+,\Omega^-,\Omega^\chi,\Omega^H),
\ee
which carry the same quantum numbers of the corresponding vector 
and scalar fields $V^n_\mu$ and $G^n$ respectively, but ghost charge
$+1$. Next, we 
implement the equations of motion of 
the background fields $\widehat{V}^n_\mu$ and $\widehat{G}^n$ at 
the quantum level 
by extending the BRST symmetry to them through the equations
\bea
s\widehat V^n_\mu=\Omega^n_\mu &\qquad&
s\Omega^n_\mu=0,
\nonumber \\
s\widehat G^n = \Omega^{G^n} & \qquad  &
s\Omega^{G^n}=0. 
\eea
Finally, in order to control the dependence of the Green's functions on the 
background fields we modify the STI functional of Eq.\r{mequ} as
\cite{Grassi:1999tp}
\bea
{\cal S}'(\g')[\Phi,\Phi^*]&=&{\cal S}(\g')[\Phi,\Phi^*]+\sum_n
\left[\Omega^n_\mu
\left(\frac{\delta\g}{\delta \widehat{V}^n_\mu}
- \frac{\delta\g}{\delta V^n_\mu}\right)
+\Omega^{G^n}
\left(\frac{\delta\g}{\delta \widehat{G^n}} - 
\frac{\delta\g}{\delta G^n}\right)\right] \nonumber \\
&=&{\cal S}(\g')[\Phi,\Phi^*]+\Omega^+_\mu
\left(\frac{\delta\g}{\delta \widehat{W}^+_\mu} - 
\frac{\delta\g}{\delta W^+_\mu}\right)+\Omega^-_\mu
\left(\frac{\delta\g}{\delta \widehat{W}^-_\mu} - 
\frac{\delta\g}{\delta W^-_\mu}\right)\nonumber \\
&+&\Omega^Z_\mu
\left(\frac{\delta\g}{\delta \widehat{Z}_\mu} - 
\frac{\delta\g}{\delta Z_\mu}\right) 
+\Omega^A_\mu
\left(\frac{\delta\g}{\delta \widehat{A}_\mu} - 
\frac{\delta\g}{\delta A_\mu}\right)+\Omega^+
\left(\frac{\delta\g}{\delta \widehat{\phi}^+} - 
\frac{\delta\g}{\delta \phi^+}\right) \nonumber \\
&+&\Omega^-
\left(\frac{\delta\g}{\delta \widehat{\phi}^-} - 
\frac{\delta\g}{\delta \phi^-}\right)+
\Omega^\chi
\left(\frac{\delta\g}{\delta \widehat{\chi}} - 
\frac{\delta\g}{\delta \chi}\right) 
+\Omega^H
\left(\frac{\delta\g}{\delta \widehat{H}} - 
\frac{\delta\g}{\delta H}\right),
\label{mequbfm}
\eea  
where $\g'$  denotes the  effective action  that depends  on the
background sources $\Omega^n_\mu$, and ${\cal S}(\g')[\Phi,\Phi^*]$ is
the  STI  functional  of   Eq.\r{mequ}.  Differentiation  of  the  STI
functional Eq.\r{mequbfm}  with respect  to the background  source and
background or quantum  fields,   will then  relate  1PI   functions  involving
background  fields  with  the   ones  involving  quantum  fields  (see
Section \ref{subBQI} below).

The final ingredient we need to know for the actual computation of STIs are 
the coupling of the anti-fields and 
background sources to the other fields of the 
theory. The former are controlled by the Lagrangian 
\bea
{\cal L}_{\rm BRST} &=&
W^{*,\pm}_{\mu}  \left\{\partial_{\mu} u^{\mp} \pm i \gw
\left(W^{\mp}_{\mu}+\widehat W^{\mp}_{\mu}\right) \left(\sw u^{A} -       
\cw u^{Z} \right) \right.\nonumber \\
&\mp& \left. i \gw u^{\mp}\left[       
\sw \left(A_{\mu}+\widehat A_{\mu}\right)  - 
\cw \left(Z_{\mu}+\widehat Z_{\mu}\right)\right] \right\} 
+u^{*,\pm} \left[\pm \frac{i \gw }{2} u^{\mp} \left( \sw u^{A} -       
\cw u^{Z} \right)\right]
\nonumber \\
& + & Z^*_\mu\left\{\partial_\mu u^Z-i\gw\cw\left[\left(W_\mu^++\widehat
W_\mu^+\right)u^-
-\left(W^-_\mu +\widehat
W_\mu^-\right)u^+\right]\right\}-u^{*,Z}\left(i\gw\cw u^-u^+\right)
\nonumber \\
& + & A^*_\mu\left\{\partial_\mu u^A+i\gw\sw\left[\left(W_\mu^++\widehat
W_\mu^+\right)u^-
-\left(W^-_\mu +\widehat
W_\mu^-\right)u^+\right]\right\}+u^{*,A}\left(i\gw\sw u^-u^+\right)
\nonumber \\
&+&
\phi^{*,\pm} \left\{ \mp \frac{i \gw}{2}\left[
\left(H+\widehat H\right) + v \mp i\left(\chi+\widehat\chi\right)\right]
u^{\mp}\right. \nonumber \\
&\pm& \left. i \gw\left(\phi^{\mp}+\widehat\phi^{\mp}\right)
\left(\sw u^{A}-\frac{\cw^2 - \sw^2}{2 \cw} u^{Z}      
\right) \right\}\nonumber \\
&+&
\chi^* \left\{\frac{\gw}{2} \left[\left(\phi^{+}+\widehat\phi^{+}\right)
 u^{-} + \left(\phi^{-}+\widehat  
\phi^{-} \right)u^{+} \right] 
-\frac{\gw}{2 \cw}\left[\left(H+\widehat H\right) + v \right] u^{Z} \right\} 
\nonumber \\
&+& H^* \left\{\frac{i \gw}{2} \left[\left(\phi^{+}+\widehat\phi^{+}\right)
 u^{-} - \left(\phi^{-}+\widehat  
\phi^{-} \right)u^{+} \right]
+ \frac{\gw}{2 \cw}\left(\chi+\widehat\chi\right)u^{Z} \right\}
\nonumber \\
&+&{\cal L}_{\rm BRST}^\psi,
\label{BRSTL}
\eea 
where ${\cal L}_{\rm BRST}^\psi$ stands for the term involving fermions
and can be found in  \cite{Grassi:1999tp}.
The Lagrangian ${\cal
L}_\Omega$ coupling the background sources with the Standard Model fields is
identical to the above one upon the replacement of the anti-fields for
background sources and ghost fields for anti-ghost fields. Notice that 
all necessary Feynman rules coming from ${\cal L}_{\rm BRST}$ and ${\cal
L}_\Omega$ are listed in Appendix~\ref{FR}.

\section{\label{sec:BI}The basic tools: STI$\rm{\bf s}$ and BQI$\rm {\bf s}$}

After  having  reviewed  the  BV   formalism  as  it  applies  to  the
electroweak sector  of the Standard  Model, we next proceed  to derive
the basic  ingredients needed for the PT  construction.  In particular
we will focus on two aspects: ({\it i}) the derivation of the STIs for
the off-shell  propagators and three-gauge-boson vertices;  as we will
see these STIs are of  central importance for the intrinsic PT method,
to  be presented  in Sections~\ref{sec:IP}  and~\ref{sec:IP2}. ({\it
ii}) the  derivation of the  BQIs relating the background  and quantum
two-   and  three-point   functions.    These  identities   facilitate
significantly the eventual comparison between the effective PT Green's
functions and the BFM Green's functions, computed at $\xi_Q = 1$.  The
crucial point  is that the conventional Green's  functions are related
to the BFM ones by means of  the same type of building blocks as those
that appear  in the STIs  of the three-gauge-boson vertex,  derived in
({\it i}), namely auxiliary, unphysical Green's functions.

\subsection{\label{subSTI} STIs}

In the BV formalism, all the STIs satisfied by the 1PI $n$-point
functions can be derived by appropriate functional differentiation of
the STI functional of Eq.\r{mequ}.

\subsubsection{Gauge-boson two-point functions}

The STI satisfied by the gauge-bosons two-point functions
$\g_{V^n_\mu V^m_\nu}$ can be obtained by considering the following
functional differentiation
\be
\left.\frac{\delta^2{\cal S}(\g)}
{\delta u^i(p_1)\delta V^j_\beta(q)}
\right|_{\Phi=0}=0 \qquad q+p_1=0,
\ee
which will provide us with the STI
\be
\sum_n\left[\g_{u^i V^{*,n}_\rho}(q)\g_{V^{n,\rho} V^j_\beta}(q)+
\g_{u^i G^{*,n}}(q)\g_{G^n V^j_\beta}(q)\right]=0,
\label{STItwopoint}
\ee
where (as always from now on) the sum over $n$ is constrained by
charge conservation.  
Different values of the indices $i$ and $j$ will determine which of
the various possible STIs implicit in Eq.\r{STItwopoint} we are
considering. For example, the STI satisfied by the $W$s two-point
function is obtained by choosing $i=\pm,\ j=\mp$, and reads
\be
\g_{u^\pm W^{*,\mp}_\rho}(q)\g_{W^{\pm,\rho} W^\mp_\beta}(q)+
\g_{u^\pm \phi^{*,\mp}}(q)\g_{\phi^\pm W^\mp_\beta}(q)=0.
\label{STItwopointW}
\ee

Choosing instead $i=Z,A$, and letting 
$V^j_\beta\equiv{\cal V}^j_\beta$ with $j=Z,A$, so that 
${\cal V}^j_\beta=\left(Z_\beta,A_\beta\right)$ with $M_{{\cal
V}^j}=(M_{\scriptscriptstyle{Z}},0)$, we obtain the STI
satisfied by the neutral gauge-bosons two-point functions, {\it i.e.},
\be
\sum_n
\g_{u^i {\cal V}^{*,n}_\rho}(q)\g_{{\cal V}^{n,\rho} {\cal V}^j_\beta}(q)+
\g_{u^i H^{*}}(q)\g_{H {\cal V}^j_\beta}(q) 
+\g_{u^i \chi^{*}}(q)\g_{\chi {\cal V}^j_\beta}(q)=0.
\label{STItwopointZA}
\ee

\subsubsection{Gauge-boson three-point functions}

The STI satisfied by the three-gauge-boson vertex, can be derived by
considering the following functional differentiation:
\be
\left.\frac{\delta^3{\cal S}(\g)}
{\delta u^k(q_1)\delta V^i_\alpha(q_2)\delta V^j_\beta(q_3)}
\right|_{\Phi=0}=0 \qquad q_1+q_2+q_3=0,
\ee
which in turn will give us the STI
\bea
& &\sum_n\left[ \g_{u^k V^{*,n}_\rho}(-q_1)\g_{V^{n,\rho}V^i_\alpha
V^j_\beta}(q_2,q_3) 
+\g_{u^k V^{*,n}_\rho V^j_\beta}(q_2,q_3)\g_{V^{n,\rho}V^i_\alpha}(q_2)\right.
\nonumber \\
&+& \g_{u^k V^{*,n}_\rho V^i_\alpha}(q_3,q_2)\g_{V^{n,\rho}V^j_\beta}(q_3)
+\g_{u^k G^{*,n}}(-q_1)\g_{G^{n}V^i_\alpha V^j_\beta}(q_2,q_3)
\nonumber \\
&+ &\left. \g_{u^k G^{*,n}V^j_\beta}(q_2,q_3)\g_{G^{n}V^i_\alpha}(q_2)
+\g_{u^k G^{*,n}V^i_\alpha}(q_3,q_2)\g_{G^{n}V^j_\beta}(q_3)\right]=0,
\label{STIthree}
\eea

Different values of the indices $i$, $j$ and $k$ will determine on
which leg (and with which four-momentum) 
we are contracting the three-gauge-boson vertex.
For example, if we choose $k=\pm,\ i=\mp$ and let
$V^j_\beta\equiv{\cal V}^j_\beta$, Eq.\r{STIthree} gives
\bea
& & \g_{u^\pm W^{*,\mp}_\rho}(-q_1)\g_{W^{\pm,\rho}W^\mp_\alpha 
{\cal V}^j_\beta}(q_2,q_3)
+\g_{u^\pm W^{*,\mp}_\rho {\cal V}^j_\beta}(q_2,q_3)
\g_{W^{\pm,\rho}W^\mp_\alpha}(q_2)
\nonumber \\
&+& \sum_n\g_{u^\pm {\cal V}^{*,n}_\rho W^\mp_\alpha}(q_3,q_2)
\g_{{\cal V}^{n,\rho}{\cal V}^j_\beta}(q_3)
+\g_{u^\pm\phi^{*,\mp}}(-q_1)\g_{\phi^{\pm}W^\mp_\alpha {\cal
V}^j_\beta}(q_2,q_3) 
\nonumber \\
&+ & \g_{u^\pm \phi^{*,\mp}{\cal
V}^j_\beta}(q_2,q_3)\g_{\phi^{\pm}W^\mp_\alpha}(q_2) 
+\g_{u^\pm H^{*}W^\mp_\alpha}(q_3,q_2)\g_{H{\cal V}^j_\beta}(q_3)
\nonumber \\
&+ &\g_{u^\pm \chi^{*}W^\mp_\alpha}(q_3,q_2)\g_{\chi{\cal V}^j_\beta}(q_3)=0,
\label{STIthreeW}
\eea
{\it i.e}, we get the STI satisfied by the three-gauge-boson vertex
when contracting from the charged $W^\pm$ legs.

The remaining two STIs, are obtained by choosing $k=Z,A$, $i=\pm$ and $j=\mp$, 
and read
\bea
& & \sum_n\g_{u^k {\cal V}^{*,n}_\rho}(-q_1)
\g_{{\cal V}^{n,\rho}W^\pm_\alpha W^\mp_\beta}(q_2,q_3)
+\g_{u^k W^{*,\pm}_\rho W^\mp_\beta}(q_2,q_3)\g_{W^{\mp,\rho}W^\pm_\alpha}(q_2)
\nonumber \\
&+& \g_{u^k W^{*,\mp}_\rho W^\pm_\alpha}(q_3,q_2)
\g_{W^{\pm,\rho}W^\mp_\beta}(q_3)
+\g_{u^k H^{*}}(-q_1)\g_{HW^\pm_\alpha W^\mp_\beta}(q_2,q_3)
\nonumber \\
&+&\g_{u^k \chi^{*}}(-q_1)\g_{\chi W^\pm_\alpha W^\mp_\beta}(q_2,q_3)
+\g_{u^k \phi^{*,\pm}W^\mp_\beta}(q_2,q_3)\g_{\phi^{\mp}W^\pm_\alpha}(q_2)
\nonumber \\
&+&\g_{u^k \phi^{*,\mp}W^\pm_\alpha}(q_3,q_2)\g_{\phi^{\pm}W^\mp_\beta}(q_3)=0,
\label{STIthreeZA}
\eea
which correspond to the STI satisfied by the three-gauge-boson vertex
when contracting on the neutral ${\cal V}^i$ legs.

\subsection{\label{subBQI} BQIs}

Standard Model BQIs where first presented in
\cite{Gambino:1999ai,Grassi:1999tp}, in the context of the 
Zinn-Justin formalism. Here we present instead the 
{\it disentangled} BQIs,
 derived in the BV formalism;
they may be derived by appropriate functional differentiation
of the BFM STI functional of~Eq.\r{mequbfm}.

\subsubsection{Gauge-boson two-point functions}

The BQIs for the two-point functions involving the gauge-bosons can be
obtained by considering the functional differentiations
\bea
\left.\frac{\delta^2{\cal S}(\g)}
{\delta \Omega^i_\alpha(p_1)\delta \widehat V^j_\beta(q)}
\right|_{\Phi=0}&=&0 \qquad q+p_1=0, \\
\left.\frac{\delta^2{\cal S}(\g)}
{\delta \Omega^i_\alpha(p_1)\delta V^j_\beta(q)}
\right|_{\Phi=0}&=&0 \qquad q+p_1=0,  \\
\left.\frac{\delta^2{\cal S}(\g)}
{\delta \Omega^i_\alpha(p_1)\delta G^j(q)}
\right|_{\Phi=0}&=&0 \qquad q+p_1=0, 
\eea
which will provide us the BQIs
\bea
\g_{\widehat V^i_\alpha \widehat
V^j_\beta}(q)&=&\sum_n\left\{\left[g_{\alpha\rho}\delta^{in}+
\g_{\Omega_\alpha^iV^{*,n}_\rho}(q)\right]\g_{V^{n,\rho} \widehat V^j_\beta}(q)
+\g_{\Omega^i_\alpha G^{*,n}}(q)\g_{G^n\widehat V^j_\beta}(q)\right\}, 
\nonumber \\
\g_{\widehat V^i_\alpha 
V^j_\beta}(q)&=&\sum_n\left\{\left[g_{\alpha\rho}\delta^{in}+
\g_{\Omega_\alpha^iV^{*,n}_\rho}(q)\right]\g_{V^{n,\rho}  V^j_\beta}(q)
+\g_{\Omega^i_\alpha G^{*,n}}(q)\g_{G^n V^j_\beta}(q)\right\}, \nonumber  \\
\g_{\widehat V^i_\alpha G^j}(q)&=&\sum_n\left\{\left[g_{\alpha\rho}\delta^{in}+
\g_{\Omega_\alpha^iV^{*,n}_\rho}(q)\right]\g_{V^{n,\rho} G^j}(q)+
\g_{\Omega^i_\alpha G^{*,n}}(q)\g_{G^nG^j}(q)\right\}. 
\eea
We can now combine these three equations in such a way that all the
two-point functions mixing background and quantum fields drop out,
therefore obtaining the BQI
\bea
\g_{\widehat V^i_\alpha \widehat V^j_\beta}(q)& = &\sum_{m,n}\left\{
\left[g_{\alpha\rho}\delta^{in} +
\g_{\Omega_\alpha^iV^{*,n}_\rho}(q)\right]\left[g_{\beta\sigma}\delta^{jm}+
\g_{\Omega_\beta^jV^{*,m}_\sigma}(q)\right]\g_{V^{m,\sigma} V^{n,\rho}}(q)
\right.\nonumber \\
& + & \left[g_{\alpha\rho}\delta^{in}+
\g_{\Omega_\alpha^iV^{*,n}_\rho}(q)\right]
\g_{\Omega^j_\beta G^{*,m}}(q)\g_{G^m V^{n,\rho}}(q) \nonumber \\
& + & \g_{\Omega^i_\alpha G^{*,n}}(q)\left[g_{\beta\rho}\delta^{jm}+
\g_{\Omega_\beta^jV^{*,m}_\rho}(q)\right]\g_{V^{m,\rho} G^n}(q) \nonumber \\
& + & \left.
\g_{\Omega^i_\alpha G^{*,n}}(q)\g_{\Omega^j_\beta G^{*,m}}(q)\g_{G^mG^n}(q)
\right\}.
\label{twopfwcase}
\eea

In what follows we will only consider the 
case of conserved massless currents. Then only the first line of the above
equation will contribute, since all the 
other terms will be proportional to $q_\alpha$ or
$q_\beta$, so that they will vanish 
when contracted with the corresponding external
current. 

The BQIs for different self-energies are obtained by different
choices of the indices $i$ and $j$ appearing in Eq.\r{twopfwcase}. For
example, choosing $i=\pm$ and $j=\mp$, we get the BQI for the $W$
propagator
\bea
\g_{\widehat W^\pm_\alpha \widehat W^\mp_\beta}(q)&=&
\left[g_{\alpha\rho}+\g_{\Omega^\pm_\alpha W^{*,\mp}_\rho}(q)\right]
\left[g_{\beta\sigma}+\g_{\Omega^\mp_\beta W^{*,\pm}_\sigma}(q)\right]
\g_{W^{\pm,\rho}W^{\mp,\sigma}}(q) \nonumber \\
&=&\g_{W^\pm_\alpha W^\mp_\beta}(q)+
\g_{\Omega^\pm_\alpha W^{*,\mp}_\rho}(q)\g_{W^{\pm,\rho}W^\mp_\beta}(q)
+\g_{\Omega^\mp_\beta W^{*,\pm}_\sigma}(q)\g_{W^\pm_\alpha W^{\mp,\sigma}}(q)
\nonumber \\
&+&\g_{\Omega^\pm_\alpha W^{*,\mp}_\rho}(q)
\g_{W^{\pm,\rho}W^{\mp,\sigma}}(q)\g_{\Omega^\mp_\beta W^{*,\pm}_\sigma}(q),
\label{tpfBQIW} 
\eea
while the BQIs involving the neutral gauge-bosons propagators are
obtained by letting $V^i_\alpha\equiv{\cal V}^i_\alpha$ and 
$V^j_\beta\equiv{\cal V}^j_\beta$, and reads
\bea
\g_{\widehat{\cal V}^i_\alpha \widehat{\cal V}^j_\beta}(q)&=&
\sum_{m,n}\left[g_{\alpha\rho}\delta^{in} +
\g_{\Omega_\alpha^i{\cal V}^{*,n}_\rho}(q)\right]
\left[g_{\beta\sigma}\delta^{jm}+
\g_{\Omega_\beta^j{\cal V}^{*,m}_\sigma}(q)\right]
\g_{{\cal V}^{n,\rho}{\cal V}^{m,\sigma}}(q) \nonumber \\
&=&\g_{{\cal V}^i_\alpha {\cal V}^j_\beta}(q)+\sum_n\left[
\g_{\Omega_\alpha^i{\cal V}^{*,n}_\rho}(q)
\g_{{\cal V}^{n,\rho}{\cal V}^j_\beta}(q)+
\g_{\Omega_\beta^j{\cal V}^{*,m}_\sigma}(q)
\g_{{\cal V}^{n,\rho}{\cal V}^i_\alpha}(q)\right] \nonumber \\
&+&\sum_{m,n}\g_{\Omega_\alpha^i{\cal V}^{*,n}_\rho}(q)
\g_{{\cal V}^{n,\rho}{\cal V}^{m,\sigma}}(q)
\g_{\Omega_\beta^j{\cal V}^{*,m}_\sigma}(q).
\label{tpfBQIZA}
\eea

\subsubsection{Gauge-boson--fermion--anti-fermion three-point functions}

For the annihilation channel (one can study equally well the elastic channel) 
we consider the following functional differentiation
\be
\left.\frac{\delta^3{\cal S}\left(\g\right)}{\delta 
\Omega^i_\alpha(q)\delta
\bar\psi(Q')\delta \psi(Q)}\right|_{\Phi=0}=0 \qquad Q'+Q+q=0, 
\ee
which will furnish the BQI
\bea
\g_{\widehat V^i_\alpha\bar\psi\psi}(Q',Q)\!&=&\!
\sum_n\left\{\left[g_{\alpha\rho}\delta^{in}+
\g_{\Omega^i_\alpha V^{*,n}_\rho}(-q)\right]
\g_{V^{n,\rho}\bar\psi\psi}(Q',Q)
+\g_{\Omega^i_\alpha G^{*,n}}(-q)
\g_{G^n\bar\psi\psi}(Q',Q)\right\} \nonumber \\
&+ & \g_{\bar\psi\psi}(-Q')
\g_{\Omega_\alpha^i \bar\psi^*\psi}(Q',Q) 
-\g_{\Omega_\alpha^i \psi^*\bar\psi}(Q,Q')
\g_{\bar\psi\psi}(Q). 
\eea
We then sandwich the above equation between on-shell spinors, and make
use of the Dirac
equation of motion to eliminate the last two terms; thus we arrive at
the on-shell BQIs
\bea
\g_{\widehat V^i_\alpha\bar\psi\psi}(Q',Q) &=&
\sum_n\left\{ \left[g_{\alpha\rho}\delta^{in}+
\g_{\Omega^i_\alpha V^{*,n}_\rho}(q)\right]
\g_{V^{n,\rho}\bar\psi\psi}(Q',Q)
+\g_{\Omega^i_\alpha G^{*,n}}(q)
\g_{G^n\bar\psi\psi}(Q',Q)\right\}. \nonumber \\
\label{tpfwcase}
\eea
The last term appearing in Eq. \r{tpfwcase} will be absent when
considering the case of massless conserved currents; moreover the BQIs
involving charged and neutral gauge-bosons background fields are
obtained, as usual, by choosing different values of the index $i$. 
Thus, for $i=\pm$ we obtain
the BQI involving the background and quantum $W$s 
\be
\g_{\widehat W^\pm_\alpha\bar\psi\psi}(Q',Q)= 
\left[g_{\alpha\rho}+
\g_{\Omega^\pm_\alpha W^{*,\mp}_\rho}(q)\right]
\g_{W^{\pm,\rho}\bar\psi\psi}(Q',Q), 
\label{thpfBQIW} 
\ee
while letting $V^i_\alpha\equiv {\cal V}^i_\alpha$ 
we get the BQIs involving background and quantum neutral
gauge-bosons, which reads
\be
\g_{\widehat {\cal V}^i_\alpha\bar\psi\psi}(Q',Q)=\sum_n
\left[g_{\alpha\rho}\delta^{in}+
\g_{\Omega^i_\alpha {\cal V}^{*,n}_\rho}(q)\right]
\g_{{\cal V}^{n,\rho}\bar\psi\psi}(Q',Q).
\label{thpfBQIZA} 
\ee

\section{\label{sec:1l}The one-loop $S$-matrix PT revisited.}

In this section we will briefly review the one-loop $S$-matrix
construction of the Electroweak sector in order to
establish the correspondence between results already 
existing in the literature and the newly introduced 
BV language. In particular we 
we will re-express the one-loop
$S$-matrix PT results in terms of the BV building blocks, 
and will familiarize ourselves with the use of the BQI. 
Notice however that the two-loop construction, which is the
main result of this paper will be carried out in the context of 
the intrinsic PT, whose one-loop preliminaries will be presented in
the next section. 

We will consider for concreteness 
the $S$-matrix element for a 2 fermion elastic scattering process  
$\psi(P)\psi(P')\to \psi(Q)\psi(Q')$ in 
the Electroweak sector of the Standard Model;
we set $q= P'-P= Q'-Q$, and $s=q^2$ is  
the square of the momentum transfer. One could equally well
study the annihilation channel of the process 
$\psi(P)\bar\psi(P')\to \psi(Q)\bar\psi(Q')$, in which case $s$ would be
the center-of-mass energy. We assume that the 
theory has been quantized using the renormalizable
($R_{\xi}$) gauges \cite{Fujikawa:fe,Abers:qs}, 
and, without loss of generality, 
we choose the Feynman gauge. Then, the only pinching 
contributions originate from the elementary three-gauge-boson vertices 
appearing inside vertex graphs. 
The bare tree-level three-gauge-boson
vertex is given by the following expression 
(all momenta are incoming, {\it i.e.}, $q+p_1+p_2 = 0$)
\bce
\bpi(0,50)(150,-20)
\Photon(15,0)(40,0){1.5}{6}
\Photon(60,20)(40,0){-1.5}{6}
\Photon(60,-20)(40,0){1.5}{6}
\Text(15,7)[l]{$\scriptstyle{V_\alpha}$}
\Text(15,-7)[l]{$\scriptstyle{q}$}
\Text(45,22)[l]{$\scriptstyle{V^1_\mu}$}
\Text(60,12)[l]{$\scriptstyle{p_1}$}
\Text(45,-22)[l]{$\scriptstyle{V^2_\nu}$}
\Text(60,-12)[l]{$\scriptstyle{p_2}$}
\Text(75,0)[l]{$=\,\g^{(0)}_{V_\alpha V^1_\mu
V^2_\nu}(q,p_1,p_2)=i\gw C\Gamma^{(0)}_{\alpha\mu\nu}(q,p_1,p_2),$} 
\epi
\ece
where 
\be
C_{{\cal V}^iW^+W^-}=\left\{
\begin{array}{ll}
\sw, &\quad {\rm if}\ \ i=A \\
-\cw, & \quad {\rm if}\ \ i=Z
\end{array}
\right.
\label{Ci}
\ee
and, finally,
\be
\Gamma_{\alpha \mu \nu}^{(0)}(q,p_1,p_2)= 
(q-p_1)_{\nu}g_{\alpha\mu} + (p_1-p_2)_{\alpha}g_{\mu\nu}
 + (p_2-q)_{\mu}g_{\alpha\nu}.
\ee
The Lorentz structure $\Gamma_{\alpha \mu \nu}^{(0)}(q,p_1,p_2)$
may be split into two parts 
\cite{Cornwall:1989gv,'tHooft:1971fh}
\be
\Gamma_{\alpha \mu \nu}^{(0)}(q,p_1,p_2) 
= 
\Gamma_{\alpha \mu \nu}^{{\rm F}}(q,p_1,p_2) + 
\Gamma_{\alpha \mu \nu}^{{\rm P}}(q,p_1,p_2),
\label{decomp}
\ee
with 
\bea
\Gamma_{\alpha \mu \nu}^{{\rm F}}(q,p_1,p_2) &=& 
(p_1-p_2)_{\alpha} g_{\mu\nu} + 2q_{\nu}g_{\alpha\mu} 
- 2q_{\mu}g_{\alpha\nu} \, , \nonumber\\
\Gamma_{\alpha \mu \nu}^{{\rm P}}(q,p_1,p_2) &=&
 p_{2\nu} g_{\alpha\mu} - p_{1\mu}g_{\alpha\nu}.  
\label{GFGP}
\eea
The vertex $\Gamma_{\alpha \mu \nu}^{{\rm F}}(q,p_1,p_2)$ 
coincides with the Feynman gauge 
BFM bare vertex involving one
background gauge-boson (carrying four-momentum
$q$) and two quantum ones (carrying four-momenta $p_1$ and $p_2$). 
The above decomposition  
allows $\Gamma_{\alpha \mu \nu}^{{\rm F}}$ to satisfy the WI
\be 
q^{\alpha} \Gamma_{\alpha \mu \nu}^{{\rm F}}(q,p_1,p_2) = 
[(p_2^2 - M_{V_2}^2) - (p_1^2 -M_{V_1}^2) 
+ ( M_{V_1}^2 - M_{V_2}^2)] g_{\mu\nu},
\label{WI2B}
\ee
The first two terms on the 
right-hand side (RHS) are the difference of 
the two-inverse
propagators appearing inside the one-loop vertex graphs
(in the renormalizable Feynman gauge);
the last term accounts for the 
difference in their masses, and is associated to the 
coupling of the corresponding would-be Goldstone bosons.
The term $\Gamma_{\alpha \mu \nu}^{{\rm P}}$
contains the pinching momenta; 
inside Feynman diagrams such 
as those of Fig.1   
they trigger
elementary WIs, 
which will eliminate the internal fermion propagator,
resulting in an effectively propagator-like contribution.
The propagator-like terms thusly generated 
are to be alloted 
to the conventional self-energy graphs, and will form part of the
effective one-loop PT gauge-boson self-energy.
On the other hand  
the remaining 
purely vertex-like parts define the effective PT 
gauge-boson-fermion-fermion 
three-point function $\widehat{\g}_{V^n_\alpha\bar\psi\psi}(Q',Q)$.  

In the next two subsections we will carry out in detail the one-loop 
PT construction for both the charged as well as
the neutral gauge-boson sector of the Standard Model.

\subsection{The charged sector}

\begin{figure}[!t]
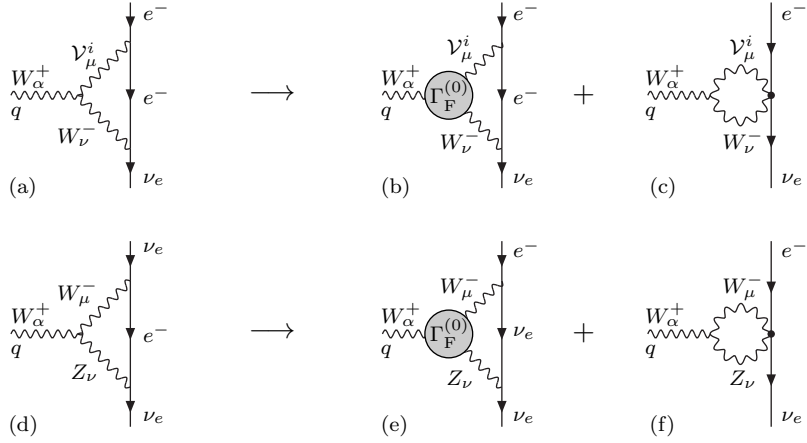

\bce
\bpi(0,170)(170,-115)

\Photon(15,0)(40,0){1.5}{6}
\Photon(60,20)(40,0){-1.5}{6}
\Photon(60,-20)(40,0){1.5}{6}
\ArrowLine(60,35)(60,20)
\ArrowLine(60,20)(60,-20)
\ArrowLine(60,-20)(60,-35)

\Photon(155,0)(180,0){1.5}{6}
\Photon(200,20)(180,0){1.5}{6}
\Photon(200,-20)(180,0){1.5}{6}
\ArrowLine(200,35)(200,20)
\ArrowLine(200,20)(200,-20)
\ArrowLine(200,-20)(200,-35)
\GCirc(180,0){9}{0.8}

\Photon(255,0)(278.5,0){1.5}{6}
\PhotonArc(290,0)(10,-8,352){1.5}{12}
\ArrowLine(301.5,35)(301.5,0)
\ArrowLine(301.5,0)(301.5,-35)
\Vertex(301.5,0){1.4}

\Photon(15,-90)(40,-90){1.5}{6}
\Photon(60,-70)(40,-90){-1.5}{6}
\Photon(60,-110)(40,-90){1.5}{6}
\ArrowLine(60,-55)(60,-70)
\ArrowLine(60,-70)(60,-110)
\ArrowLine(60,-110)(60,-125)

\Photon(155,-90)(180,-90){1.5}{6}
\Photon(200,-70)(180,-90){1.5}{6}
\Photon(200,-110)(180,-90){1.5}{6}
\ArrowLine(200,-55)(200,-70)
\ArrowLine(200,-70)(200,-110)
\ArrowLine(200,-110)(200,-125)
\GCirc(180,-90){9}{0.8}

\Photon(255,-90)(278.5,-90){1.5}{6}
\PhotonArc(290,-90)(10,-8,352){1.5}{12}
\ArrowLine(301.5,-55)(301.5,-90)
\ArrowLine(301.5,-90)(301.5,-125)
\Vertex(301.5,-90){1.4}

\Text(105,0)[l]{$\longrightarrow$}
\Text(181,0)[c]{$\scriptstyle{\Gamma^{(0)}_{\rm F}}$}
\Text(227.5,0)[l]{$+$}
\Text(15,7)[l]{$\scriptstyle{W^+_\alpha}$}
\Text(15,-7)[l]{$\scriptstyle{q}$}
\Text(48,15)[r]{$\scriptstyle{{\cal V}^i_\mu}$}
\Text(48,-15)[r]{$\scriptstyle{W^-_\nu}$}
\Text(65,0)[l]{$\scriptstyle{e^-}$}
\Text(65,32)[l]{$\scriptstyle{e^-}$}
\Text(65,-32)[l]{$\scriptstyle{\nu_e}$}
\Text(155,7)[l]{$\scriptstyle{W^+_\alpha}$}
\Text(155,-7)[l]{$\scriptstyle{q}$}
\Text(205,32)[l]{$\scriptstyle{e^-}$}
\Text(205,0)[l]{$\scriptstyle{e^-}$}
\Text(205,-32)[l]{$\scriptstyle{\nu_e}$}
\Text(255,7)[l]{$\scriptstyle{W^+_\alpha}$}
\Text(255,-7)[l]{$\scriptstyle{q}$}
\Text(185,17.5)[c]{$\scriptstyle{{\cal V}^i_\mu}$}
\Text(185,-17.5)[c]{$\scriptstyle{W^-_\nu}$}
\Text(306.5,32)[l]{$\scriptstyle{e^-}$}
\Text(306.5,-32)[l]{$\scriptstyle{\nu_e}$}
\Text(292,17.5)[c]{$\scriptstyle{{\cal V}^i_\mu}$}
\Text(292,-17.5)[c]{$\scriptstyle{W^-_\nu}$}
\Text(14,-35)[l]{\scriptsize{(a)}}
\Text(155,-35)[l]{\scriptsize{(b)}}
\Text(256.5,-35)[l]{\scriptsize{(c)}}

\Text(105,-90)[l]{$\longrightarrow$}
\Text(181,-90)[c]{$\scriptstyle{\Gamma^{(0)}_{\rm F}}$}
\Text(227.5,-90)[l]{$+$}
\Text(15,-83)[l]{$\scriptstyle{W^+_\alpha}$}
\Text(15,-97)[l]{$\scriptstyle{q}$}
\Text(48,-75)[r]{$\scriptstyle{W^-_\mu}$}
\Text(48,-105)[r]{$\scriptstyle{Z_\nu}$}
\Text(65,-90)[l]{$\scriptstyle{e^-}$}
\Text(65,-58)[l]{$\scriptstyle{\nu_e}$}
\Text(65,-122)[l]{$\scriptstyle{\nu_e}$}
\Text(155,-83)[l]{$\scriptstyle{W^+_\alpha}$}
\Text(155,-97)[l]{$\scriptstyle{q}$}
\Text(205,-58)[l]{$\scriptstyle{e^-}$}
\Text(205,-90)[l]{$\scriptstyle{\nu_e}$}
\Text(205,-122)[l]{$\scriptstyle{\nu_e}$}
\Text(255,-83)[l]{$\scriptstyle{W^+_\alpha}$}
\Text(255,-97)[l]{$\scriptstyle{q}$}
\Text(185,-72.5)[c]{$\scriptstyle{W^-_\mu}$}
\Text(185,-107.5)[c]{$\scriptstyle{Z_\nu}$}
\Text(306.5,-58)[l]{$\scriptstyle{e^-}$}
\Text(306.5,-122)[l]{$\scriptstyle{\nu_e}$}
\Text(292,-72.5)[c]{$\scriptstyle{W^-_\mu}$}
\Text(292,-107.5)[c]{$\scriptstyle{Z_\nu}$}
\Text(14,-125)[l]{\scriptsize{(d)}}
\Text(155,-125)[l]{\scriptsize{(e)}}
\Text(256.5,-125)[l]{\scriptsize{(f)}}

\epi
\ece

\caption{\label{decoW} Carrying out the fundamental vertex decomposition inside
the three-point function $\Gamma_{W^+_\alpha\psi\bar\psi}^{V^2\,(1)}$
(a), (d)
contributing to $\g_{W^+_\alpha\psi\bar\psi}^{(1)}$, gives rise to the genuine
vertex $\widehat\Gamma_{W^+_\alpha\bar\psi\psi}^{V^2\,(1)}$ (b), (e) and a 
self-energy-like
contribution $V_{\alpha\rho}^{{\rm
P}\,(1)}\frac{\gamma^\rho\PL}{\sqrt2}$ (c), (f).}
\end{figure}

In this case we will concentrate on the $S$-matrix element for the
electron-neutrino elastic scattering  process $e(P)\nu_e(P')
\to e(Q)\nu_e(Q')$.
Both the electron and its neutrino will be considered
as strictly mass-less; 
(so that we can neglect longitudinal pieces);
moreover, for definiteness, we will concentrate 
on the three-point function $\g_{W^+_\alpha\bar\psi\psi}(Q',Q)$
(exactly the same results 
hold for the three-point function involving the $W^-$ gauge-boson). 

We then start by implementing
(see Fig.\ref{decoW}a and d) the vertex decomposition of 
Eq.\r{decomp},  
with $p_{1\mu}=-k_{\mu}$, $p_{2\nu}=(k-q)_{\nu}$,
inside the $\Gamma^{V^2\,(1)}_{W^+_\alpha\bar\psi\psi}(Q',Q)$ part of 
the full one-loop three-point function 
$\g^{(1)}_{W^+_\alpha\bar\psi\psi}(Q',Q)$.
The $\Gamma^{{\rm P}}_{\alpha\mu\nu}(q,p_1,p_2)$ term triggers then
the elementary WIs
\bea
& & \ksm  = (\ksm + \Qsm) - \Qsm, 
\nonumber\\ 
& & \ksm - \qsm = (\ksm + \Qsm) - \Qpsm,
\label{PTWI}
\eea
If the external fermions have non-vanishing masses the above 
WI is slightly modified. As has been explained in detail 
\cite{Papavassiliou:1990zd,Papavassiliou:1994pr}
the resulting modification are compensated precisely by the 
contributions of the would-be Goldstone bosons, which in the 
case of massive fermions must also be considered, allowing 
the generalization of the method to the case of non-conserved 
currents.   
The first terms on the RHS of the two WI identities listed 
in Eq.(\ref{PTWI}) generate 
two self-energy like pieces (Fig.\ref{decoW}c and f), 
which are to be alloted
to the conventional self-energy. In particular,
\bea
\Gamma_{W^+_\alpha\bar\psi\psi}^{V^2\,(1)}(Q',Q) &=&
\widehat\Gamma_{W^+_\alpha\bar\psi\psi}^{V^2\,(1)}(Q',Q) 
+V_{\alpha\rho}^{{\rm P}\,(1)}(q)\frac{\gamma^{\rho}\PL}{\sqrt2}
-X^{(1)}_{1\,\alpha}(Q',Q)\Sigma^{(0)}(Q') \nonumber \\
&-& \Sigma^{(0)}(Q)X^{(1)}_{2\,\alpha}(Q',Q),
\label{PTactW}
\eea
where 
\bea
& & \widehat\Gamma_{W^+_\alpha\bar\psi\psi}^{V^2\,(1)}(Q',Q)=
i\gw^2 \int_{L_1}\Gamma^{{\rm F}}_{\alpha\mu\nu}(q,-k,k-q)
\frac{\gamma^\mu\PL}
{\sqrt 2}S^{(0)}(k+Q)\Bigg[\sw^2J_{\scriptscriptstyle{A}}(q,k)\gamma^\nu 
\nonumber \\
& & \hspace{2.6cm}
+\ J_{\scriptscriptstyle{Z}}(q,k) 
\gamma^\nu\left(\frac\PL2-\sw^2\right)\nonumber 
+J'_{\scriptscriptstyle{Z}}(q,k)\gamma^\nu\frac\PL2\Bigg],
\\
& & V^{{\rm P}\,(1)}_{\alpha\rho}(q)=2\gw^2 
g_{\alpha\rho}\sum_iC_i^2\int_{L_1}J_i(q,k)
\nonumber \\
& &\hspace{1.65cm}
= 2\gw^2  g_{\alpha\rho}\,\int_{L_1}
\left[\sw^2J_{\scriptscriptstyle{A}}(q,k)+
\cw^2J_{\scriptscriptstyle{Z}}(q,k)\right], 
\eea
$C_i\equiv C_{{\cal V}^iW^+W^-}$ is defined in Eq.\r{Ci}, and, finally,
\bea
\int_{L_1}&\equiv&\mu^{2\varepsilon}\int\!\frac{d^dk}{\left(2\pi\right)^d},
\nonumber \\
J_i(q,k) & = &\dw(k)\dnu(k-q),\nonumber \\
J'_i(q,k) & = &\dw(k-q)\dnu(k),
\eea
with
\bea
\dw(q)&=&\frac1{q^2-\Mw^2}, \nonumber \\
\dnu(q)&=&\frac1{q^2-\Mi^2}. 
\eea
Notice that the last two terms appearing in the RHS of 
Eq.\r{PTactW} vanish for on-shell external fermions, 
and will be discarded in the analysis that follows.

The (dimension-less) self-energy-like contribution $V^{{\rm
P}\,(1)}_{\alpha\rho}(q)$, together with an equal contribution coming from the
mirror vertex (not shown), after trivial manipulations  
gives rise to the dimensionful quantity
\be
\Pi^{{\rm P}\,(1)}_{\alpha\beta}(q)=2\dw^{-1}(q)
V^{{\rm P}\,(1)}_{\alpha\beta}(q),
\ee
which will be added to the conventional one-loop two-point function
$\g^{(1)}_{W^+_\alpha W^-_\beta}(q)$, to give rise to the PT two-point function
$\widehat\g^{(1)}_{W^+_\alpha W^-_\beta}(q)$:
\be
\widehat\g^{(1)}_{W^+_\alpha W^-_\beta}(q)=\g^{(1)}_{W^+_\alpha W^-_\beta}(q)
+\Pi^{{\rm P}\,(1)}_{\alpha\beta}(q).
\ee 

Correspondingly, the PT one-loop three-point function
$\widehat\g^{(1)}_{W^+_\alpha\bar\psi\psi}(Q',Q)$ will be defined as
\bea
\widehat\g^{(1)}_{W^+_\alpha\bar\psi\psi}(Q',Q) & = &
\widehat\Gamma^{V^2\,(1)}_{W^+_\alpha\bar\psi\psi}(Q',Q)+
\Gamma^{\bar\psi\psi\,(1)}_{W^+_\alpha\bar\psi\psi}(Q',Q) \nonumber \\
& = & 
\g^{(1)}_{W^+_\alpha\bar\psi\psi}(Q',Q)-
V_{\alpha\rho}^{{\rm P}\,(1)}(q)\frac{\gamma^{\rho}\PL}{\sqrt2}.
\label{1lPTthreepfW}
\eea

We can now compare these results with the ones that we 
get from the BQIs
of Eqs.\r{tpfBQIW} and \r{thpfBQIW}
found in the previous sections. At one-loop these BQIs
read
\bea
& & \g^{(1)}_{\widehat{W}^+_\alpha\widehat{W}^-_\beta}(q)=
\g^{(1)}_{W^+_\alpha W^-_\beta}(q)
+2\g^{(1)}_{\Omega^+_\alpha W^{*,-}_{\rho}}(q)\g^{(0)}_{W^{+,\rho} 
W^{-}_\beta}(q),\nonumber \\
& & \g^{(1)}_{\widehat W^+_\alpha\bar\psi\psi}(Q',Q)=
\g^{(1)}_{W^+_\alpha\bar\psi\psi}(Q',Q)+
\g^{(1)}_{\Omega^+_\alpha W^{*,-}_\rho}(q)
\g^{(0)}_{W^{+,\rho}\bar\psi\psi}(Q',Q).
\eea
Moreover perturbatively one has
\bce
\bpi(0,60)(60,-25)

\Text(-17,-0.5)[r]{$\Pi^{(1)}_{\Omega^+_\alpha W^{*,-}_\rho}(q)\,=$}

\Line(-5,0.75)(10,0.75)
\Line(-5,-0.75)(10,-0.75)
\PhotonArc(30,0)(20,0,180){-1.5}{8.5}
\DashCArc(30,0)(20,180,360){1}
\Line(50,0.75)(65,0.75)
\Line(50,-0.75)(65,-0.75)
\DashArrowLine(29.5,-20)(30.5,-20){1}

\Line(90,0.75)(105,0.75)
\Line(90,-0.75)(105,-0.75)
\PhotonArc(125,0)(20,0,180){-1.5}{8.5}
\DashCArc(125,0)(20,180,360){1}
\Line(145,0.75)(160,0.75)
\Line(145,-0.75)(160,-0.75)
\DashArrowLine(124.5,-20)(125.5,-20){1}

\Text(77.5,0)[c]{$\scriptstyle{+}$}
\Text(-5,7)[l]{$\scriptstyle{\Omega^+_\alpha}$}
\Text(90,7)[l]{$\scriptstyle{\Omega^+_\alpha}$}
\Text(75,7)[r]{$\scriptstyle{W^{*,-}_\rho}$}
\Text(170,7)[r]{$\scriptstyle{W^{*,-}_\rho}$}
\Text(30,27)[c]{$\scriptstyle{W_\mu}$}
\Text(125,27)[c]{$\scriptstyle{{\cal V}^i_\mu}$}
\Text(30,-27)[c]{$\scriptstyle{u^i}$}
\Text(125,-27)[c]{$\scriptstyle{u^+}$}

\epi
\ece
Therefore, using the Feynman rules of Appendix \ref{FR}, and observing
that $\g^{(1)}_{\Omega^+_\alpha
W^{*,-}_\rho}=i\Pi^{(1)}_{\Omega^+_\alpha W^{*,-}_\rho}$, we find
\bea
\g^{(1)}_{\Omega^+_\alpha W^{*,-}_\rho}(q)&=&2i\gw^2 
g_{\alpha\rho}\sum_iC^2_ i\int_{L_1}J_i(q,k)\nonumber \\
&=& iV^{{\rm P}\,(1)}_{\alpha\rho}(q). 
\label{r1W}
\eea
Thus, after simple algebra, we find the results
\bea
& & 2\g^{(1)}_{\Omega^+_\alpha W^{*,-}_\rho}(q)
\g^{(0)}_{W^{+,\rho} W^-_\beta}(q)=\Pi^{{\rm
P}\,(1)}_{\alpha\beta}(q),
\nonumber \\
& & \g^{(1)}_{\Omega^+_\alpha W^{*,-}_\rho}(q)
\g^{(0)}_{W^{+,\rho}\bar\psi\psi}(Q',Q)=-V^{{\rm
P}\,(1)}_{\alpha\rho}(q)\frac{\gamma^\rho\PL}{\sqrt2},
\eea
which will in turn automatically enforce the identifications 
\bea
& & \widehat\g^{(1)}_{W^+_\alpha W^-_\beta}(q) \equiv 
\g^{(1)}_{\widehat W^+_\alpha \widehat W^-_\beta}(q), \nonumber \\
& & \widehat\g_{W^+_\alpha\bar\psi\psi}^{(1)}(Q',Q)\equiv
\g_{\widehat W^+_\alpha\bar\psi\psi}^{(1)}(Q',Q).
\eea

\subsection{The neutral sector} 

\begin{figure}[!t]
\bce
\bpi(0,50)(170,-20)

\Photon(15,0)(40,0){1.5}{6}
\Photon(60,20)(40,0){-1.5}{6}
\Photon(60,-20)(40,0){1.5}{6}
\ArrowLine(60,35)(60,20)
\ArrowLine(60,20)(60,-20)
\ArrowLine(60,-20)(60,-35)

\Photon(155,0)(180,0){1.5}{6}
\Photon(200,20)(180,0){1.5}{6}
\Photon(200,-20)(180,0){1.5}{6}
\ArrowLine(200,35)(200,20)
\ArrowLine(200,20)(200,-20)
\ArrowLine(200,-20)(200,-35)
\GCirc(180,0){9}{0.8}

\Photon(255,0)(278.5,0){1.5}{6}
\PhotonArc(290,0)(10,-8,352){1.5}{12}
\ArrowLine(301.5,35)(301.5,0)
\ArrowLine(301.5,0)(301.5,-35)
\Vertex(301.5,0){1.4}

\Text(105,0)[l]{$\longrightarrow$}
\Text(181,0)[c]{$\scriptstyle{\Gamma^{(0)}_{\rm F}}$}
\Text(227.5,0)[l]{$+$}
\Text(15,7)[l]{$\scriptstyle{{\cal V}^i_\alpha}$}
\Text(15,-7)[l]{$\scriptstyle{q}$}
\Text(48,15)[r]{$\scriptstyle{W^+_\mu}$}
\Text(48,-15)[r]{$\scriptstyle{W^-_\nu}$}
\Text(65,0)[l]{$\scriptstyle{e^-}$}
\Text(65,32)[l]{$\scriptstyle{\nu_e}$}
\Text(65,-32)[l]{$\scriptstyle{e^-}$}
\Text(155,7)[l]{$\scriptstyle{{\cal V}^i_\alpha}$}
\Text(155,-7)[l]{$\scriptstyle{q}$}
\Text(205,32)[l]{$\scriptstyle{e^-}$}
\Text(205,0)[l]{$\scriptstyle{\nu_e}$}
\Text(205,-32)[l]{$\scriptstyle{e^-}$}
\Text(255,7)[l]{$\scriptstyle{{\cal V}^i_\alpha}$}
\Text(255,-7)[l]{$\scriptstyle{q}$}
\Text(185,17.5)[c]{$\scriptstyle{W^+_\mu}$}
\Text(185,-17.5)[c]{$\scriptstyle{W^-_\nu}$}
\Text(306.5,32)[l]{$\scriptstyle{e^-}$}
\Text(306.5,-32)[l]{$\scriptstyle{e^-}$}
\Text(292,17.5)[c]{$\scriptstyle{W^+_\mu}$}
\Text(292,-17.5)[c]{$\scriptstyle{W^-_\nu}$}
\Text(14,-35)[l]{\scriptsize{(a)}}
\Text(155,-35)[l]{\scriptsize{(b)}}
\Text(256.5,-35)[l]{\scriptsize{(c)}}

\epi
\ece

\caption{\label{decoZA} Carrying out the fundamental vertex decomposition inside
the three-point function $\Gamma_{{\cal V}^i_\alpha\psi\bar\psi}^{V^2\,(1)}$
(a),
contributing to $\g_{{\cal V}^i_\alpha\psi\bar\psi}^{(1)}$, gives rise
to the genuine 
vertex $\widehat\Gamma_{{\cal V}^i_\alpha\bar\psi\psi}^{V^2\,(1)}$
(b), and a self-energy-like
contribution $V_{\alpha\rho}^{i,{\rm
P}\,(1)}\frac{\gamma^\rho\PL}{\sqrt2}$ (c).}
\end{figure}

In this case we will concentrate on the $S$-matrix element for the
electron-electron elastic scattering  process $e(P)e(P')
\to e(Q)e(Q')$, where again the electrons will be treated as 
mass-less. 

As in the charged sector case, we start by implementing (see Fig.\ref{decoZA}a)
the vertex decomposition of Eq.\r{decomp},  
with $p_{1\mu}=-k_{\mu}$, $p_{2\nu}=(k-q)_{\nu}$,
inside the $\Gamma^{V^2\,(1)}_{{\cal V}^i_\alpha\bar\psi\psi}(Q',Q)$ 
part of 
the full one-loop three-point functions 
$\g^{(1)}_{{\cal V}^i_\alpha\bar\psi\psi}(Q',Q)$. 
The $\Gamma^{{\rm P}}_{\alpha\mu\nu}(q,p_1,p_2)$ term triggers then
the elementary WIs of Eq.\r{PTWI}, so that
two self-energy like pieces are generated (Figs.\ref{decoZA}c). 
In particular,
\bea
\Gamma_{{\cal V}^i_\alpha\bar\psi\psi}^{V^2\,(1)}(Q',Q) &=&
\widehat\Gamma_{{\cal V}^i_\alpha\bar\psi\psi}^{V^2\,(1)}(Q',Q) 
+V_{\alpha\rho}^{i,{\rm P}\,(1)}(q) \frac{\gamma^{\rho}\PL}{2}
-X'^{\,(1)}_{1\,\alpha}(Q',Q)\Sigma^{(0)}(Q') \nonumber \\
&-& \Sigma^{(0)}(Q)X'^{\,(1)}_{2\,\alpha}(Q',Q),
\label{PTactV}
\eea
where
\bea
& &\widehat \Gamma_{{\cal V}^i_\alpha\bar\psi\psi}^{V^2\,(1)}(Q',Q)=
i\gw^2  C_i\int_{L_1}\Gamma^{{\rm F}}_{\alpha\mu\nu}(q,-k,k-q)\gamma^\mu\PL
S^{(0)}(k+Q')\gamma^\nu\frac12\PL J_{\scriptscriptstyle{W}}(q,k),
\nonumber \\
& & V^{i,{\rm P}\, (1)}_{\alpha\rho}(q)=2\gw^2C_i g_{\alpha\rho}\int_{L_1}
J_{\scriptscriptstyle{W}}(q,k),
\label{eq1}
\eea
and, finally,
\be
J_{\scriptscriptstyle{W}}(q,k)=\dw(k)\dw(k-q).
\ee
Again the last two terms appearing in Eq.\r{PTactV} will be discarded, since
they vanish for on-shell fermions.
The (dimension-less) self-energy-like piece 
$V^{i,{\rm P}\, (1)}_{\alpha\rho}(q)$
 must be alloted to 
the conventional one-loop $ZZ$,  $AZ$, $ZA$ and 
$AA$. To accomplish that 
we notice that the effective vertex 
$(\gamma^\mu\PL / 2)$ in Eq.(\ref{PTactV})
may be written as a linear
combination of the two Standard Model tree-level vertices 
(factoring out the electroweak coupling $\gw$)
\bea
\g_{A^\mu\bar\psi\psi}^{(0)}&=&-i\sw Q_\psi\gamma^\mu ,\nonumber \\
\g_{Z^\mu\bar\psi\psi}^{(0)}&=&-i\frac1\cw \gamma^\mu\left[
\left(\sw^2Q_\psi-T^\psi_z\right)\PL+\sw^2Q_\psi\PR\right],
\eea
as follows
\be
\frac{i}{2} \gamma^\mu\PL =-\left(\frac\sw{2T^\psi_z}\right)
\g_{A^\mu\bar\psi\psi}^{(0)}+\left(\frac\cw{2T^\psi_z}\right)
\g_{Z^\mu\bar\psi\psi}^{(0)}.
\ee
When the fermion $\psi$ is an electron as in our case, $T^\psi_z=-1/2$,
and we find 
\bea
\frac{1}{2} \gamma^\mu\PL &=& -i\sw\g_{A^\mu\bar\psi\psi}^{(0)}+
\cw\g_{Z^\mu\bar\psi\psi}^{(0)}
\nonumber \\
&=&\sw(\sw\gamma^\mu)-\cw\left[-\frac1\cw\gamma^\mu
\left(\frac12\PL-\sw^2\right)\right], 
\label{lep} 
\eea
so that Eq.\r{lep} together Eq.\r{eq1} 
will fix  the PT self-energy like contributions to be
\be
V^{{\rm P}\,(1)}_{ij,\alpha\rho}(q)=2\gw^2 C_iC_jg_{\alpha\rho}\int_{L_1}
J_{\scriptscriptstyle{W}}(q,k). 
\ee
Adding an equal contribution coming from the mirror 
vertex (not shown) we find the dimensionful quantity
\be
\Pi^{{\rm P}\,(1)}_{ij,\alpha\beta}(q)=2\left[\dnu^{-1}(q)+
d_{\scriptscriptstyle{\cal V}^j}^{-1}(q)
\right]
V^{{\rm P}\,(1)}_{ij,\alpha\beta}(q),
\ee
which will be added to the conventional one-loop
two-point function 
$\g^{(1)}_{{\cal V}^i_\alpha {\cal V}^j_\beta}(q)$, 
to give rise to the PT two-point function
$\widehat\g^{(1)}_{{\cal V}^i_\alpha {\cal V}^j_\beta}(q)$:
\be
\widehat\g^{(1)}_{{\cal V}^i_\alpha{\cal
V}^j_\beta}(q)=\g^{(1)}_{{\cal V}^i_\alpha{\cal V}^j_\beta}(q)+
\Pi^{{\rm P}\,(1)}_{ij,\alpha\beta}(q).
\ee
Correspondingly, the PT one-loop three-point function
$\widehat\g^{(1)}_{{\cal V}^i_\alpha\bar\psi\psi}(Q',Q)$ 
will be defined as
\bea
\widehat\g^{(1)}_{{\cal V}^i_\alpha\bar\psi\psi}(Q',Q) & = &
\widehat\Gamma^{V^2\,(1)}_{{\cal V}^i_\alpha\bar\psi\psi}(Q',Q)+
\Gamma^{\bar\psi\psi\,(1)}_{{\cal V}^i_\alpha\bar\psi\psi}(Q',Q) \nonumber \\
& = & 
\g^{(1)}_{{\cal V}^i_\alpha\bar\psi\psi}(Q',Q)-
V_{\alpha\rho}^{i,{\rm P}\,(1)}(q)\frac{\gamma^{\rho}\PL}{2}.
\label{1lPTthreepfAZ}
\eea

We can now compare these results with those obtained
from the BQIs of Eqs.\r{tpfBQIZA} and \r{thpfBQIZA}, reported
in the previous sections. Expanding at the one-loop level these
BQIs, we find 
\bea
& & \g^{(1)}_{\widehat{\cal V}^i_\alpha\widehat{\cal V}^j_\beta}(q)=
\g^{(1)}_{{\cal V}^i_\alpha {\cal V}^j_\beta}(q)+
\g^{(1)}_{\Omega^i_\alpha{\cal V}^{*,j}_\rho}(q)
\g^{(0)}_{{\cal V}^{j,\rho}{\cal V}^j_\beta}(q)+
\g^{(1)}_{\Omega^j_\beta{\cal V}^{*,i}_\rho}(q)
\g^{(0)}_{{\cal V}^{i,\rho}{\cal V}^i_\alpha}(q),\nonumber \\
& & \g^{(1)}_{\widehat {\cal V}^i_\alpha\bar\psi\psi}(Q',Q)=
\g^{(1)}_{{\cal V}^i_\alpha\bar\psi\psi}(Q',Q)+
\sum_n\g^{(1)}_{\Omega^i_\alpha {\cal V}^{*,n}_\rho}(q)
\g^{(0)}_{{\cal V}^{n,\rho}\bar\psi\psi}(Q',Q). 
\eea

In addition, perturbatively one has (with a$=i,j$, b$=j,i$ and 
$\lambda=\alpha,\beta$)
\bce
\bpi(0,60)(50,-25)

\Text(-17,-0.5)[r]{$\Pi^{(1)}_{\Omega^{\rm a}_\lambda {\cal V}^{*,{\rm
b}}_\rho}(q)\,=$}

\Line(-5,0.75)(10,0.75)
\Line(-5,-0.75)(10,-0.75)
\PhotonArc(30,0)(20,0,180){-1.5}{8.5}
\DashCArc(30,0)(20,180,360){1}
\Line(50,0.75)(65,0.75)
\Line(50,-0.75)(65,-0.75)
\DashArrowLine(29.5,-20)(30.5,-20){1}

\Line(90,0.75)(105,0.75)
\Line(90,-0.75)(105,-0.75)
\PhotonArc(125,0)(20,0,180){-1.5}{8.5}
\DashCArc(125,0)(20,180,360){1}
\Line(145,0.75)(160,0.75)
\Line(145,-0.75)(160,-0.75)
\DashArrowLine(124.5,-20)(125.5,-20){1}

\Text(77.5,0)[c]{$\scriptstyle{+}$}
\Text(-5,7)[l]{$\scriptstyle{\Omega^{\rm a}_\lambda}$}
\Text(90,7)[l]{$\scriptstyle{\Omega^{\rm a}_\lambda}$}
\Text(75,7)[r]{$\scriptstyle{{\cal V}^{*,{\rm b}}_\rho}$}
\Text(170,7)[r]{$\scriptstyle{{\cal V}^{*,{\rm b}}_\rho}$}
\Text(30,27)[c]{$\scriptstyle{W^+_\mu}$}
\Text(125,27)[c]{$\scriptstyle{W^-_\mu}$}
\Text(30,-27)[c]{$\scriptstyle{u^-}$}
\Text(125,-27)[c]{$\scriptstyle{u^+}$}

\epi
\ece
Therefore, using the Feynman rules of Appendix \ref{FR}, and observing
that $\g^{(1)}_{\Omega^{\rm a}_\lambda
{\cal V}^{*,{\rm b}}_\rho}
=i\Pi^{(1)}_{\Omega^{\rm a}_\lambda{\cal V}^{\rm b}_\rho}$, we find
\bea
\g^{(1)}_{\Omega^{\rm a}_\lambda {\cal V}^{*,{\rm b}}_\rho}(q)&=&
2i\gw^2 C_{\rm a}C_{\rm b}g_{\alpha\rho}
\int_{L_1}J_{\scriptscriptstyle{W}}(q,k) \nonumber \\
&=& iV^{{\rm P}\,(1)}_{{\rm ab},\alpha\rho}(q).
\label{r1ZA}
\eea
Thus, after simple manipulations, we arrive at the results
\bea
&&\g^{(1)}_{\Omega^i_\alpha{\cal V}^{*,j}_\rho}(q)
\g^{(0)}_{{\cal V}^{j,\rho}{\cal V}^j_\beta}(q)+
\g^{(1)}_{\Omega^j_\beta{\cal V}^{*,i}_\rho}(q)
\g^{(0)}_{{\cal V}^{i,\rho}{\cal V}^i_\alpha}(q)=
\Pi^{{\rm P}\,(1)}_{ij,\alpha\beta}(q),\nonumber \\
&&\sum_n\g^{(1)}_{\Omega^i_\alpha {\cal V}^{*}_\rho}(q)
\g^{(0)}_{{\cal V}^{n,\rho}\bar\psi\psi}(Q',Q)
=-V^{i,{\rm P}\,(1)}_{\alpha\rho}(q)\frac{\gamma^\rho\PL}2 ,
\eea
which automatically enforce the identifications
\bea
& & \widehat\g^{(1)}_{{\cal V}^i_\alpha {\cal V}^j_\beta}(q) \equiv 
\g^{(1)}_{\widehat {\cal V}^i_\alpha \widehat {\cal V}^j_\beta}(q),\nonumber \\
& & \widehat\g_{{\cal V}^i_\alpha\bar\psi\psi}^{(1)}(Q',Q)\equiv
\g_{\widehat {\cal V}^i_\alpha\bar\psi\psi}^{(1)}(Q',Q).
\eea

\section{\label{sec:IP} Electroweak intrinsic PT at one-loop}

In the intrinsic PT construction one avoids the embedding of the PT objects
into $S$-matrix elements; of course, all  results of the intrinsic PT are
identical to those obtained in the $S$-matrix PT context.  The basic idea, is
that the pinch graphs, which are essential in  canceling the gauge dependences
of ordinary diagrams, are always missing one or more propagators corresponding
to the external legs of the improper Green's function in question. It then
follows that the gauge-dependent parts of such ordinary diagrams must also be
missing one or more external propagators. Thus the intrinsic PT construction
goal is to isolate systematically the parts of 1PI diagrams that are 
proportional to the inverse propagators of the external legs and simply discard
them.  The important point is that these inverse propagators  arise from the
STIs satisfied by ({\it i}) the three-gauge-boson vertex and --a
characteristic that distinguish the electroweak
sector of the Standard Model
 case from the QCD case-- ({\it ii}) the gauge-boson
propagators  appearing inside
appropriate sets of 
diagrams, when they will be contracted by longitudinal momenta. 
The STIs triggered are
nothing but the one appearing in Eqs.\r{STItwopoint} and \r{STIthree}. 
Of course the  momenta appearing in these STIs
will now be related to
virtual  integration  momenta appearing  in  the  quantum loop. 

In the
context of QCD, this
construction    has    been     carried    out    at    one-loop    in
\cite{Cornwall:1989gv} and recently generalized at the two-loop level in
\cite{Binosi:2002ez}.   
Here
we  present for  the first  time the  two-loop generalization  of this
construction in the electroweak sector of the Standard Model, 
employing the one-loop versions of Eqs.\r{STItwopoint} and \r{STIthree}.

The  essential feature  of  the
intrinsic  PT construction  is to  arrive at  the desired  object, for
example the effective gauge-boson  self-energy, by discarding 
in a systematic way well-defined pieces from the
conventional self-energy.  The terms  discarded originate from 
Eqs.\r{STItwopoint} and \r{STIthree}, and  
they are  all precisely  known in  terms of  physical and
unphysical  Green's  functions, appearing  in  the  theory.  Then, 
one  can  directly compare  the
result obtained by the intrinsic PT procedure to the corresponding BFM
quantity  (at $\xi_Q=1$), employing  the BQIs  of Eqs.\r{tpfBQIW} and
\r{thpfBQIZA}.

We start by reviewing the one-loop intrinsic PT construction, 
beginning again, without loss of generality, in the renormalizable
Feynman gauge.

\subsection{Charged sector}

In the charged sector case, the quantity we want to construct is the
one-loop $W$ \mbox{two-point} function.
In the absence of longitudinal momenta coming from internal gauge
boson propagators (since we work in the Feynman gauge), 
the only diagram contributing to the
$W$ self-energy that can trigger an STI, is the one containing two
three-gauge-boson vertices: 
\bea
\bpi(0,65)(-100,-45)

\Photon(63.5,0)(78.5,0){1.5}{2.5}
\PhotonArc(90,0)(10,-8,352){1.5}{12}
\Photon(101.5,0)(116.5,0){1.5}{2.5}
\Text(58,-7)[l]{$\scriptstyle{W^+_\alpha}$}
\Text(125,-7)[r]{$\scriptstyle{W^-_\beta}$}
\Text(90,17.5)[c]{$\scriptstyle{W_\sigma}$}
\Text(90,-17.5)[c]{$\scriptstyle{{\cal V}^i_\rho}$}
\epi
&=&\sum_i
\int_{L_1}J'(q,k)
\g^{(0)}_{W^+_\alpha W^-_\sigma {\cal V}^i_{\rho}}(q,-k,k-q)
\g^{(0)}_{W^-_\beta W^{+,\sigma} {\cal
V}^{i,\rho}}(q,-k,k-q). \nonumber \\
\label{cspd}
\eea

We next carry out the PT decomposition of Eq.\r{decomp} on {\it both}
the three-gauge-boson vertices appearing in Eq.\r{cspd}, {\it i.e.}, we
write
\bea
\g^{(0)}_{W^+_\alpha W^-_\sigma {\cal V}^i_{\rho}}
\g^{(0)}_{W^-_\beta W^{+,\sigma} {\cal V}^{i,\rho}}&=&
\Big[\g^{{\rm F}}_{W^+_\alpha W^-_\sigma {\cal V}^i_{\rho}}+
\g^{{\rm P}}_{W^+_\alpha W^-_\sigma {\cal V}^i_{\rho}}\Big]
\Big[\g^{{\rm F}}_{W^-_\beta W^{+,\sigma} {\cal V}^{i,\rho}}+
\g^{{\rm P}}_{W^-_\beta W^{+,\sigma} {\cal V}^{i,\rho}}\Big]
\nonumber \\
&=&
\g^{{\rm F}}_{W^+_\alpha W^-_\sigma {\cal V}^i_{\rho}}
\g^{{\rm F}}_{W^-_\beta W^{+,\sigma} {\cal V}^{i,\rho}}
+\g^{{\rm P}}_{W^+_\alpha W^-_\sigma {\cal V}^i_{\rho}}
\g^{(0)}_{W^-_\beta W^{+,\sigma}{\cal V}^{i,\rho}}\nonumber \\
&+&\g^{(0)}_{W^+_\alpha W^-_\sigma {\cal V}^i_{\rho}}
\g^{{\rm P}}_{W^-_\beta W^{+,\sigma} {\cal V}^{i,\rho}}
-\g^{{\rm P}}_{W^+_\alpha W^-_\sigma {\cal V}^i_{\rho}}
\g^{{\rm P}}_{W^-_\beta W^{+,\sigma} {\cal V}^{i,\rho}},
\label{dcW}
\eea
where, since it is important to know from which leg of the vertex we
will finally trigger the STI, we have defined
\bea
\g^{{\rm F}}_{W^\pm_\alpha W^\mp_\sigma {\cal V}^i_{\rho}}&=&\pm 
\left(i\gw C_i\right)
\Gamma^{{\rm F}}_{\alpha\sigma\rho}, \nonumber \\
\g^{{\rm P}}_{W^\pm_\alpha W^\mp_\sigma {\cal V}^i_{\rho}}&=&\pm 
\left(i\gw C_i\right)
\Gamma^{{\rm P}}_{\alpha\sigma\rho},
\eea

Of the four terms appearing in Eq.\r{dcW}, the first and the last are
left untouched, while for the second and the third we can write
\bea
&&\g^{{\rm P}}_{W^+_\alpha W^-_\sigma {\cal V}^i_{\rho}}
\g^{(0)}_{W^-_\beta W^{+,\sigma} {\cal V}^{i,\rho}}
+\g^{(0)}_{W^+_\alpha W^-_\sigma {\cal V}^i_{\rho}}
\g^{{\rm P}}_{W^-_\beta W^{+,\sigma} {\cal V}^{i,\rho}}  \nonumber \\
&=&-\left(i\gw C_i\right)
\left[k^\sigma g_{\alpha}^\rho+(k-q)^\rho g^\sigma_\alpha\right]
\g^{(0)}_{W^+_\sigma W^-_\beta{\cal V}^i_\rho}(k,-q,-k+q) \nonumber \\
&+&\left(i\gw C_i\right)
\left[k^\sigma g_{\beta}^\rho+(k-q)^\rho g^\sigma_\beta\right]
\g^{(0)}_{W^-_\sigma W^+_\alpha{\cal V}^i_\rho}(k,-q,-k+q).
\label{cspd1}
\eea

Th longitudinal momenta $k$ and $(k-q)$ appearing in the expression
above will then trigger the tree-level version of the STIs of 
Eqs.\r{STIthreeW} and \r{STIthreeZA} respectively, which read
\bea
\g^{(0)}_{u^\pm W^{*,\mp}_\sigma}(-k)\g^{(0)}_{W^{\pm,\sigma} W^\mp_\lambda
{\cal V}^i_\rho}(-q,-k+q)&=&
-\g^{(0)}_{u^\pm W^{*,\mp}_\sigma{\cal
V}^i_\rho}(-q,-k+q)\g^{(0)}_{W^{\pm,\sigma}
W^\mp_{\lambda}}(-q)\nonumber \\
&-&\sum_n\g^{(0)}_{u^\pm{\cal V}^{*,n}_\sigma W^\mp_{\lambda}}(-k+q,-q)
\g^{(0)}_{{\cal V}^{n,\sigma}
{\cal V}^i_\alpha}(-k+q)\nonumber \\
&-&\g^{(0)}_{u^\pm\phi^{*,\mp}}(-k)\g^{(0)}_{\phi^\pm W^\mp_{\lambda}{\cal
V}^i_\rho}(-q,-k+q),\nonumber\\
\sum_n\g^{(0)}_{u^i {\cal V}^{*,n}_\rho}(k-q)
\g^{(0)}_{{\cal V}^{n,\rho}W^\pm_\sigma W^\mp_{\lambda}}(k,-q)&=&-
\g^{(0)}_{u^i W^{*,\pm}_\rho
W^\mp_{\lambda}}(k,-q)\g^{(0)}_{W^{\mp,\rho}W^\pm_\sigma}(k) 
\nonumber \\
&-& \g^{(0)}_{u^i W^{*,\mp}_\rho
W^\pm_\sigma}(-q,k)\g^{(0)}_{W^{\pm,\rho}W^\mp_{\lambda}}(-q), 
\eea
where $\lambda$ can be $\beta$ or $\alpha$ depending on which term
of Eq.\r{cspd1} we are considering. Now,
from the BRST Lagrangian of Eq.\r{BRSTL} we have the relations
\bea
\g^{(0)}_{u^\pm W^{*,\mp}_\sigma}(-k)&=&-ik_\sigma, \nonumber \\
\g^{(0)}_{u^i {\cal V}^{*,n}_\rho}(k-q)&=&i(k-q)_\rho\delta^{in}, \nonumber \\
\g^{(0)}_{u^\pm\phi^{*,\mp}}(-k)&=&\pm i\Mw, 
\eea
which, together with the Feynman rules of Appendix \ref{FR}, will in
turn give us the tree-level STIs in their final form, {\it i.e.},
\bea
k^\sigma\g^{(0)}_{W^{\pm}_\sigma W^\mp_{\lambda}
{\cal V}^i_\rho}(-q,-k+q)&=&\pm
\left(\gw C_i\right)\g^{(0)}_{W^\pm_\rho W^\mp_{\lambda}}(-q)
\mp\left(\gw C_i\right)
\g^{(0)}_{{\cal V}^{i}_{\lambda}
{\cal V}^i_\rho}(-k+q)\nonumber \\
&\pm&\Mw\g^{(0)}_{\phi^\pm W^\mp_{\lambda}{\cal
V}^i_\rho}(-q,-k+q), \label{one}\\
(k-q)^\rho
\g^{(0)}_{{\cal V}^{i}_\rho W^\pm_\sigma W^\mp_{\lambda}}(k,-q)&=&
\pm\left(\gw C_i\right)
\g^{(0)}_{W^\pm_\sigma W^\mp_{\lambda}}(-q)\mp
\left(\gw C_i\right)
\g^{(0)}_{W^\mp_{\lambda} W^\pm_\sigma}(k). 
\label{two}
\eea
The first term on the RHS of these two STIs
is to be discarded  
from the $W$ self-energy. Thus, the 1PI one-loop intrinsic PT
self-energy, to be denoted as before by $\widehat\g^{(1)}_{W^+_\alpha
W^-_\beta}(q)$, is defined as 
\be
\widehat\g^{(1)}_{W^+_\alpha
W^-_\beta}(q)=\g^{(1)}_{W^+_\alpha
W^-_\beta}(q)-\Pi^{{\rm IP}\,(1)}_{\alpha\beta}(q),
\ee
where the superscript ``IP'' stands for ``intrinsic pinch'', and 
$\Pi^{{\rm IP}\,(1)}_{\alpha\beta}$ is obtained by plugging the
discarded term back into Eqs.\r{cspd1} and \r{cspd}, and  has
precisely the form 
\bea
\Pi^{{\rm
IP}\,(1)}_{\alpha\beta}(q)&=&-4i\gw^2\sum_i 
C_i^2\int_{L_1}J'_i(q,k)\g^{(0)}_{W^+_\alpha
W^-_\beta}(q) \nonumber \\
&=&-\Pi^{{\rm P}\,(1)}_{\alpha\beta}(q).
\eea

At this point, following the original IP procedure
\cite{Cornwall:1989gv}, one should combine 
the first and last term on the RHS of Eq.\r{dcW} with the
terms of the STIs of Eqs.\r{one} 
and \r{two}  which have not been discarded, and add the
remaining diagrams contributing to the $W$ self-energy, in order to
check that effectively $\widehat\g^{(1)}_{W^+_\alpha W^-_\beta}(q)$
coincides with $\g^{(1)}_{\widehat W^+_\alpha \widehat
W^-_\beta}(q)$. However in light of the BQI of Eq.\r{tpfBQIW}, this last
identification is more immediate, in the sense that no further
manipulation of the answer is needed: 
the difference  between ${\widehat\g}^{(1)}_{W^+_\alpha W^-_\beta}(q)$ and
$\g^{(1)}_{W^+_\alpha W^-_\beta}(q)$ is the same as the difference between 
$\g^{(1)}_{\widehat W^+_\alpha \widehat W^-_\beta}(q)$ and 
$\g^{(1)}_{W^+_\alpha W^-_\beta}(q)$, as given by the BQI. Already at
the one-loop level we can see that the use of the BQIs constitutes a
definite technical advantage. 

\subsection{Neutral sector}

In the neutral sector case, our aim is to construct the one-loop
${\cal V}^i$ two-point functions. As before, in the absence of
longitudinal momenta coming from the internal gauge-boson propagators,
and recalling that we are working in the mass-less conserved current
case, the only diagram contributing to the ${\cal V}^i$ self-energy
that can trigger an STI, is the one containing two three-gauge-boson
vertices, {\it i.e.},
\bea
\bpi(0,65)(-100,-45)

\Photon(63.5,0)(78.5,0){1.5}{2.5}
\PhotonArc(90,0)(10,-8,352){1.5}{12}
\Photon(101.5,0)(116.5,0){1.5}{2.5}
\Text(58,-7)[l]{$\scriptstyle{{\cal V}^i_\alpha}$}
\Text(125,-7)[r]{$\scriptstyle{{\cal V}^j_\beta}$}
\Text(90,17.5)[c]{$\scriptstyle{W_\sigma}$}
\Text(90,-17.5)[c]{$\scriptstyle{W_\rho}$}
\epi
&= &\int_{L_1}J_{\scriptscriptstyle{W}}(q,k)
\g^{(0)}_{{\cal V}^i_{\alpha} W^-_\sigma W^+_\rho}(q,-k,k-q)
\g^{(0)}_{{\cal V}^j_\beta W^{+,\sigma} W^{-,\rho}}(q,-k,k-q). \nonumber\\
\label{cspdZA}
\eea

As in the charged sector case, 
we next carry out the PT decomposition of Eq.\r{decomp} on both
the three-gauge-boson vertices appearing in Eq.\r{cspd}, concentrating
only on the term
\bea
& & \g^{{\rm P}}_{{\cal V}^i_{\alpha} W^-_\sigma W^+_\rho}
\g^{(0)}_{{\cal V}^j_\beta W^{+,\sigma} W^{-,\rho}}
+\g^{(0)}_{{\cal V}^i_{\alpha} W^-_\sigma W^+_\rho}
\g^{{\rm P}}_{{\cal V}^j_\beta W^{+,\sigma} W^{-,\rho}} \nonumber \\
&=&\left(i\gw C_i\right)\left[k^\sigma g^\rho_\alpha+(k-q)^\rho
g^\sigma_\alpha\right]\g^{(0)}_{ W^+_\sigma
W^-_\rho{\cal V}^j_\beta}(k,-k+q,-q) \nonumber \\
&-&\left(i\gw C_j\right) 
\left[k^\sigma g^\rho_\beta+(k-q)^\rho g^\sigma_\beta\right]
\g^{(0)}_{ W^-_\sigma W^+_\rho{\cal V}^i_{\alpha}}(k,-k+q,-q).
\eea

The longitudinal momenta $k$ and $(k-q)$ appearing in the above
equation will trigger in this case the following tree-level STIs
\bea
k^\sigma\g^{(0)}_{W^\pm_\sigma W^\mp_\rho{\cal V}^{\rm a}_{\rm
x}}(-k+q,-q)&=&\mp \left(\gw C_{\rm a}\right)
\g^{(0)}_{{\cal V}^{\rm a}_\rho{\cal V}^{\rm a}_{\lambda}}(-q)\pm
\left(\gw C_{\rm a}\right)\g^{(0)}_{W^\pm_{\rm
x}W^\mp_{\rho}}(-k+q)\nonumber \\
&\pm&\Mw\g^{(0)}_{\phi^\pm W^\mp{\cal V}^{\rm a}_{\lambda}}(-k+q,-q),\nonumber\\
(k-q)^\rho\g^{(0)}_{W^\mp_\rho W^\pm_\sigma{\cal V}^{\rm a}_{\rm
x}}(k,-q)&=&\mp \left(\gw C_{\rm a}\right)
\g^{(0)}_{{\cal V}^{\rm a}_\sigma{\cal V}^{\rm a}_{\lambda}}(-q)\pm
\left(\gw C_{\rm a}\right)\g^{(0)}_{W^\mp_{\rm
x}W^\pm_{\sigma}}(k)\nonumber \\
&\pm&\Mw\g^{(0)}_{\phi^\mp W^\pm{\cal V}^{\rm a}_{\lambda}}(k,-q),
\label{twoZA}
\eea
where a$=j,i$ and $\lambda=\beta,\alpha$ depending on which term of
Eq.\r{cspdZA} we are considering.

As before, the first terms appearing in the RHS of these STIs
will be discarded from the self-energy.
Thus the 1PI one-loop intrinsic PT
self-energy, to be denoted as before by $\widehat\g^{(1)}_{{\cal V}^i_\alpha
{\cal V}^j_\beta}(q)$, is defined as 
\be
\widehat\g^{(1)}_{{\cal V}^i_\alpha
{\cal V}^j_\beta}(q)=\g^{(1)}_{{\cal V}^i_\alpha
{\cal V}^j_\beta}(q)-\Pi^{{\rm IP}\,(1)}_{ij,\alpha\beta}(q),
\ee
where the quantity 
$\Pi^{{\rm IP}\,(1)}_{ij,\alpha\beta}(q)$ is obtained by plugging the
discarded terms back into Eq.\r{cspdZA},
and  has precisely the form
\be
\Pi^{{\rm IP}\,(1)}_{ij,\alpha\beta}(q)=
-2i\gw^2C_iC_j\int_{L_1}J_{{\scriptscriptstyle W}}(q,k)\left[
\g^{(0)}_{{\cal V}^i_\alpha
{\cal V}^i_\beta}(q)+\g^{(0)}_{{\cal V}^j_\alpha
{\cal V}^j_\beta}(q)\right],
\ee
{\it i.e.}, after trivial algebra, we find the identity 
\be
\Pi^{{\rm IP}\,(1)}_{ij,\alpha\beta}(q)=
-\Pi^{{\rm P}\,(1)}_{ij,\alpha\beta}(q).
\ee

Again, since the difference between $\widehat\g^{(1)}_{{\cal V}^i_\alpha
{\cal V}^j_\beta}(q)$ and $\g^{(1)}_{{\cal V}^i_\alpha
{\cal V}^j_\beta}(q)$ is the same as the difference between
$\g^{(1)}_{\widehat{\cal V}^i_\alpha \widehat{\cal V}^j_\beta}(q)$
and $\g^{(1)}_{{\cal V}^i_\alpha {\cal V}^j_\beta}(q)$ as given by the
BQI of Eq.\r{tpfBQIZA}, we conclude that the PT result coincides with
the BFM one.

\begin{figure}[!t]
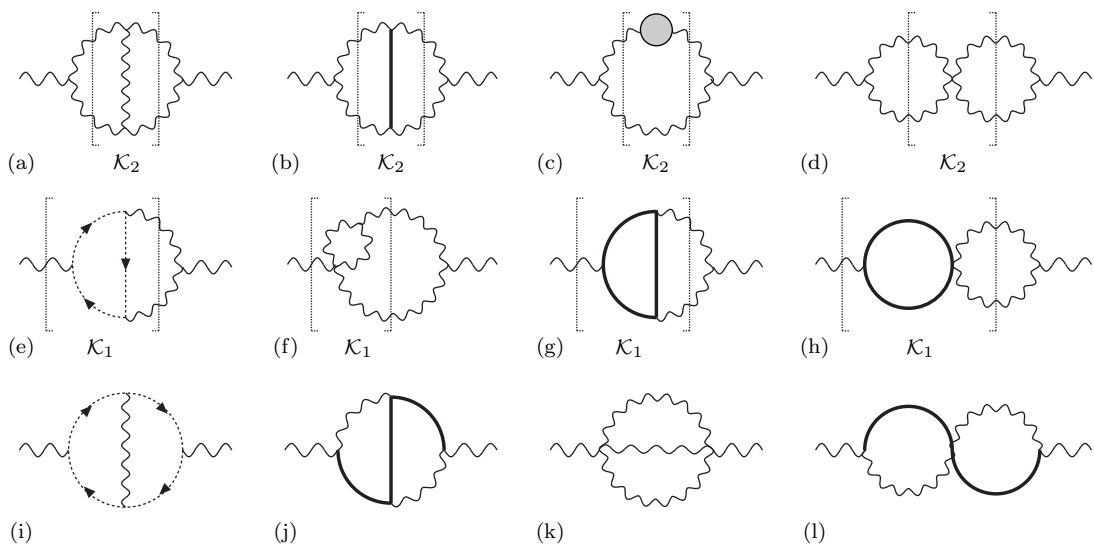

\bce
\bpi(0,200)(0,-155)

\Photon(-210,0)(-191.5,0){2.5}{2}
\PhotonArc(-170,0)(20,-6,354){1.5}{18}
\Photon(-170,-18.5)(-170,18.5){1.5}{6}
\Photon(-148.5,0)(-130,0){-2.5}{2}
\DashLine(-182.5,-25)(-182.5,25){0.5}
\DashLine(-182.5,25)(-180,25){0.5}
\DashLine(-182.5,-25)(-180,-25){0.5}
\DashLine(-157.5,-25)(-157.5,25){0.5}
\DashLine(-157.5,25)(-160,25){0.5}
\DashLine(-157.5,-25)(-160,-25){0.5}
\Text(-170,-32)[c]{$\scriptstyle {\cal K}_2$}

\Photon(-109,0)(-91,0){2.5}{2}
\PhotonArc(-70,0)(20,-6,354){1.5}{18}
\SetWidth{1.3}
\Line(-70,18.5)(-70,-18.5)
\SetWidth{0.5}
\Photon(-48.5,0)(-30,0){-2.5}{2}
\DashLine(-57.5,-25)(-57.5,25){0.5}
\DashLine(-57.5,25)(-60,25){0.5}
\DashLine(-57.5,-25)(-60,-25){0.5}
\DashLine(-82.5,-25)(-82.5,25){0.5}
\DashLine(-82.5,25)(-80,25){0.5}
\DashLine(-82.5,-25)(-80,-25){0.5}
\Text(-70,-32)[c]{$\scriptstyle {\cal K}_2$}

\Photon(-10,0)(8.5,0){2.5}{2}
\PhotonArc(30,0)(20,-6,354){1.5}{18}
\GCirc(30,18.5){6}{0.8}
\Photon(50.5,0)(70,0){-2.5}{2}
\DashLine(17.5,-25)(17.5,25){0.5}
\DashLine(17.5,-25)(20,-25){0.5}
\DashLine(17.5,25)(20,25){0.5}
\DashLine(42.5,-25)(42.5,25){0.5}
\DashLine(42.5,25)(40,25){0.5}
\DashLine(42.5,-25)(40,-25){0.5}
\Text(30,-32)[c]{$\scriptstyle {\cal K}_2$}

\Photon(90,0)(108.5,0){2.5}{2}
\PhotonArc(125,0)(15,-6,354){1.5}{14}
\PhotonArc(158,0)(15,-6,354){1.5}{14}
\Photon(174.5,0)(194,0){-2.5}{2}
\DashLine(125,-25)(125,25){0.5}
\DashLine(125,25)(127.5,25){0.5}
\DashLine(125,-25)(127.5,-25){0.5}
\DashLine(158,-25)(158,25){0.5}
\DashLine(158,25)(155.5,25){0.5}
\DashLine(158,-25)(155.5,-25){0.5}
\Text(142.5,-32)[c]{$\scriptstyle {\cal K}_2$}

\Photon(-210,-70)(-190,-70){2.5}{2}
\PhotonArc(-170,-70)(20,270,90){1.5}{9}
\DashCArc(-170,-70)(20,90,270){1}
\DashArrowLine(-170,-50)(-170,-90){1}
\Photon(-148.5,-71)(-130,-71){-2.5}{2}
\DashArrowLine(-184.5,-56)(-184,-55.5){1}
\DashArrowLine(-184,-84.5)(-184.5,-84){1}
\DashLine(-157.5,-95)(-157.5,-45){0.5}
\DashLine(-157.5,-45)(-160,-45){0.5}
\DashLine(-157.5,-95)(-160,-95){0.5}
\DashLine(-200,-45)(-200,-95){0.5}
\DashLine(-200,-95)(-197.5,-95){0.5}
\DashLine(-200,-45)(-197.5,-45){0.5}
\Text(-180,-102)[c]{$\scriptstyle {\cal K}_1$}

\Photon(-109,-70)(-91,-70){2.5}{2}
\PhotonArc(-70,-70)(20,181,489){1.5}{16}
\PhotonArc(-86.3,-62.6)(8,2,362){1.5}{8}
\Photon(-48.5,-70)(-30,-70){-2.5}{2}
\DashLine(-70,-95)(-70,-45){0.5}
\DashLine(-70,-45)(-72.5,-45){0.5}
\DashLine(-70,-95)(-72.5,-95){0.5}
\DashLine(-100,-95)(-100,-45){0.5}
\DashLine(-100,-45)(-97.5,-45){0.5}
\DashLine(-100,-95)(-97.5,-95){0.5}
\Text(-82.5,-102)[c]{$\scriptstyle {\cal K}_1$}

\Photon(-10,-70)(10,-70){2.5}{2}
\PhotonArc(30,-70)(20,270,90){1.5}{9}
\SetWidth{1.3}
\CArc(30,-70)(20,90,270)
\Line(30,-49.5)(30,-90)
\SetWidth{0.5}
\Photon(51.5,-71)(70,-71){-2.5}{2}
\DashLine(42.5,-95)(42.5,-45){0.5}
\DashLine(42.5,-45)(40,-45){0.5}
\DashLine(42.5,-95)(40,-95){0.5}
\DashLine(0,-45)(0,-95){0.5}
\DashLine(0,-95)(2.5,-95){0.5}
\DashLine(0,-45)(2.5,-45){0.5}
\Text(20,-102)[c]{$\scriptstyle {\cal K}_1$}

\Photon(90,-70)(108.5,-70){2.5}{2}
\PhotonArc(158,-70)(15,-6,354){1.5}{14}
\SetWidth{1.3}
\CArc(125,-70)(16.5,-6,354)
\SetWidth{0.5}
\Photon(174.5,-70)(194,-70){-2.5}{2}
\DashLine(100,-95)(100,-45){0.5}
\DashLine(100,-45)(102.5,-45){0.5}
\DashLine(100,-95)(102.5,-95){0.5}
\DashLine(158,-95)(158,-45){0.5}
\DashLine(158,-45)(155.5,-45){0.5}
\DashLine(158,-95)(155.5,-95){0.5}
\Text(130,-102)[c]{$\scriptstyle {\cal K}_1$}

\Photon(-210,-140)(-191.5,-140){2.5}{2}
\DashCArc(-170,-140)(21.5,0,360){1}
\Photon(-170,-161.5)(-170,-118.5){1.5}{6}
\Photon(-148.5,-140)(-130,-140){-2.5}{2}
\DashArrowLine(-156,-124.5)(-155.5,-125){1}
\DashArrowLine(-155.5,-155)(-156,-155.5){1}
\DashArrowLine(-184.5,-125)(-184,-124.5){1}
\DashArrowLine(-184,-155.5)(-184.5,-155){1}

\Photon(-110,-140)(-90.1,-140){2.5}{2}
\PhotonArc(-70,-140)(20,90,180){1.5}{4}
\PhotonArc(-70,-140)(20,270,360){1.5}{4}
\Photon(-49.9,-140)(-30,-140){-2.5}{2}
\SetWidth{1.3}
\CArc(-70,-140)(20,0,90)
\CArc(-70,-140)(20,180,270)
\Line(-70,-160)(-70,-120)
\SetWidth{0.5}

\Photon(-10,-140)(8.5,-140){2.5}{2}
\PhotonArc(30,-140)(20,-6,354){1.5}{18}
\Photon(8.5,-139.5)(51.5,-140){1.5}{4.5}
\Photon(51.5,-140)(70,-140){-2.5}{2}

\Photon(90,-140)(108.5,-140){2.5}{2}
\PhotonArc(125,-140)(16,180,0){1.5}{7.5}
\PhotonArc(158,-140)(16,360,180){1.5}{7.5}
\SetWidth{1.3}
\CArc(158,-140)(16.5,180,360)
\CArc(125,-140)(16.5,0,180)
\SetWidth{0.5}
\Photon(174.5,-140)(194,-140){-2.5}{2}

\Text(-210,-32)[c]{\scriptsize{(a)}}
\Text(-110,-32)[c]{\scriptsize{(b)}}
\Text(-10,-32)[c]{\scriptsize{(c)}}
\Text(90,-32)[c]{\scriptsize{(d)}}

\Text(-210,-102)[c]{\scriptsize{(e)}}
\Text(-110,-102)[c]{\scriptsize{(f)}}
\Text(-10,-102)[c]{\scriptsize{(g)}}
\Text(90,-102)[c]{\scriptsize{(h)}}

\Text(-210,-172)[c]{\scriptsize{(i)}}
\Text(-110,-172)[c]{\scriptsize{(j)}}
\Text(-10,-172)[c]{\scriptsize{(k)}}
\Text(90,-172)[c]{\scriptsize{(l)}}
\epi
\ece
\caption{\label{figdiag} 
Some of the Feynman diagrams contributing to the conventional two-loop
gauge-boson self-energies $\g^{(2)}_{V^nV^m}$ 
in the $R_\xi$ gauges. In the square brackets we indicate the
associated kernel as explained in the text: thus diagrams (a)--(d)
belongs to set ({\it i}), diagrams (e)--(h) to set ({\it ii}), and
diagrams (i)--(l) to set ({\it iii}). The blob in diagram (c)
denotes  the one-loop correction to the gauge-boson propagator.}
\end{figure}

\section{\label{sec:IP2} Electroweak intrinsic PT at two-loops}

In the next three sections we  
will generalize the intrinsic PT construction presented above
to two loops. The results presented 
are completely new,
since the PT (intrinsic or $S$-matrix) has never been applied 
to the purely bosonic part of the 
Electroweak sector beyond one loop. The only two-loop  
results existing in the literature involve 
the special subset of 
contributions containing a fermion loop
\cite{Bauberger:1994nk}. In this section we will outline the 
general formalism necessary for carrying out the two-loop 
construction in a concise and expeditious way, 
and in the next two sections we will present 
detailed  constructions for the cases of the 
charged and neutral sectors.

The 1PI Feynman diagram contributing to the conventional
two-loop gauge-boson self-energy both in the charged and neutral
sector can be separated in three distinct sets
\cite{Binosi:2002ez}: 
({\it i}) the set of diagrams that contains two external (tree-level)
three-gauge-boson vertices, and thus can be written schematically
(suppressing Lorentz indices) as 
$\g^{(0)}_{\scriptscriptstyle{VVV}}\left[{\cal
K}_2\right]\g^{(0)}_{\scriptscriptstyle{VVV}}$,  with ${\cal K}_2$ a 
kernel associated to these  
diagrams; ({\it ii}) the set of diagrams which has  only one
external (tree-level) three-gauge-boson vertex, and that therefore can
be written as  $\g^{(0)}_{\scriptscriptstyle{VVV}}\left[{\cal
K}_1\right]$ or $\left[{\cal 
K}_1\right]\g^{(0)}_{\scriptscriptstyle{VVV}}$;  ({\it iii}) all the remaining
diagrams, containing no external three-gauge-boson vertices.
Notice that the diagrams belonging to this last set are simple 
``spectators'' as far as this construction is concerned, 
since, due to the fact that they do not
contain any longitudinal momenta, they cannot trigger any STI:
therefore they will be left untouched. 
Out of all the 20
possible 1PI topologies (including seagull and tadpoles diagrams) that 
contribute to the two-loop gauge-boson self-energies
\cite{Weiglein:1994hd}, the only ones that furnish diagrams
belonging to the sets ({\it i}) and ({\it ii}) are the following
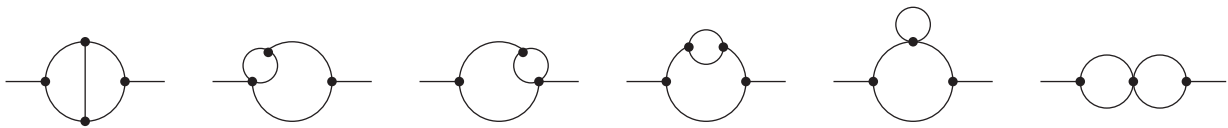
\bce
\bpi(0,50)(230,-25)

\Line(0,0)(15,0)
\CArc(30,0)(15,0,360)
\Line(45,0)(60,0)
\Line(30,-15)(30,15)
\Vertex(15,0){1.8}
\Vertex(30,-15){1.8}
\Vertex(30,+15){1.8}
\Vertex(45,0){1.8}

\Line(78,0)(93,0)
\CArc(108,0)(15,180,128)
\Line(123,0)(138,0)
\CArc(96,6)(6.5,0,360)
\Vertex(93,0){1.8}
\Vertex(123,0){1.8}
\Vertex(99,11){1.8}

\Line(156,0)(171,0)
\CArc(186,0)(15,52,360)
\Line(201,0)(216,0)
\CArc(198,6)(6.5,0,360)
\Vertex(171,0){1.8}
\Vertex(201,0){1.8}
\Vertex(195,11){1.8}

\Line(234,0)(249,0)
\CArc(264,0)(15,116,64)
\Line(279,0)(294,0)
\CArc(264,13)(6.5,0,360)
\Vertex(249,0){1.8}
\Vertex(279,0){1.8}
\Vertex(270.5,13){1.8}
\Vertex(257.5,13){1.8}

\Line(312,0)(327,0)
\CArc(342,0)(15,0,360)
\Line(357,0)(372,0)
\CArc(342,21.5)(6.5,0,360)
\Vertex(327,0){1.8}
\Vertex(357,0){1.8}
\Vertex(342,15){1.8}

\Line(390,0)(405,0)
\CArc(415,0)(10,0,360)
\CArc(435,0)(10,0,360)
\Line(445,0)(460,0)
\Vertex(405,0){1.8}
\Vertex(445,0){1.8}
\Vertex(425,0){1.8}

\epi
\ece
Some of the diagrams arising from the above topologies, together with
their associated kernels, are shown in Fig.\ref{figdiag}.

At this point we  make the following observation \cite{Binosi:2002ez}:
if one  were to  carry out the  decomposition of Eq.\r{decomp}  to the
pair of external  vertices appearing in the diagrams  of set ({\it i})
and  to the external  vertex appearing  in the  diagrams of  set ({\it
ii}), the longitudinal momenta stemming  from the pinching part of the
vertices $\gp$ would be triggering the one-loop version of the STIs of
Eqs.\r{STIthreeW} or \r{STIthreeZA}, just  as in the one-loop case one
has been  triggering the tree-level  version of these STIs.   The only
exception  are those  diagrams  of  the set  ({\it  i}) which  contain
one-loop corrections to the  internal propagators, such as the diagram
of Fig.\ref{figdiag}c:  the final vertex  STIs triggered in  this case
are   still   the   tree-level   version   of   Eqs.\r{STIthreeW}   or
\r{STIthreeZA}.  There is however  an important difference between the
QCD and the Electroweak case regarding the role of such graphs.  In the
case of QCD, out of  the two possible longitudinal momenta originating
from  (either) $\gp$  only one  will reach the  other side  of the
diagram, thus  triggering the corresponding  (tree-level) STI, whereas
the  other  one  will  vanish  when  contracted  with  the  transverse
(one-loop)  gluon self-energy.  In  the Electroweak  case however,  the
corresponding (one-loop) gauge-boson self-energies are not transverse;
thus the  second longitudinal  momentum will also  reach the other
side, after first triggering the one-loop version of the corresponding
STIs,  Eqs.\r{STItwopointW}  and~\r{STItwopointZA}.  Thus,  additional
pinching  contributions will  be  generated, which  must be  carefully
determined.

The first step in the construction is to carry
out the usual PT decomposition of the (tree-level) three-gauge-boson
vertex of Eq.\r{decomp} to the diagrams of sets~({\it
i}) and~({\it ii}). For the diagrams belonging to the set ({\it i}) we
will generically write
\bea
\g^{(0)}_{\scriptscriptstyle{VVV}}\left[{\cal K}_2\right]
\g^{(0)}_{\scriptscriptstyle{VVV}} &=&
\g^{\rm F}_{\scriptscriptstyle{VVV}}
\left[{\cal K}_2\right]\g^{\rm F}_{\scriptscriptstyle{VVV}}+
\g^{\rm P}_{\scriptscriptstyle{VVV}}
\left[{\cal K}_2\right]\g^{(0)}_{\scriptscriptstyle{VVV}}+
\g^{(0)}_{\scriptscriptstyle{VVV}}
\left[{\cal K}_2\right]\g^{\rm P}_{\scriptscriptstyle{VVV}} \nonumber \\
&-& \g^{\rm P}_{\scriptscriptstyle{VVV}}
\left[{\cal K}_2\right]\g^{\rm P}_{\scriptscriptstyle{VVV}},
\label{generic}
\eea
and, as in the one-loop case, of the four terms appearing in the above
equation, the first and the last are left untouched, and constitute
part of the PT answer. Instead, to the second and third term of
Eq.\r{generic} corresponding to all the kernels ${\cal K}_2$ that {\it
do not} contain one-loop corrections to the 
internal (gauge-boson) propagators (such as the ones associated to the
diagrams of Fig.\ref{figdiag}a, b and d), we add
the pinching part of all the diagrams belonging to set~({\it ii}), for
which we generically write
\bea
\g^{(0)}_{\scriptscriptstyle{VVV}}\left[{\cal K}_1\right]&=&
\g^{\rm F}_{\scriptscriptstyle{VVV}}\left[{\cal K}_1\right]+
\g^{\rm P}_{\scriptscriptstyle{VVV}}\left[{\cal K}_1\right], \nonumber
\\
\left[{\cal K}_1\right]\g^{(0)}_{\scriptscriptstyle{VVV}}&=&
\left[{\cal K}_1\right]\g^{\rm F}_{\scriptscriptstyle{VVV}}+
\left[{\cal K}_1\right]\g^{\rm P}_{\scriptscriptstyle{VVV}}.
\eea
Then, one arrives at the equation
\be
\bpi(0,60)(120,-35)
\Text(-100,0)[l]{$\left\{\g^{(0)}_{\scriptscriptstyle{VVV}}\left[{\cal K}_2\right]
\g^{(0)}_{\scriptscriptstyle{VVV}}+
\g^{(0)}_{\scriptscriptstyle{VVV}}\left[{\cal K}_1\right]+\left[{\cal
K}_1\right]\g^{(0)}_{\scriptscriptstyle{VVV}} 
\right\}^{\rm P}=$}

\Text(136,-30)[l]{$\equiv\,{\cal C}_{2L},$}

\Photon(163.5,0)(178.5,0){1.5}{2.5}
\PhotonArc(193,0)(13,-8,352){1.5}{12}
\Photon(207.5,0)(222.5,0){1.5}{2.5}
\GCirc(207.5,0){6}{0.8}
\Text(180,-7)[r]{$\scriptstyle{\g^{\rm P}}$}
\Vertex(178.5,0){1.8}

\Text(235.5,0)[c]{$+$}

\Photon(248.5,0)(263.5,0){1.5}{2.5}
\PhotonArc(278,0)(13,-8,352){1.5}{12}
\Photon(292.5,0)(307.5,0){1.5}{2.5}
\GCirc(263.5,0){6}{0.8}
\Text(295,-7)[l]{$\scriptstyle{\g^{\rm P}}$}
\Vertex(292.5,0){1.8}

\epi
\label{start1}
\ee
with the blobs representing the full one-loop three-gauge-boson
vertex.

For the second and third terms of Eq.\r{generic}, corresponding to the
remaining kernels ${\cal K}_2$ that {\it do} contain one-loop corrections
to the internal gauge-boson propagators (such as the one associated to
the diagram of Fig.\ref{figdiag}c), we instead directly get 
\be
\bpi(0,60)(120,-35)
\Text(-100,0)[l]{$\left\{\g^{(0)}_{\scriptscriptstyle{VVV}}\left[{\cal
K}_2\right] 
\g^{(0)}_{\scriptscriptstyle{VVV}}\right\}^{\rm P}=$}

\Text(7,-30)[l]{$\equiv\,{\cal C}_{1L},$}

\Photon(33.5,0)(48.5,0){1.5}{2.5}
\PhotonArc(63,0)(13,-8,352){1.5}{12}
\Photon(77.5,0)(92.5,0){1.5}{2.5}
\Vertex(48.5,0){1.8}
\Text(50,-7)[r]{$\scriptstyle{\g^{\rm P}}$}
\GCirc(63,13){5.5}{0.8}
\Text(105.5,0)[c]{$+$}

\Photon(118.5,0)(133.5,0){1.5}{2.5}
\PhotonArc(148,0)(13,-8,352){1.5}{12}
\Photon(162.5,0)(177.5,0){1.5}{2.5}
\Vertex(162.5,0){1.8}
\Text(164,-7)[l]{$\scriptstyle{\g^{\rm P}}$}
\GCirc(148,13){5.5}{0.8}
\Text(190.5,0)[c]{$+$}

\Photon(203.5,0)(218.5,0){1.5}{2.5}
\PhotonArc(233,0)(13,-8,352){1.5}{12}
\Photon(247.5,0)(262.5,0){1.5}{2.5}
\Vertex(218.5,0){1.8}
\Text(220,-7)[r]{$\scriptstyle{\g^{\rm P}}$}
\GCirc(233,-13){5.5}{0.8}
\Text(275.5,0)[c]{$+$}

\Photon(288.5,0)(303.5,0){1.5}{2.5}
\PhotonArc(318,0)(13,-8,352){1.5}{12}
\Photon(332.5,0)(347.5,0){1.5}{2.5}
\Vertex(332.5,0){1.8}
\Text(334,-7)[l]{$\scriptstyle{\g^{\rm P}}$}
\GCirc(318,-13){5.5}{0.8}

\epi
\label{start2}
\ee
where now the blobs denote one-loop corrections to the gauge-boson
propagator. 

To carry out the generalization of the
intrinsic PT to two-loops we must next isolate from   
Eqs.\r{start1} and \r{start2} 
those terms stemming 
from the triggering of the STIs  of Eqs.\r{STIthreeW},
\r{STIthreeZA}, 
\r{STItwopointW} and \r{STItwopointZA} 
which are proportional to 
$\left[\Delta^{(-1)\,\rho}_{\alpha}(q)\right]^{(n)}$, with  $n=0,1$; we will
denote such contributions generically by $\Pi_{\alpha\beta}^{{\rm
IP}\,(2)}(q)$ in the charged sector case and $\Pi_{ij,\alpha\beta}^{{\rm
IP}\,(2)}(q)$ in the neutral sector case.
Thus the 1PI diagrams contributing to the two-loop gauge-boson
self-energies, can be cast respectively in the form
\bea
\g^{(2)}_{W^+_\alpha W^-_\beta}(q)&=&G^{(2)}_{W^+_\alpha W^-_\beta}(q)
+\Pi_{\alpha\beta}^{{\rm IP}\,(2)}(q), 
\label{GW} \\
\g^{(2)}_{{\cal V}^i_\alpha{\cal V}^j_\beta}(q)&=&
G^{(2)}_{{\cal V}^i_\alpha{\cal V}^j_\beta}(q)+\Pi_{ij,\alpha\beta}^{{\rm
IP}\,(2)}(q). 
\label{GZA}
\eea

Notice however that  
the 1PR set of one-loop self-energy diagrams, 
must also be rearranged following the intrinsic PT procedure, and be
converted into the equivalent  string involving PT one-loop
self-energies (which are known objects from the one-loop results). 
This treatment of the 1PR string will
give rise, in addition to the PT strings, 
to ({\it i}) a set of contributions which are proportional to the
inverse propagator of the external legs, and ({\it ii}) 
a set of contributions which is {\it
effectively} 1PI, and therefore  also belongs to the definition of the 1PI
two-loop PT gauge-boson self-energies; we will denote these two sets
of contributions collectively by
$S^{{\rm IP}\,(2)}_{\alpha\beta}(q)$ (charged sector)
and $S^{{\rm IP}\,(2)}_{ij,\alpha\beta}(q)$ (neutral sector). 

Thus, the sum of the 1PI and 1PR contributions to the conventional
two-loop gauge-boson self-energies can be cast in the form
\bea
\g^{(2)}_{W^+_\alpha W^-_\beta}(q)-i
\g^{(1)}_{W^+_\alpha W^-_\rho}(q)\dw(q)
\g^{(1)}_{W^{+,\rho}
W^-_\beta}(q)&=&G^{(2)}_{W^+_\alpha W^-_\beta}(q)
-i\widehat\g^{(1)}_{W^+_\alpha W^-_\rho}(q)\dw(q)
\widehat\g^{(1)}_{W^{+,\rho}
W^-_\beta}(q)\nonumber \\
&+&\Pi_{\alpha\beta}^{{\rm IP}\,(2)}(q)+S^{{\rm IP}\,(2)}_{\alpha\beta}(q), 
\label{GWc} \\
\g^{(2)}_{{\cal V}^i_\alpha{\cal V}^j_\beta}(q)-i\sum_{n}
\g^{(1)}_{{\cal V}^i_\alpha{\cal V}^n_\rho}(q)
d_{\scriptscriptstyle{{\cal V}^n}}(q)
\g^{(1)}_{{\cal V}^{n,\rho}{\cal V}^j_\beta}(q)
&=&
G^{(2)}_{{\cal V}^i_\alpha{\cal V}^j_\beta}(q)-i
\sum_{n}\widehat
\g^{(1)}_{{\cal V}^i_\alpha{\cal V}^n_\rho}(q)
d_{\scriptscriptstyle{{\cal V}^n}}(q)\widehat
\g^{(1)}_{{\cal V}^{n,\rho}{\cal V}^j_\beta}(q) \nonumber \\
&+&\Pi_{ij,\alpha\beta}^{{\rm
IP}\,(2)}(q)+S^{{\rm IP}\,(2)}_{ij,\alpha\beta}(q). 
\label{GZAc}
\eea

By definition of the intrinsic PT procedure, we will now discard from
the above expressions all the terms which are proportional to the
inverse propagator of the external legs ({\it i.e.}, 
$d_{\scriptscriptstyle{W}}^{-1}(q)$ or 
$d_{\scriptscriptstyle{{\cal V}^i}}^{-1}(q)$ and
$d_{\scriptscriptstyle{{\cal V}^j}}^{-1}(q)$ in the
charged, respectively neutral, sector), thus defining the quantities
\bea
R_{\alpha\beta}^{{\rm IP}\,(2)}(q)&=&
\Pi'^{{\rm IP}\,(2)}_{\alpha\beta}(q)+S'^{{\rm IP}\,(2)}_{\alpha\beta}(q), 
\label{RW} \\
R^{{\rm IP}\,(2)}_{ij,\alpha\beta}(q)
&=&\Pi'^{{\rm
IP}\,(2)}_{ij,\alpha\beta}(q)+S'^{{\rm IP}\,(2)}_{ij,\alpha\beta}(q),
\label{RZA}
\eea
where the primed functions are obtained from the unprimed ones
appearing in Eqs.\r{GWc} and \r{GZAc} by discarding the aforementioned terms. 

Thus, making use of Eqs.\r{GW}--\r{RZA}, the
1PI two-loop 
intrinsic PT gauge-boson self-energies, to be denoted as before by
$\widehat\g^{(2)}_{W^+_\alpha W^-_\beta}(q)$ and
$\widehat\g^{(2)}_{{\cal V}^i_\alpha {\cal V}^j_\beta}(q)$, will be defined
as
\bea
\widehat\g^{(2)}_{W^+_\alpha W^-_\beta}(q)&=&G^{(2)}_{W^+_\alpha W^-_\beta}(q)
+R^{{\rm IP}\,(2)}_{\alpha\beta}(q) \nonumber \\
&=&\g^{(2)}_{W^+_\alpha W^-_\beta}(q)-\Pi_{\alpha\beta}^{{\rm
IP}\,(2)}(q)+R^{{\rm IP}\,(2)}_{\alpha\beta}(q), \label{ipdefW} \\
\widehat\g^{(2)}_{{\cal V}^i_\alpha {\cal V}^j_\beta}(q)&=&
G^{(2)}_{{\cal V}^i_\alpha {\cal V}^j_\beta}(q)+
R^{{\rm IP}\,(2)}_{ij,\alpha\beta}(q) \nonumber \\
&=&\g^{(2)}_{{\cal V}^i_\alpha {\cal V}^j_\beta}(q)-\Pi_{ij,\alpha\beta}^{{\rm
IP}\,(2)}(q)+R^{{\rm IP}\,(2)}_{ij,\alpha\beta}(q) 
\label{ipdefZA}.
\eea  

We next proceed to give the details of the two-loop construction in
both the charged as well as the neutral gauge-boson sector.  

\section{\label{sec:2lcs}Two-loop Charged sector}

As explained in the previous section, the starting point for the IP
construction are
Eqs.\r{start1} and \r{start2}: from them, by using the two- and 
three-point functions STIs, we will isolate the 1PI parts that are
proportional to the inverse propagator of the external legs, and simply
discard them. In the charged gauge-boson sector
case these equations read 
\bea
{\cal C}_{\scriptscriptstyle{2L}}
&=&-i\gw\sum_i C_i\int_{L_1}
J'_i(q,k)
\Bigg\{
\left[k^\sigma g^\rho_\alpha+(k-q)^\rho
g^\sigma_\alpha\right] 
\g^{(1)}_{W^+_\sigma W^-_\beta {\cal V}^i_\rho}(k,-q,-k+q)\nonumber \\ 
&-& \left[k^\sigma g^\rho_\beta+(k-q)^\rho
g^\sigma_\beta\right]
\g^{(1)}_{W^-_\sigma W^+_\alpha {\cal V}^i_\rho}(k,-q,-k+q) \Bigg\},
\label{start1W}\\
{\cal C}_{\scriptscriptstyle{1L}}^{\rm c}&=&
-\gw\sum_i C_i\int_{L_1}J'_i(q,k)\dw(k)\times\nonumber \\
&\times&
\Bigg\{\left[k^\sigma g^\rho_\alpha+(k-q)^\rho g^\sigma_\alpha\right]
\g^{(1)}_{W^+_\sigma W^-_\mu}(k)
\g^{(0)}_{W^{+,\mu} W^-_\beta {\cal V}^i_\rho}(k,-q,-k+q) \nonumber \\
&-&\left[k^\sigma g^\rho_\beta+(k-q)^\rho g^\sigma_\beta\right]
\g^{(1)}_{W^-_\sigma W^+_\mu}(k)
\g^{(0)}_{W^{-,\mu} W^+_\alpha {\cal V}^i_\rho}(k,-q,-k+q)\Bigg\},
\label{start2Wc} \\
{\cal C}_{\scriptscriptstyle{1L}}^{\rm n}&=&\gw\sum_{i,j}
\int_{L_1}J_i(q,k)d_{\scriptscriptstyle{{\cal V}^j}}(k)
\Bigg\{ C_i\left[k^\mu
g^\sigma_\alpha+(k-q)^\sigma g^\mu_\alpha\right] 
\g^{(1)}_{{\cal V}^i_\mu{\cal V}^j_\rho}(k)\g^{(0)}_{W^+_\sigma W^-_\beta
{\cal V}^{j,\rho}}(-k+q,-q,k)\nonumber \\
&-& C_j\left[k^\mu g^\sigma_\beta+(k-q)^\sigma
g^\mu_\beta\right] \g^{(1)}_{{\cal V}^j_\mu{\cal V}^i_\rho}(k)
\g^{(0)}_{W^-_\sigma W^+_\alpha
{\cal V}^{i,\rho}}(-k+q,-q,k)\Bigg\},
\label{start2Wn}
\eea
where ${\cal C}_{1L}={\cal C}_{1L}^{\rm c}+{\cal C}^{\rm n}_{1L}$ and 
the superscript ``c'' and ``n'' stands for
``charged'' and ``neutral'', depending on which one-loop propagator the
longitudinal momentum is hitting.

Let us start from the analysis of the ${\cal C}_{2L}$ contribution, 
Eq.\r{start1W}. For the two terms proportional to the 
longitudinal momentum $k$, the STI triggered will be the one-loop
version of the STI of Eq.\r{STIthreeW}; writing only the terms that we
are going to discard (as always from now on), this STI reads
\bea
k^\sigma \g^{(1)}_{W^\pm_\sigma W^\mp_{\lambda} {\cal V}^i_\rho}(-q,-k+q)&=&
-i\g^{(1)}_{u^\pm W^{*,\mp}_\sigma}(-k)
\g^{(0)}_{W^{\pm,\sigma} W^\mp_{\lambda} {\cal V}^i_\rho}(-q,-k+q)\nonumber \\
&-&i
\g^{(1)}_{u^\pm W^{*,\mp}_\sigma{\cal
V}^i_\rho}(-q,-k+q)\g^{(0)}_{W^{\pm,\sigma} W^\mp_{\lambda}}(-q)\nonumber \\
&-&i\g^{(0)}_{u^\pm W^{*,\mp}_\sigma{\cal
V}^i_\rho}(-q,-k+q)\g^{(1)}_{W^{\pm,\sigma} W^\mp_{\lambda}}(-q),
\label{STI1lW}
\eea
where $\lambda$ can be $\beta$ or $\alpha$ depending on which of the
two terms we are considering.

Next, making use of the following (mutually inverse) relations
\bea
\g_{u^\pm W^{*,\mp}_\sigma}(k)&=&ik_\sigma\g_{u^\pm W^{*,\mp}}(k),\nonumber \\
\g_{u^\pm W^{*,\mp}}(k)&=&-i\frac{k^\sigma}{k^2}\g_{u^\pm W^{*,\mp}_\sigma}(k),
\label{r1}
\eea 
and observing that
\be
\g^{(1)}_{u^\pm W^{*,\mp}}(k)
=-\frac i{k^2}L^{u\,(1)}_{\scriptscriptstyle{V}}(k),
\label{r2}
\ee
where $L^{u\,(1)}_{\scriptscriptstyle{V}}(k)$ 
denotes the part of the one-loop  $u^\pm\bar u^\pm$ ghost self-energy
which involves an internal gauge-boson propagator, we find
\be
-i\g^{(1)}_{u^\pm W^{*,\mp}_\sigma}(-k)
\g^{(0)}_{W^\pm_\sigma W^\mp_{\lambda} {\cal V}^i_\rho}(-q,-k+q)
=\pm\left(i\gw C_i\right)\frac1{k^2}L^{u\,(1)}_{\scriptscriptstyle{V}}(k)
\g^{(0)}_{W^\pm_\rho W^\mp_{\lambda}}(-q).
\ee
This result, together with the Feynman rules listed in Appendix~\ref{FR},
will finally give us the one-loop STI of Eq.\r{STI1lW} in its
final form, {\it i.e.},
\bea
k^\sigma \g^{(1)}_{W^\pm_\sigma W^\mp_{\lambda} {\cal V}^i_\rho}(-q,-k+q)&=&
\pm\left(i\gw C_i\right)\frac1{k^2}L^{u\,(1)}_{\scriptscriptstyle{V}}(k)
\g^{(0)}_{W^\pm_\rho W^\mp_{\lambda}}(q) \nonumber \\
&-&i\g^{(1)}_{u^\pm W^{*,\mp}_\sigma{\cal
V}^i_\rho}(-q,-k+q)\g^{(0)}_{W^{\pm,\sigma} W^\mp_{\lambda}}(q)\nonumber \\
&\pm&\left(\gw C_i\right)\g^{(1)}_{W^\pm_\rho W^\mp_{\lambda}}(q).
\eea

For the remaining two ${\cal C}_{2L}$ terms, proportional to
the longitudinal momentum $\left(k-q\right)$, the STI triggered will 
instead  be the one-loop
version of the STI of Eq.\r{STIthreeZA}, {\it i.e.},
\bea
\left(k-q\right)^\rho\g^{(1)}_{{\cal V}^i_\rho W^\pm_\sigma
W^\mp_\beta}(k,-q)&=&
i\sum_n\g^{(1)}_{u^i{\cal V}^{*,n}_\rho}(k-q)\g^{(0)}_{{\cal V}^{n,\rho}
W^\pm_\sigma 
W^\mp_{\lambda}}(k,-q) \nonumber\\
&+&i\g^{(1)}_{u^i W^{*,\mp}_\rho
W^\pm_\sigma}(-q,k)\g^{(0)}_{W^{\pm,\rho}W^\mp_{\lambda}}(-q) \nonumber \\
&+&i\g^{(0)}_{u^i W^{*,\mp}_\rho
W^\pm_{\sigma}}(-q,k)\g^{(1)}_{W^{\pm,\rho}W^\mp_{\lambda}}(-q).
\label{STI1lZA}
\eea

We next make use of the following relations
\bea
\g_{u^i{\cal V}^{*,n}_\rho}(k-q)&=&i(k-q)_\rho\g_{u^i{\cal
V}^{*,n}}(k-q), \nonumber \\
\g_{u^i{\cal V}^{*,n}}(k-q)&=&-i\frac{(k-q)^\rho}{(k-q)^2}\g_{u^i{\cal
V}^{*,n}_\rho}(k-q),
\eea
and observe that 
\be
\g^{(1)}_{u^i{\cal V}^{*,n}}(k-q)=-\frac
i{(k-q)^2}L^{in\,(1)}_{\scriptscriptstyle{V}} (k-q),
\ee
where $L^{in\,(1)}_V(k-q)$ represents
the part of the one-loop $u^i\bar u^n$ ghost
self-energy which involves an internal gauge-boson propagator, to find
\be
i\sum_n\g^{(1)}_{u^i{\cal V}^{*,n}_\rho}(k-q)\g^{(0)}_{{\cal V}^{n,\rho}
W^\pm_\sigma 
W^\mp_{\lambda}}(k,-q)=\pm\left(i\gw
C_n\right)\frac1{(k-q)^2}L^{in\,(1)}_{\scriptscriptstyle{V}}(k-q)
\g^{(0)}_{W^\pm_\sigma W^\mp_{\lambda}}(q).
\ee

This result, together with the Feynman rules listed in
Appendix~\ref{FR}, will finally give us the STI of Eq.\r{STI1lZA} in
its final form, {\it i.e.},
\bea
\left(k-q\right)^\rho\g^{(1)}_{{\cal V}^i_\rho W^\pm_\sigma
W^\mp_\beta}(k,-q)&=&\pm i\gw
\sum_n C_n\frac1{(k-q)^2}L^{in\,(1)}_{\scriptscriptstyle{V}}(k-q)
\g^{(0)}_{W^\pm_\sigma W^\mp_{\lambda}}(q) \nonumber \\
&-&
i\g^{(1)}_{u^i W^{*,\mp}_\rho
W^\pm_\sigma}(-q,k)\g^{(0)}_{W^{\pm,\rho}W^\mp_{\lambda}}(q) \nonumber \\
&\pm&\left(\gw C_i\right)
\g^{(1)}_{W^{\pm,\rho}W^\mp_{\lambda}}(q).
\eea

Using then the following properties of the BRST vertices
\bea
\g^{(1)}_{u^\pm W^{*,\mp}_\sigma{\cal
V}^i_\rho}(-q,-k+q)&=&\g^{(1)}_{u^\pm W^{*,\mp}{\cal
V}^i}(-q,-k+q)g_{\sigma\rho},\nonumber \\
\g^{(1)}_{u^i W^{*,\mp}_\rho
W^\pm_\sigma}(-q,k)&=&\g^{(1)}_{u^i W^{*,\mp}
W^\pm}(-q,k)g_{\rho\sigma},\nonumber \\
\g^{(1)}_{u^+ W^{*,-}{\cal
V}^i}(-q,-k+q)&=&-
\g^{(1)}_{u^- W^{*,+}{\cal
V}^i}(-q,-k+q),\nonumber \\
\g^{(1)}_{u^i W^{*,-}
W^+}(-q,k)&=&-\g^{(1)}_{u^i W^{*,+}
W^-}(-q,k),
\eea
we can write the discarded terms coming from the ${\cal C}_{2L}$
contribution as
\bea
{\cal C}_{2L}&=&
2\gw^2\sum_i C_i\int_{L_1}J'_i(q,k)
\left[C_i\frac1{k^2}L^{u\,(1)}_{\scriptscriptstyle{V}}(k)+\sum_j
C_j\frac1{(k-q)^2}L^{ij\,(1)}_{\scriptscriptstyle{V}} (k-q)
\right]\g^{(0)}_{W^+_\alpha W^-_\beta}(q) \nonumber \\
&-&2\gw\sum_i C_i\int_{L_1}J'_i(q,k)\left[\g^{(1)}_{u^+ W^{*,-}_\sigma{\cal
V}^i_\alpha}(-q,-k+q)+\g^{(1)}_{u^i W^{*,-}_\sigma
W^+_\alpha}(-q,k)\right]\g^{(0)}_{W^{+,\sigma} W^-_\beta}(q) \nonumber
\\
&-&4i\gw^2\sum_iC_i^2\int_{L_1}J'_i(q,k)\g^{(1)}_{W^+_\alpha W^-_\beta}(q).
\eea
  
We next consider the lower order corrections, {\it i.e.}, the
contributions coming from the ${\cal C}_{1L}$ terms, beginning from
the charged contribution ${\cal C}^{\rm c}_{1L}$ of Eq.\r{start2Wc}.

For the two terms proportional to the longitudinal momentum $k$, the
STI triggered will be the one-loop version of Eq.\r{STItwopointW},
which reads
\bea
k^\sigma\g^{(1)}_{W^\pm_\sigma W^\mp_\mu}(k)&=&
i\g^{(1)}_{u^\pm
W^{*,\mp}_\sigma}(k)\g^{(0)}_{W^{\pm,\sigma}W^\mp_\mu}(k)
+i\g^{(1)}_{u^\pm\phi^{*,\mp}}(k)\g^{(0)}_{\phi^\pm
W^\mp_\mu}(k)\nonumber \\
&+&i\g^{(0)}_{u^\pm\phi^{*,\mp}}(k)\g^{(1)}_{\phi^\pm
W^\mp_\mu}(k).
\label{b1}
\eea
Using then the tree-level version of the STIs of Eqs.\r{STItwopointW}
and~\r{STIthreeW}, we find
\be
i\g^{(1)}_{u^\pm
W^{*,\mp}_\sigma}(k)\g^{(0)}_{W^{\pm,\sigma}W^\mp_\mu}(k)
\g^{(0)}_{W^{\pm,\mu}W^\mp_{\lambda}{\cal V}^i_\rho}(k,-q,-k+q)=
\mp\left(\gw
C_i\right)\Mw^2\frac1{k^2}L^{u\,(1)}_{\scriptscriptstyle{V}}(k)
\g^{(0)}_{W^\pm_\rho W^{\mp}_{\lambda}}(q).
\ee

Moreover observing that 
\be
\Mw\g^{(1)}_{u^\pm\phi^{*,\mp}}=\mp L_{\scriptscriptstyle{G}}^{u\,(1)}(k),
\ee
where $L_{\scriptscriptstyle{G}}^{u\,(1)}(k)$ is the part of the one-loop
$u^\pm\bar u^\pm$ ghost self-energy involving an internal Goldstone
boson propagator, we find
\be
i\g^{(1)}_{u^\pm\phi^{*,\mp}}(k)\g^{(0)}_{\phi^\pm
W^\mp_\mu}(k)
\g^{(0)}_{W^{\pm,\mu}W^\mp_{\lambda}{\cal V}^i_\rho}(k,-q,-k+q)=
\mp\left(\gw C_i\right)L^{u\,(1)}_{\scriptscriptstyle{G}}(k)
\g^{(0)}_{W^\pm_\rho W^{\mp}_{\lambda}}(q).
\ee
Finally, the last term appearing in Eq.~\r{b1} is part of the PT
answer, and it is precisely the term responsible for converting the
conventional 
$W^\pm\phi^\mp{\cal V}^i$ vertex into the corresponding BFM vertex 
 $\widehat{W}^\pm\phi^\mp{\cal V}^i$ .

For the remaining two ${\cal C}^{\rm c}_{1L}$ terms proportional to
the longitudinal momentum $(k-q)$, the STI triggered will simply be
the one of Eq.~\r{two}; thus, the discarded term stemming from
${\cal C}^{\rm c}_{1L}$, reads
\bea
{\cal C}^{\rm c}_{1L}&=&2\gw^2\sum_i
C^2_i\int_{L_1}J'_i(q,k)\dw(k)
\left[\frac{\Mw^2}{k^2}L^{u\,(1)}_{\scriptscriptstyle{V}}(k)+
L^{u\,(1)}_{\scriptscriptstyle{G}}(k)\right]\g^{(0)}_{W^+_\alpha
W^-_\beta}(q)
\nonumber \\
&-& 2\gw^2\sum_i
C^2_i\int_{L_1}J'_i(q,k)\dw(k)
\g^{(1)}_{W^+_\alpha W^-_\mu}(k)
\g^{(0)}_{W^{+,\mu} W^-_\beta}(q).
\eea

In the case of the neutral contributions ${\cal C}^{\rm n}_{1L}$, the
STI triggered by the two terms proportional to the longitudinal
momentum $k$, will be the one-loop version of
Eq.\r{STItwopointZA}. Taking into account that the Green's functions
$\g_{{\cal V}^n_\rho H}$ vanish at tree-, one- and two-loop level as
they violate {\it CP} invariance, we arrive at the one-loop STI
\bea
k^\mu\g^{(1)}_{{\cal V}^{\rm a}_\mu{\cal V}^{\rm b}_\rho}(k)&=&
i\sum_n\g^{(1)}_{u^{\rm a}{\cal V}^{*,n}_\mu}(k)\g^{(0)}_{{\cal V}^{n,\mu}
{\cal V}^{\rm b}_\rho}(k)
+i\g^{(1)}_{u^{\rm a}\chi^*}(k)\g^{(0)}_{\chi{\cal V}^{\rm b}_\rho}(k)
\nonumber \\
&+& i\g^{(0)}_{u^{\rm a}\chi^*}(k)\g^{(1)}_{\chi{\cal V}^{\rm b}_\rho}(k),
\label{b2}
\eea
where a$=i,j$, b$=j,i$ and depending on which of the
two terms we are considering.

Using then the tree-level version of the STIs of Eqs.\r{STItwopointZA}
and~\r{STIthreeZA}, we find
\be
i\sum_n\g^{(1)}_{u^{\rm a}{\cal V}^{*,n}_\mu}(k)
\g^{(0)}_{{\cal V}^{n,\mu}{\cal V}^{\rm b}_\rho}(k)
\g^{(0)}_{W^\pm_\sigma W^\mp_{\lambda}{\cal V}^{\rm b}_\rho}
(-k+q,-q,k)=
\pm\left(\gw C_{\rm b}\right)M^2_{\scriptscriptstyle{{\cal
V}^{\rm b}}}\frac1{k^2}\g^{(0)}_{W^\pm_\sigma W^\mp_{\lambda}}(q).
\ee
Moreover, observing that
\be
iM_{{\cal V}^{\rm b}}\g^{(1)}_{u^{\rm a}\chi^*}(k)=L^{{\rm
ab}\,(1)}_{\scriptscriptstyle{G}}(k),
\ee
being $L_{\scriptscriptstyle{G}}^{{\rm ab}\,(1)}(k)$ the part of the one-loop
$u^{\rm a}\bar u^{\rm b}$ ghost self-energy involving an internal Goldstone
boson propagator, we find
\be
i\g^{(1)}_{u^{\rm a}\chi^*}(k)\g^{(0)}_{\chi{\cal V}^{\rm b}_\rho}(k)
\g^{(0)}_{W^\pm_\sigma W^\mp_{\lambda}{\cal V}^{\rm b}_\rho}
(-k+q,-q,k)=\pm\left(\gw C_{\rm b}\right)
L_{\scriptscriptstyle{G}}^{{\rm ab}\,(1)}(k)\g^{(0)}_{W^\pm_\sigma
W^\mp_{\lambda}}(q).
\ee
Finally, the last term appearing in Eq.~\r{b2} is part of the PT
answer, and it is precisely the term responsible for generating 
the characteristic BFM vertex 
$\chi\widehat{W}^\pm W^\mp$ (recall that the tree-level 
coupling $\chi W^\pm W^\mp$ does not exist 
in the conventional $R_\xi$ gauges).

For the remaining two ${\cal C}^{\rm n}_{1L}$ terms proportional to
the longitudinal momentum $(k-q)$, the STI triggered will simply be
\be
\left(k-q\right)^\sigma\g^{(0)}_{W^\pm_\sigma W^\mp_{\lambda}{\cal
V}^{\rm a}_\rho}(-q,k)=\mp\left(\gw C_{\rm a}\right)
\g^{(0)}_{W^\pm_\rho W^\mp_{\lambda}}(q).
\ee

Thus, the discarded term originating from
${\cal C}^{\rm n}_{1L}$, reads
\bea
{\cal C}^{\rm n}_{1L}&=&2\gw^2\sum_{i,j}C_i C_j\int_{L_1} J_i(q,k)
d_{\scriptscriptstyle{{\cal V}^j}}(k)
\left[
\frac{M^2_{\scriptscriptstyle{{\cal
V}^j}}}{k^2}L^{ij\,(1)}_{\scriptscriptstyle{V}}(k)
+L^{ij\,(1)}_{\scriptscriptstyle{G}}(k)\right]\g^{(0)}_{W^+_\alpha
W^-_\beta}(q) \nonumber \\
&-&2\gw^2\sum_{i,j}C_i C_j\int_{L_1} J_i(q,k)
d_{\scriptscriptstyle{{\cal V}^j}}(k)
\g^{(1)}_{{\cal V}^i_\alpha {\cal
V}^j_\rho}(k)\g^{(0)}_{W^{+,\rho}W^-_\beta}(q).
\eea

Having analyzed in detail all the contributions, we can now add them
up to construct the quantity $\Pi^{{\rm IP}\,(2)}_{\alpha\beta}(q)$;
after some algebra, and using the relations 
\bea
\frac{M^2_{\scriptscriptstyle{V}}}{k^2\left(k^2-
M^2_{\scriptscriptstyle{V}}\right)}+\frac1{k^2}&=&
\frac1{k^2-M^2_{\scriptscriptstyle{V}}}, \nonumber \\
L^{u\,(1)}_{\scriptscriptstyle{V}}(k)+L^{u\,(1)}_{\scriptscriptstyle{G}}(k)
&=&L^{u\,(1)}(k),\nonumber \\
L^{ij\,(1)}_{\scriptscriptstyle{V}}(k)+L^{ij\,(1)}_{\scriptscriptstyle{G}}(k)
&=&L^{ij\,(1)}(k),
\eea
the discarded terms furnished by the intrinsic PT algorithm 
will give rise to the quantity
\bea
\Pi^{\rm IP\,(2)}_{\alpha\beta}(q)
&=& 2\gw^2\sum_iC_i\int_{L_1}\left[
C_iJ'_i(q,k)\dw(k)
L^{u\,(1)}(k)+
\sum_j C_jJ_i(q,k)d_{\scriptscriptstyle{{\cal V}^j}}(k)
L^{ij\,(1)}(k)\right]
\times
\nonumber \\ &\times&
\g^{(0)}_{W^+_\alpha W^-_\beta}(q)\nonumber \\
&-&2\gw^2\sum_iC_i\int_{L_1}\left[
C_iJ'_i(q,k)\dw(k)
\g^{(1)}_{W^+_\alpha W^-_\mu}(k)+
\sum_j C_jJ_i(q,k)d_{\scriptscriptstyle{{\cal V}^j}}(k)
\g^{(1)}_{{\cal
V}^i_\alpha{\cal V}^j_\mu}(k)\right]
\times\nonumber \\
&\times&\g^{(0)}_{W^{+,\mu} W^-_\beta}(q)
\nonumber\\
&-&2\gw\sum_i C_i\int_{L_1}J'_i(q,k)\left[\g^{(1)}_{u^+ W^{*,-}_\sigma{\cal
V}^i_\alpha}(-q,-k+q)+\g^{(1)}_{u^i W^{*,-}_\sigma
W^+_\alpha}(-q,k)\right]\times\nonumber \\
&\times&\g^{(0)}_{W^{+,\sigma} W^-_\beta}(q) \nonumber
\\
&-&4i\gw^2\sum_iC_i^2\int_{L_1}J'_i(q,k)\g^{(1)}_{W^+_\alpha W^-_\beta}(q).
\label{piIPW}
\eea

\begin{figure}[t]
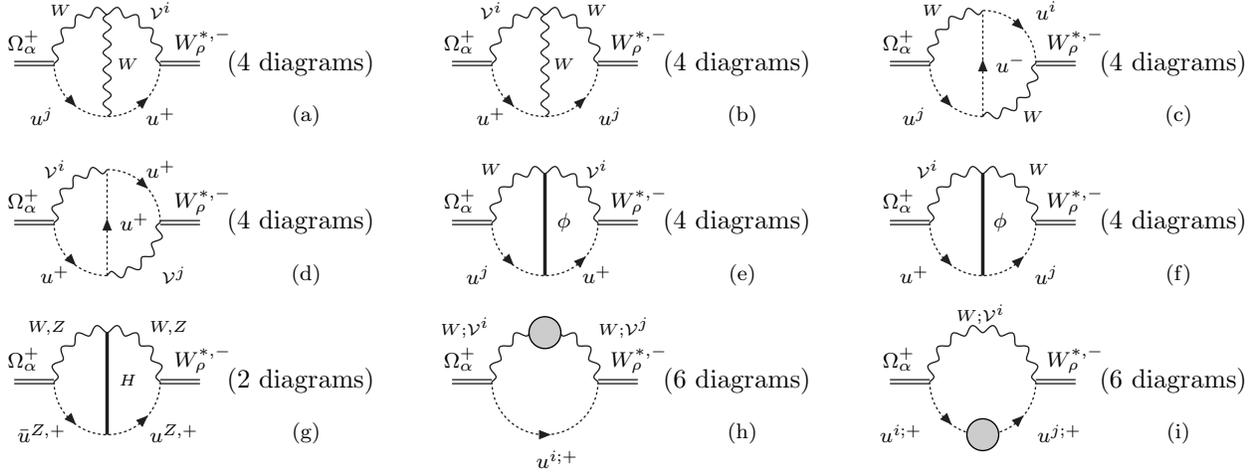

\bce
\bpi(0,160)(130,-125)

\Line(-105,0.75)(-90,0.75)
\Line(-105,-0.75)(-90,-0.75)
\PhotonArc(-70,0)(20,0,180){-1.5}{8.5}
\DashCArc(-70,0)(20,180,360){1}
\Line(-50,0.75)(-35,0.75)
\Line(-50,-0.75)(-35,-0.75)
\Photon(-70,18.5)(-70,-20){-1.5}{6}
\DashArrowLine(-57,-15)(-55.5,-13.6){1}
\DashArrowLine(-84,-13.8)(-82.5,-15.4){1}

\Text(-25,0)[l]{\footnotesize{(4 diagrams)}}
\Text(5,-20)[c]{\scriptsize{(a)}}

\Line(60,0.75)(75,0.75)
\Line(60,-0.75)(75,-0.75)
\PhotonArc(95,0)(20,0,180){-1.5}{8.5}
\DashCArc(95,0)(20,180,360){1}
\Photon(95,18.5)(95,-20){-1.5}{6}
\Line(115,0.75)(130,0.75)
\Line(115,-0.75)(130,-0.75)
\DashArrowLine(108,-15)(109.5,-13.6){1}
\DashArrowLine(81,-13.8)(82.5,-15.4){1}

\Text(140,0)[l]{\footnotesize{(4 diagrams)}}
\Text(170,-20)[c]{\scriptsize{(b)}}

\Line(225,0.75)(240,0.75)
\Line(225,-0.75)(240,-0.75)
\PhotonArc(260,0)(20,90,180){-1.5}{4.5}
\PhotonArc(260,0)(20,270,360){-1.5}{4.5}
\DashCArc(260,0)(20,0,90){1}
\DashCArc(260,0)(20,180,270){1}
\DashArrowLine(260,-20)(260,18.5){1}
\Line(280,0.75)(295,0.75)
\Line(280,-0.75)(295,-0.75)
\DashArrowLine(273,15)(274.5,13.6){1}
\DashArrowLine(246,-13.8)(247.5,-15.4){1}

\Text(305,0)[l]{\footnotesize{(4 diagrams)}}
\Text(335,-20)[c]{\scriptsize{(c)}}

\Line(-105,-59.25)(-90,-59.25)
\Line(-105,-60.75)(-90,-60.75)
\PhotonArc(-70,-60)(20,90,180){-1.5}{4.5}
\PhotonArc(-70,-60)(20,270,360){-1.5}{4.5}
\DashCArc(-70,-60)(20,0,90){1}
\DashCArc(-70,-60)(20,180,270){1}
\Line(-50,-59.25)(-35,-59.25)
\Line(-50,-60.75)(-35,-60.75)
\DashArrowLine(-70,-80)(-70,-41.5){1}
\DashArrowLine(-57,-45)(-55.5,-46.4){1}
\DashArrowLine(-84,-73.8)(-82.5,-75.4){1}

\Text(-25,-60)[l]{\footnotesize{(4 diagrams)}}
\Text(5,-80)[c]{\scriptsize{(d)}}

\Line(60,-59.25)(75,-59.25)
\Line(60,-60.75)(75,-60.75)
\PhotonArc(95,-60)(20,0,180){-1.5}{8.5}
\DashCArc(95,-60)(20,180,360){1}
\SetWidth{1.3}
\Line(95,-41.5)(95,-80)
\SetWidth{0.5}
\Line(115,-59.25)(130,-59.25)
\Line(115,-60.75)(130,-60.75)
\DashArrowLine(108,-75)(109.5,-73.6){1}
\DashArrowLine(81,-73.8)(82.5,-75.4){1}

\Text(140,-60)[l]{\footnotesize{(4 diagrams)}}
\Text(170,-80)[c]{\scriptsize{(e)}}

\Line(225,-59.25)(240,-59.25)
\Line(225,-60.75)(240,-60.75)
\PhotonArc(260,-60)(20,0,180){-1.5}{8.5}
\DashCArc(260,-60)(20,180,360){1}
\SetWidth{1.3}
\Line(260,-41.5)(260,-80)
\SetWidth{0.5}
\Line(280,-59.25)(295,-59.25)
\Line(280,-60.75)(295,-60.75)
\DashArrowLine(273,-75)(274.5,-73.6){1}
\DashArrowLine(246,-73.8)(247.5,-75.4){1}

\Text(305,-60)[l]{\footnotesize{(4 diagrams)}}
\Text(335,-80)[c]{\scriptsize{(f)}}

\Line(-105,-119.25)(-90,-119.25)
\Line(-105,-120.75)(-90,-120.75)
\PhotonArc(-70,-120)(20,0,180){-1.5}{8.5}
\DashCArc(-70,-120)(20,180,360){1}
\SetWidth{1.3}
\Line(-70,-101.5)(-70,-140)
\SetWidth{0.5}
\Line(-50,-119.25)(-35,-119.25)
\Line(-50,-120.75)(-35,-120.75)
\DashArrowLine(-57,-135)(-55.5,-133.6){1}
\DashArrowLine(-84,-133.8)(-82.5,-135.4){1}

\Text(-25,-120)[l]{\footnotesize{(2 diagrams)}}
\Text(5,-140)[c]{\scriptsize{(g)}}

\Line(60,-119.25)(75,-119.25)
\Line(60,-120.75)(75,-120.75)
\PhotonArc(95,-120)(20,0,180){-1.5}{8.5}
\DashCArc(95,-120)(20,180,360){1}
\Line(115,-119.25)(130,-119.25)
\Line(115,-120.75)(130,-120.75)
\DashArrowLine(94.5,-140)(95.5,-140){1}
\GCirc(95,-101.5){6}{0.8}

\Text(140,-120)[l]{\footnotesize{(6 diagrams)}}
\Text(170,-140)[c]{\scriptsize{(h)}}

\Line(225,-119.25)(240,-119.25)
\Line(225,-120.75)(240,-120.75)
\PhotonArc(260,-120)(20,0,180){-1.5}{8.5}
\DashCArc(260,-120)(20,180,360){1}
\Line(280,-119.25)(295,-119.25)
\Line(280,-120.75)(295,-120.75)
\DashArrowLine(273,-135)(274.5,-133.6){1}
\DashArrowLine(246,-133.8)(247.5,-135.4){1}
\GCirc(260,-140){6}{0.8}

\Text(305,-120)[l]{\footnotesize{(6 diagrams)}}
\Text(335,-140)[c]{\scriptsize{(i)}}

\Text(-108,8)[l]{$\scriptstyle{\Omega^+_\alpha}$}
\Text(57,8)[l]{$\scriptstyle{\Omega^+_\alpha}$}
\Text(223,8)[l]{$\scriptstyle{\Omega^+_\alpha}$}
\Text(-108,-52)[l]{$\scriptstyle{\Omega^+_\alpha}$}
\Text(57,-52)[l]{$\scriptstyle{\Omega^+_\alpha}$}
\Text(223,-52)[l]{$\scriptstyle{\Omega^+_\alpha}$}
\Text(-108,-112)[l]{$\scriptstyle{\Omega^+_\alpha}$}
\Text(57,-112)[l]{$\scriptstyle{\Omega^+_\alpha}$}
\Text(223,-112)[l]{$\scriptstyle{\Omega^+_\alpha}$}

\Text(-45,8)[l]{$\scriptstyle{W^{*,-}_\rho}$}
\Text(120,8)[l]{$\scriptstyle{W^{*,-}_\rho}$}
\Text(285,8)[l]{$\scriptstyle{W^{*,-}_\rho}$}
\Text(-45,-52)[l]{$\scriptstyle{W^{*,-}_\rho}$}
\Text(120,-52)[l]{$\scriptstyle{W^{*,-}_\rho}$}
\Text(285,-52)[l]{$\scriptstyle{W^{*,-}_\rho}$}
\Text(-45,-112)[l]{$\scriptstyle{W^{*,-}_\rho}$}
\Text(120,-112)[l]{$\scriptstyle{W^{*,-}_\rho}$}
\Text(285,-112)[l]{$\scriptstyle{W^{*,-}_\rho}$}

\Text(-88,20)[c]{$\scriptscriptstyle{W}$}
\Text(-50,20)[c]{$\scriptscriptstyle{{\cal V}^i}$}
\Text(-62.5,0)[c]{$\scriptscriptstyle{W}$}
\Text(-95,-20)[c]{$\scriptstyle{u^j}$}
\Text(-50,-20)[c]{$\scriptstyle u^+$}

\Text(115,20)[c]{$\scriptscriptstyle{W}$}
\Text(75,20)[c]{$\scriptscriptstyle{{\cal V}^i}$}
\Text(102.5,0)[c]{$\scriptscriptstyle{W}$}
\Text(120,-20)[c]{$\scriptstyle{u^j}$}
\Text(75,-20)[c]{$\scriptstyle u^+$}

\Text(75,-40)[c]{$\scriptscriptstyle{W}$}
\Text(115,-40)[c]{$\scriptscriptstyle{{\cal V}^i}$}
\Text(102.5,-60)[c]{$\scriptstyle{\phi}$}
\Text(70,-80)[c]{$\scriptstyle{u^j}$}
\Text(115,-80)[c]{$\scriptstyle u^+$}

\Text(282,-40)[c]{$\scriptscriptstyle{W}$}
\Text(240,-40)[c]{$\scriptscriptstyle{{\cal V}^i}$}
\Text(267.5,-60)[c]{$\scriptstyle{\phi}$}
\Text(285,-80)[c]{$\scriptstyle{u^j}$}
\Text(235,-80)[c]{$\scriptstyle u^+$}

\Text(-93,-100)[c]{$\scriptscriptstyle{W,Z}$}
\Text(-47,-100)[c]{$\scriptscriptstyle{W,Z}$}
\Text(-62.5,-120)[c]{$\scriptscriptstyle{H}$}
\Text(-95,-140)[c]{$\scriptstyle{\bar u^{Z,+}}$}
\Text(-45,-140)[c]{$\scriptstyle u^{Z,+}$}

\Text(242,20)[c]{$\scriptscriptstyle{W}$}
\Text(285,20)[c]{$\scriptstyle{u^i}$}
\Text(271.5,0)[c]{$\scriptstyle{u^-}$}
\Text(235,-20)[c]{$\scriptstyle{u^j}$}
\Text(280,-20)[c]{$\scriptscriptstyle{W}$}

\Text(-50,-40)[c]{$\scriptstyle{u^+}$}
\Text(-90,-40)[c]{$\scriptscriptstyle{{\cal V}^i}$}
\Text(-60,-60)[c]{$\scriptstyle{u^+}$}
\Text(-45,-80)[c]{$\scriptscriptstyle{{\cal V}^j}$}
\Text(-90,-80)[c]{$\scriptstyle u^+$}

\Text(65,-100)[c]{$\scriptscriptstyle{W;{\cal V}^i}$}
\Text(125,-100)[c]{$\scriptscriptstyle{W;{\cal V}^j}$}
\Text(100,-150)[c]{$\scriptstyle{u^{i;+}}$}

\Text(230,-140)[c]{$\scriptstyle{u^{i;+}}$}
\Text(290,-140)[c]{$\scriptstyle{u^{j;+}}$}
\Text(260,-95)[c]{$\scriptscriptstyle{W;{\cal V}^i}$}

\epi
\ece

\caption{\label{figxx} Two loop contributions to the two-point function
$\g^{(2)}_{\Omega^+_\alpha W^{*,-}_\rho}(q)$. The blobs represent one-loop
corrections to the corresponding propagator.}

\end{figure}

The last step consists in comparing this result to the 
one coming from the BQI of Eq.\r{tpfBQIW}. 
At two loops this BQI reads
\bea
\g^{(2)}_{\widehat{W}^+_\alpha\widehat{W}^-_\beta}(q)&=&
\g^{(2)}_{W^+_\alpha W^-_\beta}(q)
+2\g^{(2)}_{\Omega^+_\alpha W^{*,-}_{\rho}}(q)\g^{(0)}_{W^{+,\rho} 
W^{-}_\beta}(q)+2\g^{(1)}_{\Omega^+_\alpha W^{*,-}_{\rho}}(q)
\g^{(1)}_{W^{+,\rho} W^{-}_\beta}(q) \nonumber \\
&+&\g^{(1)}_{\Omega^+_\alpha W^{*,-}_\rho}(q)
\g^{(0)}_{W^{+,\rho} W^{-,\sigma}}(q)
\g^{(1)}_{\Omega^-_\beta W^{*,+}_\sigma}(q),
\label{BQIc2l}
\eea
and the diagrams contributing to the two-loop two-point function
$\g^{(2)}_{\Omega^+_\alpha W^{*,-}_{\rho}}$ are shown in
Fig.\ref{figxx}. Using the one-loop result of Eq.\r{r1W}, 
it is then easy to prove that
\be
2\g^{(1)}_{\Omega^+_\alpha W^{*,-}_{\rho}}(q)
\g^{(1)}_{W^{+,\rho} W^{-}_\beta}(q)\equiv
4i\gw^2\sum_iC_i^2\int_{L_1}J'_i(q,k)\g^{(1)}_{W^+_\alpha W^-_\beta}(q),
\ee
while it is a long but straightforward exercise to demonstrate that
\bea
& & \hspace{5.0cm}
2\g^{(2)}_{\Omega^+_\alpha W^{*,-}_{\rho}}(q)\g^{(0)}_{W^{+,\rho} 
W^{-}_\beta}(q)\nonumber \\
&\equiv&
-2\gw^2\sum_iC_i\int_{L_1}\left[
C_iJ'_i(q,k)\dw(k)
L^{u\,(1)}(k)+
\sum_j C_jJ_i(q,k)d_{\scriptscriptstyle{{\cal V}^j}}(k)
L^{ij\,(1)}(k)\right]\times
\nonumber \\
&\times&\g^{(0)}_{W^+_\alpha W^-_\beta}(q)\nonumber \\
&+&2\gw^2\sum_iC_i\int_{L_1}\left[
C_iJ'_i(q,k)\dw(k)
\g^{(1)}_{W^+_\alpha W^-_\mu}(k)+
\sum_j C_jJ_i(q,k)d_{\scriptscriptstyle{{\cal V}^j}}(k)
\g^{(1)}_{{\cal
V}^i_\alpha{\cal V}^j_\mu}(k)\right]\times\nonumber \\
&\times&\g^{(0)}_{W^{+,\mu} W^-_\beta}(q)
\nonumber\\
&+&2\gw\sum_i C_i\int_{L_1}J'_i(q,k)\left[\g^{(1)}_{u^+ W^{*,-}_\sigma{\cal
V}^i_\alpha}(-q,-k+q)+\g^{(1)}_{u^i W^{*,-}_\sigma
W^+_\alpha}(-q,k)\right]\g^{(0)}_{W^{+,\sigma}
W^-_\beta}(q). \nonumber \\
\eea
In particular the first term on the RHS of the above equation gives
the diagrams of Fig.\ref{figxx}i, the second term the ones of
Fig.\ref{figxx}h, and the last term all the remaining diagrams of
Fig.\ref{figxx} (a--g). 

The last term appearing in Eq.\r{BQIc2l}
will be finally generated in the conversion of the 1PR strings into the
1PR PT strings, as follows. After treating the conventional 1PR
diagrams involving charged gauge-boson self-energies along the same
lines explained in \cite{Binosi:2002ez} for the QCD case, one
arrives, after discarding the terms proportional to the inverse
propagator of the external legs, to the result 
\be
S'^{{\rm IP}\,(2)}_{\alpha\beta}(q)=2iV^{{\rm P}\,(1)}_{\alpha\rho}(q)
\g^{(1)}_{W^{+,\rho}W^-_\beta}(q)+\g^{(1)}_{\Omega^+_\alpha
W^{*,-}_\rho}(q)\g^{(0)}_{W^{+,_\rho}W^{-,\sigma}}(q)
\g^{(1)}_{\Omega^-_\beta W^{*,+}_\sigma}(q).
\label{p1}
\ee
On the other hand, from Eq.\r{piIPW} we have
\be
\Pi'^{{\rm IP}\,(2)}_{\alpha\beta}(q)=-2iV^{{\rm P}\,(1)}_{\alpha\rho}(q)
\g^{(1)}_{W^{+,\rho}W^-_\beta}(q),
\label{p2}
\ee
so that adding by parts these two equations we obtain
\be
R^{{\rm IP}\,(2)}_{\alpha\beta}(q)=\g^{(1)}_{\Omega^+_\alpha
W^{*,-}_\rho}(q)\g^{(0)}_{W^{+,_\rho}W^{-,\sigma}}(q)
\g^{(1)}_{\Omega^-_\beta W^{*,+}_\sigma}(q).
\ee

Thus, finally, the quantity $-\Pi^{{\rm IP}\,(2)}_{\alpha\beta}(q)
+R^{{\rm IP}\,(2)}_{\alpha\beta}(q)$ will provide precisely all the
terms appearing in the two-loop version of the relevant BQI [Eq.\r{BQI2lZA}],
{\it i.e.}, one has 
\bea
-\Pi^{{\rm IP}\,(2)}_{\alpha\beta}(q)
+R^{{\rm IP}\,(2)}_{\alpha\beta}(q)&=&
2\g^{(2)}_{\Omega^+_\alpha W^{*,-}_{\rho}}(q)\g^{(0)}_{W^{+,\rho} 
W^{-}_\beta}(q)+
2\g^{(1)}_{\Omega^+_\alpha W^{*,-}_{\rho}}(q)
\g^{(1)}_{W^{+,\rho} W^{-}_\beta}(q) \nonumber \\
&+&\g^{(1)}_{\Omega^+_\alpha W^{*,-}_\rho}(q)
\g^{(0)}_{W^{+,\rho} W^{-,\sigma}}(q)
\g^{(1)}_{\Omega^-_\beta W^{*,+}_\sigma}(q). \nonumber \\
\eea
Then, the difference between $\widehat\g^{(2)}_{W^+_\alpha W^-_\beta}(q)$
and $\g^{(2)}_{W^+_\alpha W^-_\beta}(q)$ given by the intrinsic PT
definition of Eq.\r{ipdefW}, is the same as the difference between 
$\g^{(2)}_{\widehat W^+_\alpha\widehat W^-_\beta}(q)$ and
$\g^{(2)}_{W^+_\alpha W^-_\beta}(q)$ given by the BQI, thus proving
that
\be
\widehat\g^{(2)}_{W^+_\alpha W^-_\beta}(q)\equiv
\g^{(2)}_{\widehat W^+_\alpha\widehat W^-_\beta}(q).
\ee

\section{\label{sec:2lns}Two-loop Neutral sector}

In the neutral gauge-boson sector the starting point will be the
following expressions
\bea
{\cal C}_{2L}^{ij}&=&i\gw\int_{L_1}
J_{\scriptscriptstyle{W}}(q,k)
\left\{C_i\left[k^\sigma g_\alpha^\rho+(k-q)^\rho g_\alpha^\sigma\right]
\g^{(1)}_{W^+_\sigma W^-_\rho {\cal V}^j_\beta}(k,-k+q,-q)\right. \nonumber \\
&-&\left. C_j\left[k^\sigma g_\beta^\rho+(k-q)^\rho g_\beta^\sigma\right]
\g^{(1)}_{W^-_\sigma W^+_\rho {\cal V}^i_\alpha}(k,-k+q,-q)\right\}, 
\label{start1ZA}\\
{\cal C}_{1L}^{ij}&=&2\gw\int_{L_1}J_{\scriptscriptstyle{W}}(q,k)\dw(q)
\times \nonumber \\
&\times& 
\left\{C_i\left[k^\sigma g_\alpha^\rho+(k-q)^\rho
g_\alpha^\sigma\right]\g^{(1)}_{W^+_\sigma W^-_\mu}(k)
\g^{(0)}_{W^{+,\mu} W^-_\rho {\cal V}^j_\beta}(k,-k+q,-q)\right.
\nonumber \\
&-&\left. C_j\left[k^\sigma g_\beta^\rho+(k-q)^\rho
g_\beta^\sigma\right]
\g^{(1)}_{W^-_\sigma W^+_\mu}(k)
\g^{(1)}_{W^{-,\mu} W^+_\rho {\cal V}^i_\alpha}(k,-k+q,-q)\right\}. 
\label{start2ZA}
\eea

We then start the analysis from the ${\cal C}^{ij}_{2L}$
contributions, Eq.\r{start1ZA}. For the two terms proportional to the
longitudinal momentum $k$, the STI triggered will be the one-loop
version of the STI of Eq.\r{STIthreeW}, which, making use of Eqs.\r{r1}
and \r{r2} and the Feynman rules of Appendix \ref{FR}, in this case reads
\bea
k^\sigma\g^{(1)}_{W^{\pm,\sigma} W^\mp_\rho
{\cal V}^{\rm a}_{\lambda}}(-k+q,-q)&=&
\mp \left(i\gw C_{\rm
a}\right)\frac1{k^2}L^{u\,(1)}_{\scriptscriptstyle V}(k)
\g^{(0)}_{{\cal V}^{\rm a}_\rho{\cal V}^{\rm a}_{\lambda}}(q)
\nonumber \\
&-&i\g^{(1)}_{u^\pm{\cal V}^{*,{\rm a}}_\sigma W^\mp_\rho}(-q,-k+q)
\g^{(0)}_{{\cal V}^{{\rm a},\sigma} {\cal V}^{\rm a}_{\lambda}}(q)
\nonumber \\
&\mp&\sum_{n}\left(\gw C_n\right)
\g^{(1)}_{{\cal V}^n_\rho{\cal V}^{\rm a}_{\lambda}}(q). 
\eea

For the two remaining ${\cal C}^{ij}_{2L}$ terms, proportional to the
longitudinal momentum $(k-q)$ the STI triggered will read instead
\bea
(k-q)^\rho\g^{(1)}_{W^{\mp,\rho} W^\pm_\sigma
{\cal V}^{\rm a}_{\lambda}}(k,-q)&=&
\mp \left(i\gw C_{\rm
a}\right)\frac1{(k-q)^2}L^{u\,(1)}_{\scriptscriptstyle V}(k-q)
\g^{(0)}_{{\cal V}^{\rm a}_\sigma{\cal V}^{\rm a}_{\lambda}}(q)
\nonumber \\
&+&i\g^{(1)}_{u^\mp{\cal V}^{*,{\rm a}}_\rho W^\pm_\sigma}(-q,k)
\g^{(0)}_{{\cal V}^{{\rm a},\rho} {\cal V}^{\rm a}_{\lambda}}(q)
\nonumber \\
&\mp&\sum_{n}\left(\gw C_n\right)
\g^{(1)}_{{\cal V}^n_\sigma{\cal V}^{\rm a}_{\lambda}}(q).
\eea

Thus, the discarded terms stemming from the ${\cal C}^{ij}_{2L}$
contributions, can be written as
\bea
{\cal C}^{ij}_{2L}&=&2\gw^2C_iC_j\int_{L_1}J_{\scriptscriptstyle{W}}(q,k)
\frac1{k^2}L^{u\,(1)}_{\scriptscriptstyle V}(k)
\left[\g^{(0)}_{{\cal V}^i_\alpha{\cal V}^i_\beta}(q)+
\g^{(0)}_{{\cal V}^j_\alpha{\cal V}^j_\beta}(q)\right]\nonumber \\
&+&2\gw\int_{L_1}J_{\scriptscriptstyle{W}}(q,k)
\left[C_i\g^{(1)}_{u^+{\cal V}^{*,j}_\sigma W^-_\alpha}(-q,-k+q)
\g^{(0)}_{{\cal V}^{j,\sigma}{\cal V}^j_\beta}(q)\right.\nonumber\\
&-&\left.
C_j\g^{(1)}_{u^-{\cal V}^{*,i}_\sigma W^+_\beta}(-q,-k+q)
\g^{(0)}_{{\cal V}^i,\alpha{\cal V}^{i,\sigma}}(q)\right]\nonumber \\
&-&2i\gw^2\int_{L_1}J_{\scriptscriptstyle{W}}(q,k)\sum_nC_n
\left[C_i\g^{(1)}_{{\cal V}^n_\alpha{\cal V}^j_\beta}(q)
+C_j\g^{(1)}_{{\cal V}^i_\alpha{\cal V}^n_\beta}(q)
\right].
\eea

We next consider the lower order corrections coming from the ${\cal
C}^{ij}_{1L}$ contributions, Eq.\r{start2ZA}. For the two terms
proportional to the longitudinal momentum $k$, the STI triggered will
be precisely the one appearing in Eq.\r{b1}. However, for the first
two terms of this STI we now find the following results
\bea
i\g^{(1)}_{u^\pm
W^{*,\mp}_\sigma}(k)\g^{(0)}_{W^{\pm,\sigma}W^\mp_\mu}(k)
\g^{(0)}_{W^{\pm,\mu} W^{\mp}_\rho{\cal V}^{\rm a}_{\rm
x}}(k,-k+q,-q)&=&
\pm\left(\gw C_{\rm a}\right)\Mw^2\frac1{k^2}
L^{u\,(1)}_{\scriptscriptstyle{V}}(k)\g^{(0)}_{{\cal V}^{\rm a}_\rho 
{\cal V}^{\rm a}_{\lambda}}(q), \nonumber \\
i \g^{(1)}_{u^\pm\phi^{*,\mp}}(k)\g^{(0)}_{\phi^\pm W^\mp_\mu}(k)
\g^{(0)}_{W^{\pm,\mu} W^{\mp}_\rho{\cal V}^{\rm a}_{\rm
x}}(k,-k+q,-q)&=&\pm\left(\gw C_{\rm
a}\right)L^{u\,(1)}_{\scriptscriptstyle G}(k)\g^{(0)}_{{\cal V}^{\rm a}_\rho 
{\cal V}^{\rm a}_{\lambda}}(q), 
\eea
while, as before, the last term of Eq.\r{b1} will part of the PT
answer. 

Finally, for the remaining two ${\cal C}^{ij}_{1L}$ terms
proportional to the longitudinal momentum \mbox{$(k-q)$}, 
the STI triggered will simply be the
one of Eq.\r{twoZA}; thus, the discarded terms stemming from the
${\cal C}^{ij}_{1L}$ contributions, will read
\bea
{\cal C}^{ij}_{1L}&=&2\gw^2C_iC_j\int_{L_1}J_{\scriptscriptstyle
W}(q,k)\dw(k)
\left[\frac{\Mw^2}{k^2}
L^{u\,(1)}_{\scriptscriptstyle{W}}(k)+
L^{u\,(1)}_{\scriptscriptstyle{G}}(k) \right]\left[
\g^{(0)}_{{\cal V}^i_\alpha{\cal V}^i_\beta}(q)+
\g^{(0)}_{{\cal V}^j_\alpha{\cal V}^j_\beta}(q)\right]
\nonumber \\
&-&2\gw^2C_iC_j\int_{L_1}J_{\scriptscriptstyle
W}(q,k)\dw(k)
\g^{(1)}_{W^+_\alpha W^-_\mu}(k)
\left[
\g^{(0)}_{{\cal V}^i_\alpha{\cal V}^i_\beta}(q)+
\g^{(0)}_{{\cal V}^j_\alpha{\cal V}^j_\beta}(q)\right].
\eea

The two contributions analyzed will then add up to give the
quantity $\Pi^{{\rm IP}\,(2)}_{ij,\alpha\beta}(q)$ which reads, after
trivial manipulations,
\bea
\Pi^{{\rm IP}\,(2)}_{ij,\alpha\beta}(q)&=&
2\gw^2C_iC_j\int_{L_1}J_{\scriptscriptstyle
W}(q,k)\dw(k)
L^{u\,(1)}(k)\left[
\g^{(0)}_{{\cal V}^i_\alpha{\cal V}^i_\beta}(q)+
\g^{(0)}_{{\cal V}^j_\alpha{\cal V}^j_\beta}(q)\right]
\nonumber \\
&-&2\gw^2C_iC_j\int_{L_1}J_{\scriptscriptstyle
W}(q,k)\dw(k)
\g^{(1)}_{W^+_\alpha W^-_\mu}(k)
\left[
\g^{(0)}_{{\cal V}^i_\alpha{\cal V}^i_\beta}(q)+
\g^{(0)}_{{\cal V}^j_\alpha{\cal V}^j_\beta}(q)\right]\nonumber \\
&+&2\gw\int_{L_1}J_{\scriptscriptstyle{W}}(q,k)
\left[C_i\g^{(1)}_{u^+{\cal V}^{*,j}_\sigma W^-_\alpha}(-q,-k+q)
\g^{(0)}_{{\cal V}^{j,\sigma}{\cal V}^j_\beta}(q)\right.\nonumber\\
&-&\left.
C_j\g^{(1)}_{u^-{\cal V}^{*,i}_\sigma W^+_\beta}(-q,-k+q)
\g^{(0)}_{{\cal V}^i,\alpha{\cal V}^{i,\sigma}}(q)\right]\nonumber \\
&-&2i\gw^2\int_{L_1}J_{\scriptscriptstyle{W}}(q,k)\sum_nC_n
\left[C_i\g^{(1)}_{{\cal V}^n_\alpha{\cal V}^j_\beta}(q)
+C_j\g^{(1)}_{{\cal V}^i_\alpha{\cal V}^n_\beta}(q)
\right].
\label{piIPZA}
\eea

\begin{figure}[!t]
\bce
\bpi(0,105)(130,-75)

\Line(-105,0.75)(-90,0.75)
\Line(-105,-0.75)(-90,-0.75)
\PhotonArc(-70,0)(20,0,180){-1.5}{8.5}
\DashCArc(-70,0)(20,180,360){1}
\Line(-50,0.75)(-35,0.75)
\Line(-50,-0.75)(-35,-0.75)
\Photon(-70,18.5)(-70,-20){-1.5}{6}
\DashArrowLine(-57,-15)(-55.5,-13.6){1}
\DashArrowLine(-84,-13.8)(-82.5,-15.4){1}

\Text(-25,0)[l]{\footnotesize{(4 diagrams)}}
\Text(5,-20)[c]{\scriptsize{(a)}}

\Line(60,0.75)(75,0.75)
\Line(60,-0.75)(75,-0.75)
\PhotonArc(95,0)(20,90,180){-1.5}{4.5}
\PhotonArc(95,0)(20,270,360){-1.5}{4.5}
\DashCArc(95,0)(20,0,90){1}
\DashCArc(95,0)(20,180,270){1}
\DashArrowLine(95,-20)(95,18.5){1}
\Line(115,0.75)(130,0.75)
\Line(115,-0.75)(130,-0.75)
\DashArrowLine(108,15)(109.5,13.6){1}
\DashArrowLine(81,-13.8)(82.5,-15.4){1}

\Text(140,0)[l]{\footnotesize{(4 diagrams)}}
\Text(170,-20)[c]{\scriptsize{(b)}}

\Line(225,0.75)(240,0.75)
\Line(225,-0.75)(240,-0.75)
\PhotonArc(260,0)(20,0,180){-1.5}{8.5}
\DashCArc(260,0)(20,180,360){1}
\SetWidth{1.3}
\Line(260,18.5)(260,-20)
\SetWidth{0.5}
\Line(280,0.75)(295,0.75)
\Line(280,-0.75)(295,-0.75)
\DashArrowLine(273,-15)(274.5,-13.6){1}
\DashArrowLine(246,-13.8)(247.5,-15.4){1}

\Text(305,0)[l]{\footnotesize{(2 diagrams)}}
\Text(335,-20)[c]{\scriptsize{(c)}}

\Line(-105,-59.25)(-90,-59.25)
\Line(-105,-60.75)(-90,-60.75)
\PhotonArc(-70,-60)(20,0,180){-1.5}{8.5}
\DashCArc(-70,-60)(20,180,360){1}
\Line(-50,-59.25)(-35,-59.25)
\Line(-50,-60.75)(-35,-60.75)
\DashArrowLine(-70.5,-80)(-69.5,-80){1}
\GCirc(-70,-41.5){6}{0.8}

\Text(-25,-60)[l]{\footnotesize{(2 diagrams)}}
\Text(5,-80)[c]{\scriptsize{(d)}}

\Line(60,-59.25)(75,-59.25)
\Line(60,-60.75)(75,-60.75)
\PhotonArc(95,-60)(20,0,180){-1.5}{8.5}
\DashCArc(95,-60)(20,180,360){1}
\Line(115,-59.25)(130,-59.25)
\Line(115,-60.75)(130,-60.75)
\DashArrowLine(108,-75)(109.5,-73.6){1}
\DashArrowLine(81,-73.8)(82.5,-75.4){1}
\GCirc(95,-80){6}{0.8}

\Text(140,-60)[l]{\footnotesize{(2 diagrams)}}
\Text(170,-80)[c]{\scriptsize{(e)}}

\Text(-108,8)[l]{$\scriptstyle{\Omega^{\rm a}_{\lambda}}$}
\Text(57,8)[l]{$\scriptstyle{\Omega^{\rm a}_{\lambda}}$}
\Text(223,8)[l]{$\scriptstyle{\Omega^{\rm a}_{\lambda}}$}
\Text(-108,-52)[l]{$\scriptstyle{\Omega^{\rm a}_{\lambda}}$}
\Text(57,-52)[l]{$\scriptstyle{\Omega^{\rm a}_{\lambda}}$}

\Text(-45,8)[l]{$\scriptstyle{{\cal V}^{*,{\rm b}}_\rho}$}
\Text(120,8)[l]{$\scriptstyle{{\cal V}^{*,{\rm b}}_\rho}$}
\Text(285,8)[l]{$\scriptstyle{{\cal V}^{*,{\rm b}}_\rho}$}
\Text(-45,-52)[l]{$\scriptstyle{{\cal V}^{*,{\rm b}}_\rho}$}
\Text(120,-52)[l]{$\scriptstyle{{\cal V}^{*,{\rm b}}_\rho}$}

\Text(-88,20)[c]{$\scriptscriptstyle{W}$}
\Text(-62.5,0)[c]{$\scriptscriptstyle{{\cal V}^n}$}
\Text(-50,20)[c]{$\scriptscriptstyle{W}$}
\Text(-88,-20)[c]{$\scriptstyle{u}$}
\Text(-50,-20)[c]{$\scriptstyle u$}

\Text(77,20)[c]{$\scriptscriptstyle{W}$}
\Text(115,20)[c]{$\scriptstyle{u}$}
\Text(106.5,0)[c]{$\scriptstyle{u^n}$}
\Text(77,-20)[c]{$\scriptstyle{u}$}
\Text(115,-20)[c]{$\scriptscriptstyle{W}$}

\Text(242,20)[c]{$\scriptscriptstyle{W}$}
\Text(280,20)[c]{$\scriptscriptstyle{W}$}
\Text(267.5,0)[c]{$\scriptscriptstyle{H}$}
\Text(242,-20)[c]{$\scriptstyle{u}$}
\Text(280,-20)[c]{$\scriptstyle u$}

\Text(-88,-40)[c]{$\scriptscriptstyle{W}$}
\Text(-50,-40)[c]{$\scriptscriptstyle{W}$}
\Text(-70,-87)[c]{$\scriptstyle{u}$}

\Text(95,-33)[c]{$\scriptscriptstyle{W}$}
\Text(77,-80)[c]{$\scriptstyle{u}$}
\Text(115,-80)[c]{$\scriptstyle{u}$}

\epi
\ece

\caption{\label{figx} Two loop contributions to the two-point function
$\g^{(2)}_{\Omega^{\rm a}_{\lambda} {\cal V}^{*,{\rm b}}_\rho}(q)$. The
blobs represent one-loop corrections to the corresponding propagator.} 
\end{figure}

The last step is then to compare this result to the 
one coming from the BQI of Eq.\r{tpfBQIZA}. 
At two loops this BQI reads
\bea
\g^{(2)}_{\widehat{\cal V}^i_\alpha\widehat{\cal V}^j_\beta}(q)&=&
\g^{(2)}_{{\cal V}^i_\alpha{\cal V}^j_\beta}(q)+
\g^{(2)}_{\Omega^i_\alpha{\cal V}^{*,j}_\rho}(q)\g^{(0)}_{{\cal
V}^{j,\rho}{\cal V}^j_\beta}(q)+
\g^{(2)}_{\Omega^j_\beta{\cal V}^{*,i}_\rho}(q)\g^{(0)}_{{\cal
V}^{i,\rho}{\cal V}^i_\alpha}(q)\nonumber \\
&+&\sum_n\left[\g^{(1)}_{\Omega^i_\alpha{\cal V}^{*,n}_\rho}(q)\g^{(1)}_{{\cal
V}^{n,\rho}{\cal V}^j_\beta}(q)+
\g^{(1)}_{\Omega^j_\beta{\cal V}^{*,n}_\rho}(q)\g^{(1)}_{{\cal
V}^{n,\rho}{\cal V}^i_\alpha}(q)\right]\nonumber \\
&+&\sum_n
\g^{(1)}_{\Omega^i_\alpha{\cal V}^{*,n}_\rho}(q)
\g^{(0)}_{{\cal V}^{n,\rho}{\cal V}^{n,\sigma}}(q)
\g^{(1)}_{\Omega^j_\beta{\cal V}^{*,n}_\sigma}(q),
\label{BQI2lZA}
\eea
and the diagrams contributing to the two-loop two-point function
$\g^{(2)}_{\Omega^{\rm a}_{\lambda}{\cal V}^{\rm b}_\rho}(q)$ are shown
in Fig.\ref{figx}. Using the one-loop result of Eq.\r{r1ZA}, 
it is then easy to show that
\bea
& & \sum_n\left[\g^{(1)}_{\Omega^i_\alpha{\cal V}^{*,n}_\rho}(q)\g^{(1)}_{{\cal
V}^{n,\rho}{\cal V}^j_\beta}(q)+
\g^{(1)}_{\Omega^j_\beta{\cal V}^{*,n}_\rho}(q)\g^{(1)}_{{\cal
V}^{n,\rho}{\cal V}^i_\alpha}(q)\right]\nonumber \\
&\equiv&
2i\gw^2\int_{L_1}J_{\scriptscriptstyle{W}}(q,k)\sum_nC_n
\left[C_i\g^{(1)}_{{\cal V}^n_\alpha{\cal V}^j_\beta}(q)
+C_j\g^{(1)}_{{\cal V}^i_\alpha{\cal V}^n_\beta}(q)
\right],
\eea
while it is a long but straightforward exercise to check that
\bea
& & \hspace{1.0cm}
\g^{(2)}_{\Omega^i_\alpha{\cal V}^{*,j}_\rho}(q)\g^{(0)}_{{\cal
V}^{j,\rho}{\cal V}^j_\beta}(q)+
\g^{(2)}_{\Omega^j_\beta{\cal V}^{*,i}_\rho}(q)\g^{(0)}_{{\cal
V}^{i,\rho}{\cal V}^i_\alpha}(q)\nonumber \\
&\equiv&-2\gw^2C_iC_j\int_{L_1}J_{\scriptscriptstyle
W}(q,k)\dw(k)
L^{u\,(1)}(k)\left[
\g^{(0)}_{{\cal V}^i_\alpha{\cal V}^i_\beta}(q)+
\g^{(0)}_{{\cal V}^j_\alpha{\cal V}^j_\beta}(q)\right]
\nonumber \\
&+&2\gw^2C_iC_j\int_{L_1}J_{\scriptscriptstyle
W}(q,k)\dw(k)
\g^{(1)}_{W^+_\alpha W^-_\mu}(k)
\left[
\g^{(0)}_{{\cal V}^i_\alpha{\cal V}^i_\beta}(q)+
\g^{(0)}_{{\cal V}^j_\alpha{\cal V}^j_\beta}(q)\right]\nonumber \\
&-&2\gw\int_{L_1}J_{\scriptscriptstyle{W}}(q,k)
\left[C_i\g^{(1)}_{u^+{\cal V}^{*,j}_\sigma W^-_\alpha}(-q,-k+q)
\g^{(0)}_{{\cal V}^{j,\sigma}{\cal V}^j_\beta}(q)\right.\nonumber\\
&-&\left.
C_j\g^{(1)}_{u^-{\cal V}^{*,i}_\sigma W^+_\beta}(-q,-k+q)
\g^{(0)}_{{\cal V}^i,\alpha{\cal V}^{i,\sigma}}(q)\right].
\eea
In particular the first term on the RHS of the above equation gives
the diagrams of Fig.\ref{figx}e, the second term the ones of
Fig.\ref{figx}d, and the last term all the remaining diagrams of
Fig.\ref{figx} (a--c). 

The last term appearing in Eq.\r{BQI2lZA}
will be finally generated in the conversion of the 1PR strings into the
1PR PT strings, as follows. After treating the conventional 1PR
diagrams involving neutral gauge-bosons self-energies along the same lines
explained in \cite{Binosi:2002ez} for the QCD case, one
arrives, after discarding the terms proportional to the inverse
propagator of the external legs, to the result 
\bea
S'^{{\rm IP}\,(2)}_{ij,\alpha\beta}(q)&=&i\sum_n\left[
V^{{\rm P}\,(1)}_{in,\alpha\rho}(q)
\g^{(1)}_{{\cal V}^{n,\rho}{\cal V}^j_\beta}(q)+
V^{{\rm P}\,(1)}_{jn,\beta\rho}(q)
\g^{(1)}_{{\cal V}^{n,\rho}{\cal V}^i_\alpha}(q)\right]\nonumber \\
&+&\sum_n\g^{(1)}_{\Omega^i_\alpha{\cal V}^{*,n}_\rho}(q)
\g^{(0)}_{{\cal V}^{n,\rho}{\cal V}^{n,\sigma}}(q)
\g^{(1)}_{\Omega^j_\beta{\cal V}^{*,n}_\sigma}(q).
\label{p3}
\eea
On the other hand from Eq.\r{piIPZA} we have
\be
\Pi'^{{\rm IP}\,(2)}_{ij\alpha\beta}(q)=-i\sum_n\left[
V^{{\rm P}\,(1)}_{in,\alpha\rho}(q)
\g^{(1)}_{{\cal V}^{n,\rho}{\cal V}^j_\beta}(q)+
V^{{\rm P}\,(1)}_{jn,\beta\rho}(q)
\g^{(1)}_{{\cal V}^{n,\rho}{\cal V}^i_\alpha}(q)\right]
\label{p4}
\ee
so that adding by parts these two equations we obtain
\be
R^{{\rm IP}\,(2)}_{ij,\alpha\beta}(q)=
\sum_n\g^{(1)}_{\Omega^i_\alpha{\cal V}^{*,n}_\rho}(q)
\g^{(0)}_{{\cal V}^{n,\rho}{\cal V}^{n,\sigma}}(q)
\g^{(1)}_{\Omega^j_\beta{\cal V}^{*,n}_\sigma}(q).
\ee

Thus, finally, the quantity $-\Pi^{{\rm IP}\,(2)}_{ij,\alpha\beta}(q)
+R^{{\rm IP}\,(2)}_{ij,\alpha\beta}(q)$ will provide precisely all the
terms appearing in the two-loop version of the relevant BQI [Eq.\r{BQIc2l}],
{\it i.e.}, one has 
\bea
-\Pi^{{\rm IP}\,(2)}_{ij,\alpha\beta}(q)
+R^{{\rm IP}\,(2)}_{ij,\alpha\beta}(q)&=&
\g^{(2)}_{\Omega^i_\alpha{\cal V}^{*,j}_\rho}(q)\g^{(0)}_{{\cal
V}^{j,\rho}{\cal V}^j_\beta}(q)+
\g^{(2)}_{\Omega^j_\beta{\cal V}^{*,i}_\rho}(q)\g^{(0)}_{{\cal
V}^{i,\rho}{\cal V}^i_\alpha}(q)\nonumber \\
&+&\sum_n\left[\g^{(1)}_{\Omega^i_\alpha{\cal V}^{*,n}_\rho}(q)\g^{(1)}_{{\cal
V}^{n,\rho}{\cal V}^j_\beta}(q)+
\g^{(1)}_{\Omega^j_\beta{\cal V}^{*,n}_\rho}(q)\g^{(1)}_{{\cal
V}^{n,\rho}{\cal V}^i_\alpha}(q)\right]\nonumber \\
&+&\sum_n
\g^{(1)}_{\Omega^i_\alpha{\cal V}^{*,n}_\rho}(q)
\g^{(0)}_{{\cal V}^{n,\rho}{\cal V}^{n,\sigma}}(q)
\g^{(1)}_{\Omega^j_\beta{\cal V}^{*,n}_\sigma}(q).
\eea
Then, the difference between $\widehat\g^{(2)}_{{\cal V}^i_\alpha{\cal
V}^j_\beta}(q)$ 
and $\g^{(2)}_{{\cal V}^i_\alpha {\cal V}^j_\beta}(q)$ 
as given by the intrinsic PT definition of Eq.\r{ipdefZA}, 
is the same as the difference between 
$\g^{(2)}_{\widehat {\cal V}^i_\alpha\widehat {\cal V}^j_\beta}(q)$ and
$\g^{(2)}_{{\cal V}^i_\alpha {\cal V}^j_\beta}(q)$ 
as given by the BQI, thus proving that
\be
\widehat\g^{(2)}_{{\cal V}^i_\alpha {\cal V}^j_\beta}(q)\equiv
\g^{(2)}_{\widehat {\cal V}^i_\alpha\widehat {\cal V}^j_\beta}(q).
\ee

\section{\label{sec:concl}Discussion and Conclusions}

In this  paper we  have extended the  two-loop PT construction  to the
Electroweak  sector of  the Standard  Model.  This  generalization has
been a  pending problem,  mainly due to  the proliferation  of Feynman
diagrams as compared  to the QCD two-loop case, as well  as due to the
conceptual  complication arising  from the  non-transversality  of the
massive  gauge-boson  one-loop   self-energies  appearing  inside  the
two-loop diagrams.   The aforementioned logistical  problems have been
solved  by   resorting  to   the  recently  introduced   intrinsic  PT
construction by  means of the STI satisfied by the
one-loop three-gauge boson vertices;  the latter are nested inside the
two-loop Feynman  graphs determining the  gauge-boson self-energies at
the   same  order.   Thus,   instead  of   manipulating  algebraically
individual  Feynman  diagrams,  entire  classes  of  diagrams  may  be
simultaneously  addressed.   In the  construction  we have  restricted
ourselves  to  the  operationally  simpler case  of  massless-external
fermions; thus,  only the parts  of the self-energies  proportional to
$g_{\mu\nu}$ have  been considered.  For the  same reason longitudinal
contributions  to  the STIs  employed have  been
consistently  discarded   throughout.   The  final   outcome  of  this
construction are gauge-independent two-loop self-energies for both the
charged ($W$) and neutral gauge-bosons ($A,Z$).

The comparison of  the resulting PT expressions with  those of the BFM
in the Feynman gauge, constitutes an almost obligatory exercise, given
the well-known correspondence established  in the literature. The task
of carrying out this comparison is greatly simplified by employing the
BQIs; the  latter relate the
BFM $n$-point functions with the conventional (quantum) ones, by means
of a set of well-defined auxiliary Green's functions, definable in the
BV framework.   The  non-trivial step  in  this
exercise is to establish that the pieces removed from the conventional
self-energy, following the strict rules of PT procedure, are precisely
those accounting  for the difference between the  BFM and conventional
two-point functions  as captured in the corresponding  BQI.  Thus, the
correspondence between the PT and the BFM results in the Feynman gauge
persists   unaltered  in   the  case   of  the   two-loop  Electroweak
construction.

The generalization  to the  case of massive  external fermions  of the
construction  presented here,  {\it  i.e.},  the  case of  non-conserved
external  currents, is  technically  more involved  for the  following
reasons   \cite{Papavassiliou:1990zd,Papavassiliou:1994pr}:  First  of
all, in the  case of the $S$-matrix construction,  the tree-level WIs
listed  in Eq.(\ref{PTWI})  are modified  by the  presence  of masses,
giving   rise  to   additional   terms.  These   terms  will   combine
non-trivially  with  the  additional  graphs containing  the  would-be
Goldstone   bosons,  in   order  to   give  rise   to   the  necessary
cancellations.   In addition,  the  propagator-like corrections  which
will be obtained from the  vertices must be judiciously alloted to not
only  the corresponding  gauge-boson  self-energies, but  also to  the
self-energies  describing the  various higher  order mixings,  such as
$\g_{\phi^\pm   W^\mp_\beta}(q)$,  $\g_{\phi^\pm  \phi^\mp}(q)$,
$\g_{\chi   {\cal  V}^j_\beta}(q)$,   $\g_{H   {\cal  V}^j_\beta}(q)$,
$\g_{\chi \chi}(q)$,  and $\g_{H  \chi}(q)$, whose PT  counterparts to
the given order  must also be constructed \cite{Papavassiliou:1994pr}.
In  the  case of  the  intrinsic PT,  where  the  vertex diagrams  are
essentially inert,  the complications from the fact  that the currents
are   not-conserved  are  mainly   due  to   the  appearance   of  the
aforementioned    mixing    self-energies    in   the    corresponding
three-gauge-boson STIs of Eq.(\ref{STIthree}).  Moreover, in both the
$S$-matrix  PT  and  the  intrinsic  PT the  presence  of  the  mixing
self-energies     in     the     relevant    BQIs     [{\it     viz.}
Eq.(\ref{twopfwcase})] further complicates the final comparison of the
results  between the  PT and  the  BFM.  The  point we  would like  to
emphasize  however,  is  that,  despite  all  these  technical  issues
discussed above, no additional conceptual difficulties are expected in
the non-conserved current case.

The distinction  between the $S$-matrix  PT and intrinsic  PT warrants
some further comments. In this  paper we have focused on the intrinsic
PT construction,  because of the realization that  the parts discarded
correspond to very precise terms  in the STI satisfied by the one-loop
three-gauge-boson vertex; thus the algorithm presented here  
constitutes the  natural generalization to two-loops of the 
one-loop    intrinsic   PT    construction   presented    in   Section
\ref{sec:IP2}. On the  other hand we have refrained  from carrying out
the two-loop  $S$-matrix PT construction {\it  explicitly}, {\it i.e.},
by directly rearranging  the two-loop vertex graphs, as  has been done
in  \cite{Papavassiliou:1999az,Binosi:2002ez}; at  present  this would
constitute an  arduous diagrammatic task, since the  equivalent of the
three-gauge-boson vertex STI, whose  use has been crucial in obtaining
compact results in the intrinsic PT case, still eludes us.  This issue
is  currently  under investigation,  and  we  hope  to report  further
progress in the near future.
  
At  the  phenomenological  level,  and  especially  in  the  field  of
precision  Electroweak physics,  the  two-loop construction  presented
here may serve as a starting point for complementing existing two-loop
calculations   \cite{Freitas:2002ja}.   In  particular,   the  current
accuracy in  the measurement of $\Mw$  is $\Mw= 80.451  \pm 0.033$ GeV
\cite{Charlton:2001am}, and  it is supposed to be  further improved in
the final  LEP II  analysis and  the Tevatron Run  II, each  giving an
error  of $\delta\Mw  \approx 30$  MeV.  Furthermore,  at the  LHC the
error  is  expected  to  be  as  low as  $\delta\Mw\approx  15  $  MeV
\cite{Haywood:1999qg}.  Even more impressively, high-luminosity linear
colliders  operating at  the $W^{+}W^{-}$  threshold could  reduce the
error          to           $\delta\Mw\approx          6$          MeV
\cite{Aguilar-Saavedra:2001rg,Abe:2001nq}.   In  order  to  match  the
expected experimental precision, quantities  such as $\Delta r$ or the
two-loop  $\rho$ parameter  must be  determined with  high theoretical
accuracy \cite{Freitas:2002ja}; in particular, purely bosonic two-loop
corrections  may have  to be  calculated eventually.   The theoretical
framework  put forward  in the  present paper  sets up  the  stage for
carrying out  such a  task in a  systematic way \cite{RHO}.   Aside of
these possibilities however,  the two-loop construction presented here
renders  the various  one-loop  results  of the  past  (listed in  the
Introduction)  conceptually far  more robust,  demonstrating  that the
special field-theoretic properties achieved  by means of the PT method
are not a fortuitous one-loop accident.

\begin{acknowledgments}

We  happily acknowledge  valuable correspondence  with  G.~Barnich and
P.A.~Grassi.   The work  of D.B.  is  supported by  the Ministerio  of
Ciencia  y Tecnolog\'\i a,  Spain, under  Grant BFM2001-0262,  and the
research  of   J.P.  is  supported   by  CICYT,  Spain,   under  Grant
AEN-99/0692.

\end{acknowledgments}

\appendix

\section{\label{FR} Feynman rules}

We report here  all necessary Feynman rules for  computing the various
Green's  functions  appearing  in  the  BQIs,  in  the  Feynman  gauge
$\xi_{\scriptscriptstyle{W}}=\xi_{\scriptscriptstyle{Z}}
=\xi_{\scriptscriptstyle{A}}=1$.

\subsection{Anti-fields}

\subsubsection{Gauge-boson sector}

\bce
\bpi(0,125)(220,-95)

\Line(20,0.75)(40,0.75)
\Line(20,-0.75)(40,-0.75)
\Photon(60,20)(40,0){1.5}{6}
\DashArrowLine(60,-20)(40,0){1}

\Text(15,-8)[l]{\footnotesize{$W^{*,\pm}_\alpha$}}
\Text(62.5,20.5)[l]{\footnotesize{$W^\mp_\beta$}}
\Text(62.5,-19.5)[l]{\footnotesize{$u^A$}}
\Text(70,0)[l]{$=\pm i\gw\sw g_{\alpha\beta}$}

\Line(170,0.75)(190,0.75)
\Line(170,-0.75)(190,-0.75)
\Photon(190,0)(210,20){1.5}{6}
\DashArrowLine(210,-20)(190,0){1}

\Text(165,-8)[l]{\footnotesize{$W^{*,\pm}_\alpha$}}
\Text(212.5,20.5)[l]{\footnotesize{$W^\mp_\beta$}}
\Text(212.5,-19.5)[l]{\footnotesize{$u^Z$}}
\Text(220,0)[l]{$=\mp i\gw\cw g_{\alpha\beta}$}

\Line(320,0.75)(340,0.75)
\Line(320,-0.75)(340,-0.75)
\Photon(360,20)(340,0){1.5}{6}
\DashArrowLine(360,-20)(340,0){1}

\Text(315,-8)[l]{\footnotesize{$W^{*,\pm}_\alpha$}}
\Text(362.5,20.5)[l]{\footnotesize{$A_\beta$}}
\Text(362.5,-19.5)[l]{\footnotesize{$u^\mp$}}
\Text(370,0)[l]{$=\mp i \gw\sw g_{\alpha\beta}$}

\Line(20,-69.25)(40,-69.25)
\Line(20,-70.75)(40,-70.75)
\Photon(60,-50)(40,-70){1.5}{6}
\DashArrowLine(60,-90)(40,-70){1}

\Text(15,-78)[l]{\footnotesize{$W^{*,\pm}_\alpha$}}
\Text(62.5,-49.5)[l]{\footnotesize{$Z_\beta$}}
\Text(62.5,-89.5)[l]{\footnotesize{$u^\mp$}}
\Text(70,-70)[l]{$=\pm i\gw\cw g_{\alpha\beta}$}

\Line(170,-69.25)(190,-69.25)
\Line(170,-70.75)(190,-70.75)
\Photon(210,-50)(190,-70){1.5}{6}
\DashArrowLine(210,-90)(190,-70){1}

\Text(165,-78)[l]{\footnotesize{$Z^{*}_\alpha$}}
\Text(212.5,-49.5)[l]{\footnotesize{$W^\pm_\beta$}}
\Text(212.5,-89.5)[l]{\footnotesize{$u^\mp$}}
\Text(220,-70)[l]{$=\mp i\gw\cw g_{\alpha\beta}$}

\Line(320,-69.25)(340,-69.25)
\Line(320,-70.75)(340,-70.75)
\Photon(360,-50)(340,-70){1.5}{6}
\DashArrowLine(360,-90)(340,-70){1}

\Text(315,-78)[l]{\footnotesize{$A^{*}_\alpha$}}
\Text(362.5,-49.5)[l]{\footnotesize{$W^\pm_\beta$}}
\Text(362.5,-89.5)[l]{\footnotesize{$u^\mp$}}
\Text(370,-70)[l]{$=\pm i\gw\sw g_{\alpha\beta}$}

\epi
\ece

\subsubsection{Scalar sector}

\bce
\bpi(0,190)(220,-160)

\Line(20,0.75)(40,0.75)
\Line(20,-0.75)(40,-0.75)
\SetWidth{1.3}
\Line(60,20)(40,0)
\SetWidth{0.5}
\DashArrowLine(60,-20)(40,0){1}

\Text(15,-8)[l]{\footnotesize{$\phi^{*,\pm}$}}
\Text(62.5,20.5)[l]{\footnotesize{$\phi^\mp$}}
\Text(62.5,-19.5)[l]{\footnotesize{$u^A$}}
\Text(70,0)[l]{$=\pm i\gw\sw$}

\Line(170,0.75)(190,0.75)
\Line(170,-0.75)(190,-0.75)
\SetWidth{1.3}
\Line(190,0)(210,20)
\SetWidth{0.5}
\DashArrowLine(210,-20)(190,0){1}

\Text(165,-8)[l]{\footnotesize{$\phi^{*,\pm}$}} 
\Text(212.5,20.5)[l]{\footnotesize{$\phi^\mp$}}
\Text(212.5,-19.5)[l]{\footnotesize{$u^Z$}}
\Text(220,0)[l]{$=\mp i\gw\frac{\cw^2-\sw^2}{2\cw}$}

\Line(320,0.75)(340,0.75)
\Line(320,-0.75)(340,-0.75)
\SetWidth{1.3}
\Line(360,20)(340,0)
\SetWidth{0.5}
\DashArrowLine(360,-20)(340,0){1}

\Text(315,-8)[l]{\footnotesize{$\phi^{*,\pm}$}}
\Text(362.5,20.5)[l]{\footnotesize{$\chi$}}
\Text(362.5,-19.5)[l]{\footnotesize{$u^\mp$}}
\Text(370,0)[l]{$=- \frac\gw2$}

\Line(20,-69.25)(40,-69.25)
\Line(20,-70.75)(40,-70.75)
\SetWidth{1.3}
\Line(60,-50)(40,-70)
\SetWidth{0.5}
\DashArrowLine(60,-90)(40,-70){1}

\Text(15,-78)[l]{\footnotesize{$\phi^{*,\pm}$}}
\Text(62.5,-49.5)[l]{\footnotesize{$H$}}
\Text(62.5,-89.5)[l]{\footnotesize{$u^\mp$}}
\Text(70,-70)[l]{$=\mp i\frac\gw2$}

\Line(170,-69.25)(190,-69.25)
\Line(170,-70.75)(190,-70.75)
\SetWidth{1.3}
\Line(210,-50)(190,-70)
\SetWidth{0.5}
\DashArrowLine(210,-90)(190,-70){1}

\Text(165,-78)[l]{\footnotesize{$\chi^{*}$}}
\Text(212.5,-49.5)[l]{\footnotesize{$\phi^\pm$}}
\Text(212.5,-89.5)[l]{\footnotesize{$u^\mp$}}
\Text(220,-70)[l]{$=\frac\gw2$}

\Line(320,-69.25)(340,-69.25)
\Line(320,-70.75)(340,-70.75)
\SetWidth{1.3}
\Line(360,-50)(340,-70)
\SetWidth{0.5}
\DashArrowLine(360,-90)(340,-70){1}

\Text(315,-78)[l]{\footnotesize{$\chi^{*}$}}
\Text(362.5,-49.5)[l]{\footnotesize{$H$}}
\Text(362.5,-89.5)[l]{\footnotesize{$u^Z$}}
\Text(370,-70)[l]{$=-\frac\gw{2\cw}$}

\Line(20,-139.25)(40,-139.25)
\Line(20,-140.75)(40,-140.75)
\SetWidth{1.3}
\Line(60,-120)(40,-140)
\SetWidth{0.5}
\DashArrowLine(60,-160)(40,-140){1}

\Text(15,-148)[l]{\footnotesize{$H^{*}$}}
\Text(62.5,-119.5)[l]{\footnotesize{$\phi^\pm$}}
\Text(62.5,-159.5)[l]{\footnotesize{$u^\mp$}}
\Text(70,-140)[l]{$=\pm i\frac\gw2$}

\Line(170,-139.25)(190,-139.25)
\Line(170,-140.75)(190,-140.75)
\SetWidth{1.3}
\Line(210,-120)(190,-140)
\SetWidth{0.5}
\DashArrowLine(210,-160)(190,-140){1}

\Text(165,-148)[l]{\footnotesize{$H^{*}$}}
\Text(212.5,-119.5)[l]{\footnotesize{$\chi$}}
\Text(212.5,-159.5)[l]{\footnotesize{$u^Z$}}
\Text(220,-140)[l]{$=\frac\gw{2\cw}$}

\epi
\ece

\subsection{Background sources}

The Feynman rules for the vertices involving the background sources 
$\Omega^n_\mu$ and $\Omega^{G^n}$ 
are obtained from the above one by trading the 
ghost field for an anti-ghost field (and thus reversing the arrow of the 
ghost line).

\subsubsection{Gauge-boson sector}

\bce
\bpi(0,125)(220,-95)

\Line(20,0.75)(40,0.75)
\Line(20,-0.75)(40,-0.75)
\Photon(60,20)(40,0){1.5}{6}
\DashArrowLine(40,0)(60,-20){1}

\Text(15,-8)[l]{\footnotesize{$\Omega^{\pm}_\alpha$}}
\Text(62.5,20.5)[l]{\footnotesize{$W^\mp_\beta$}}
\Text(62.5,-19.5)[l]{\footnotesize{$\bar u^A$}}
\Text(70,0)[l]{$=\pm i\gw\sw g_{\alpha\beta}$}

\Line(170,0.75)(190,0.75)
\Line(170,-0.75)(190,-0.75)
\Photon(190,0)(210,20){1.5}{6}
\DashArrowLine(190,0)(210,-20){1}

\Text(165,-8)[l]{\footnotesize{$\Omega^{\pm}_\alpha$}}
\Text(212.5,20.5)[l]{\footnotesize{$W^\mp_\beta$}}
\Text(212.5,-19.5)[l]{\footnotesize{$\bar u^Z$}}
\Text(220,0)[l]{$=\mp i\gw\cw g_{\alpha\beta}$}

\Line(320,0.75)(340,0.75)
\Line(320,-0.75)(340,-0.75)
\Photon(360,20)(340,0){1.5}{6}
\DashArrowLine(340,0)(360,-20){1}

\Text(315,-8)[l]{\footnotesize{$\Omega^{\pm}_\alpha$}}
\Text(362.5,20.5)[l]{\footnotesize{$A_\beta$}}
\Text(362.5,-19.5)[l]{\footnotesize{$\bar u^\pm$}}
\Text(370,0)[l]{$=\mp i\gw\sw g_{\alpha\beta}$}

\Line(20,-69.25)(40,-69.25)
\Line(20,-70.75)(40,-70.75)
\Photon(60,-50)(40,-70){1.5}{6}
\DashArrowLine(40,-70)(60,-90){1}

\Text(15,-78)[l]{\footnotesize{$\Omega^{\pm}_\alpha$}}
\Text(62.5,-49.5)[l]{\footnotesize{$Z_\beta$}}
\Text(62.5,-89.5)[l]{\footnotesize{$\bar u^\pm$}}
\Text(70,-70)[l]{$=\pm i\gw\cw g_{\alpha\beta}$}

\Line(170,-69.25)(190,-69.25)
\Line(170,-70.75)(190,-70.75)
\Photon(210,-50)(190,-70){1.5}{6}
\DashArrowLine(190,-70)(210,-90){1}

\Text(165,-78)[l]{\footnotesize{$\Omega^{Z}_\alpha$}}
\Text(212.5,-49.5)[l]{\footnotesize{$W^\pm_\beta$}}
\Text(212.5,-89.5)[l]{\footnotesize{$\bar u^\pm$}}
\Text(220,-70)[l]{$=\mp i\gw\cw g_{\alpha\beta}$}

\Line(320,-69.25)(340,-69.25)
\Line(320,-70.75)(340,-70.75)
\Photon(360,-50)(340,-70){1.5}{6}
\DashArrowLine(340,-70)(360,-90){1}

\Text(315,-78)[l]{\footnotesize{$\Omega^{A}_\alpha$}}
\Text(362.5,-49.5)[l]{\footnotesize{$W^\pm_\beta$}}
\Text(362.5,-89.5)[l]{\footnotesize{$\bar u^\pm$}}
\Text(370,-70)[l]{$=\pm i\gw\sw g_{\alpha\beta}$}

\epi
\ece

\subsubsection{Scalar sector}

\bce
\bpi(0,190)(220,-160)

\Line(20,0.75)(40,0.75)
\Line(20,-0.75)(40,-0.75)
\SetWidth{1.3}
\Line(60,20)(40,0)
\SetWidth{0.5}
\DashArrowLine(40,0)(60,-20){1}

\Text(15,-8)[l]{\footnotesize{$\Omega^{\pm}$}}
\Text(62.5,20.5)[l]{\footnotesize{$\phi^\mp$}}
\Text(62.5,-19.5)[l]{\footnotesize{$\bar u^A$}}
\Text(70,0)[l]{$=\pm i\gw\sw$}

\Line(170,0.75)(190,0.75)
\Line(170,-0.75)(190,-0.75)
\SetWidth{1.3}
\Line(190,0)(210,20)
\SetWidth{0.5}
\DashArrowLine(190,0)(210,-20){1}

\Text(165,-8)[l]{\footnotesize{$\Omega^{\pm}$}} 
\Text(212.5,20.5)[l]{\footnotesize{$\phi^\mp$}}
\Text(212.5,-19.5)[l]{\footnotesize{$\bar u^Z$}}
\Text(220,0)[l]{$=\mp i\gw\frac{\cw^2-\sw^2}{2\cw}$}

\Line(320,0.75)(340,0.75)
\Line(320,-0.75)(340,-0.75)
\SetWidth{1.3}
\Line(360,20)(340,0)
\SetWidth{0.5}
\DashArrowLine(340,0)(360,-20){1}

\Text(315,-8)[l]{\footnotesize{$\Omega^{\pm}$}}
\Text(362.5,20.5)[l]{\footnotesize{$\chi$}}
\Text(362.5,-19.5)[l]{\footnotesize{$\bar u^\pm$}}
\Text(370,0)[l]{$=- \frac\gw2$}

\Line(20,-69.25)(40,-69.25)
\Line(20,-70.75)(40,-70.75)
\SetWidth{1.3}
\Line(60,-50)(40,-70)
\SetWidth{0.5}
\DashArrowLine(40,-70)(60,-90){1}

\Text(15,-78)[l]{\footnotesize{$\Omega^{\pm}$}}
\Text(62.5,-49.5)[l]{\footnotesize{$H$}}
\Text(62.5,-89.5)[l]{\footnotesize{$\bar u^\pm$}}
\Text(70,-70)[l]{$=\mp i\frac\gw2$}

\Line(170,-69.25)(190,-69.25)
\Line(170,-70.75)(190,-70.75)
\SetWidth{1.3}
\Line(210,-50)(190,-70)
\SetWidth{0.5}
\DashArrowLine(190,-70)(210,-90){1}

\Text(165,-78)[l]{\footnotesize{$\Omega^{\chi}$}}
\Text(212.5,-49.5)[l]{\footnotesize{$\phi^\pm$}}
\Text(212.5,-89.5)[l]{\footnotesize{$\bar u^\pm$}}
\Text(220,-70)[l]{$=\frac\gw2$}

\Line(320,-69.25)(340,-69.25)
\Line(320,-70.75)(340,-70.75)
\SetWidth{1.3}
\Line(360,-50)(340,-70)
\SetWidth{0.5}
\DashArrowLine(340,-70)(360,-90){1}

\Text(315,-78)[l]{\footnotesize{$\Omega^{\chi}$}}
\Text(362.5,-49.5)[l]{\footnotesize{$H$}}
\Text(362.5,-89.5)[l]{\footnotesize{$\bar u^Z$}}
\Text(370,-70)[l]{$=-\frac\gw{2\cw}$}

\Line(20,-139.25)(40,-139.25)
\Line(20,-140.75)(40,-140.75)
\SetWidth{1.3}
\Line(60,-120)(40,-140)
\SetWidth{0.5}
\DashArrowLine(40,-140)(60,-160){1}

\Text(15,-148)[l]{\footnotesize{$\Omega^{H}$}}
\Text(62.5,-119.5)[l]{\footnotesize{$\phi^\pm$}}
\Text(62.5,-159.5)[l]{\footnotesize{$\bar u^\pm$}}
\Text(70,-140)[l]{$=\pm i\frac\gw2$}

\Line(170,-139.25)(190,-139.25)
\Line(170,-140.75)(190,-140.75)
\SetWidth{1.3}
\Line(210,-120)(190,-140)
\SetWidth{0.5}
\DashArrowLine(190,-140)(210,-160){1}

\Text(165,-148)[l]{\footnotesize{$\Omega^{H}$}}
\Text(212.5,-119.5)[l]{\footnotesize{$\chi$}}
\Text(212.5,-159.5)[l]{\footnotesize{$\bar u^Z$}}
\Text(220,-140)[l]{$=\frac\gw{2\cw}$}

\epi
\ece

\end{document}